\documentclass[preprint]{elsarticle}

\usepackage[margin=2.5cm]{geometry}
\usepackage{graphicx}
\usepackage{amssymb}
\usepackage{float}
\usepackage[switch]{lineno}
\usepackage{soul}
\usepackage{csquotes}
\usepackage{tabularx,booktabs}
\newcolumntype{C}{>{\centering\arraybackslash}X}
\biboptions{numbers,sort&compress}

\usepackage{svg}
\usepackage{amsmath}
\usepackage{dblfloatfix} 
\usepackage{wrapfig} 
\usepackage{nomencl}
\makenomenclature
\usepackage{etoolbox}
\renewcommand\nomgroup[1]{%
  \item[\bfseries
  \ifstrequal{#1}{A}{Subscript}{%
  \ifstrequal{#1}{B}{Abbreviations}{%
  \ifstrequal{#1}{C}{Variable}{}}}%
]}

\usepackage{adjustbox}  
\usepackage{multirow}   
\usepackage{url} 
\usepackage{eurosym}
\usepackage{amsfonts} 
\usepackage{dblfloatfix} 
\usepackage{hyperref} 

\usepackage[colorinlistoftodos]{todonotes}
\usepackage[utf8]{inputenc} 
\usepackage{multirow}
\usepackage{array}

\usepackage[nolist]{acronym} 			
\usepackage{subfigure} 

\usepackage[shortcuts]{extdash} 
\journal{TBD}

\begin{document}

\begin{frontmatter}

\title{PyPSA\=/Earth. A New Global Open Energy System Optimization Model Demonstrated in Africa.}

\author[uoe]{Maximilian~Parzen\corref{cor1}}
\ead{max.parzen@ed.ac.uk}
\author[eig]{Hazem~Abdel-Khalek}
\author[ir]{Ekaterina~Fedorova}
\author[uoe]{Matin~Mahmood}
\author[kit]{Martha~Maria~Frysztacki}
\author[jlug]{Johannes~Hampp}
\author[uoe]{Lukas~Franken}
\author[oth,tub]{Leon~Schumm}
\author[tub]{Fabian~Neumann}
\author[unipi]{Davide~Poli}
\author[uoe]{Aristides~Kiprakis}
\author[unipi]{Davide~Fioriti\corref{cor1}}
\ead{davide.fioriti@unipi.it}

\cortext[cor1]{Corresponding authors}
\address[uoe]{University of Edinburgh, Institute for Energy Systems, EH9 3DW Edinburgh, United Kingdom}
\address[eig]{Fraunhofer Research Institution for Energy Infrastructures and Geothermal Systems IEG, Cottbus, Germany}
\address[ir]{Kutuzovskaya 25 263, 143005, Odintsovo, Moscow region, Russia}
\address[jlug]{Justus-Liebig-University Gießen, Center for international Development and Environmental Research, Gießen, Germany}
\address[kit]{Karlsruhe Institute of Technology, Institute for Automation and Applied Informatics,  76344 Eggenstein-Leopoldshafen, Germany}
\address[unipi]{University of Pisa, Department of Energy, Systems, Territory and Construction Engineering, Largo Lucio Lazzarino, 56122 Pisa, Italy}
\address[tub]{Department of Digital Transformation in Energy Systems, Institute of Energy Technology, Technische Universität Berlin, Fakultät III, Einsteinufer 25 (TA 8), 10587 Berlin, Germany}
\address[oth]{Research Center on Energy Transmission and Storage (FENES), Faculty of Electrical and Information Technology, University of Applied Sciences (OTH) Regensburg, Seybothstrasse 2, 93053 Regensburg, Germany}



\begin{abstract}

Macro-energy system modelling is used by decision-makers to steer the global energy transition toward an affordable, sustainable and reliable future.
Closed-source models are the current standard for most policy and industry decisions. However, open models have proven to be competitive alternatives that promote science, robust technical analysis, collaboration and transparent policy decision making. Yet, two issues slow the adoption: open models are often designed with limited geographic scope, hindering synergies to collaborate, or are based on low spatially resolved data, limiting their use.
Here we introduce PyPSA\=/Earth, the first open-source global energy system model with data in high spatial and temporal resolution. It enables large-scale collaboration by providing a tool that can model the world energy system or any subset of it.
This work is derived from the European PyPSA\=/Eur model using new data and functions. It is suitable for operational as well as combined generation, storage and transmission expansion studies.
The model provides two main features: (1) customizable data extraction and preparation scripts with global coverage and (2) a PyPSA energy modelling framework integration. The data includes electricity demand, generation and medium to high-voltage networks from open sources, yet additional data can be further integrated. A broad range of clustering and grid meshing strategies help adapt the model to computational and practical needs.
A data validation for the entire African continent is performed and the optimization features are tested with a 2060 net-zero planning study for Nigeria. The demonstration shows that the presented developments can build a highly detailed energy system model for energy planning studies to support policy and technical decision-making.
We anticipate that PyPSA\=/Earth can represent an open reference model and we welcome to join forces to address the challenges of the energy transition together.



\end{abstract}



\begin{highlights}
\item New open energy system model with global coverage and high-resolution data
\item Create model-ready data and integrated energy planning study for any set of countries
\item Novel use of OpenStreetMap network and generation data
\item Africa and Nigeria-specific data validation and net-zero carbon emission scenarios
\item PyPSA meets Earth: research initiative to foster open data, tools and solvers
\end{highlights}

\begin{keyword}
Macro-Energy Systems \sep Optimization \sep OpenStreetMap \sep PyPSA\=/Earth \sep PyPSA-Africa \sep PyPSA meets Earth 
\end{keyword}
\end{frontmatter}



\tableofcontents

\section{Introduction}\label{sec:introduction}

\subsection{Motivation}

Energy system planning models are broadly adopted around the world. They are used as instruments to inform policy and investment decision-making, such as operational, supply diversification, and long-term infrastructure planning studies. Inscrutable ‘black-box’ models, even though being under criticism in academia~\cite{Pfenninger2017EnergyWorkings, Pfenninger2017TheBehind}, are still the standard for high-impact modelling such as the African Continental Power System Plan \cite{IRENA2021}. This prevents transparent decision-making while having other major drawbacks, as described in \cite{Pfenninger2017EnergyWorkings, Pfenninger2017TheBehind}.
Open-source models evolved to overcome these typical black-box model problems and can perform equivalent or even more tasks, but at no charge, while additionally supporting transparent and robust analyses \cite{Brown2018PyPSA:Analysis}. In many examples, the European Commission applies open tools and requests their use in funded projects, proving its belief in the benefits of openness and transparency \cite{EC2020a}. Now with the encouraging rise of more than 31 models in 2019 \cite{Groissbock2019AreUse}, simultaneously concerns of failed collaboration and duplication are arising that cost taxpayer money \cite{Juha2022}.
As a result, it becomes increasingly important to avoid duplication and provide modelling solutions that allow global united efforts. For these reasons, in this study, we propose an open-source community-backed flexible energy system model able to represent any arbitrarily large region of the world energy system in high spatial and temporal resolution that leverages other existing open-source projects to serve industry, policymakers, and researchers.

\subsection{Literature analysis}

In general, models are idealised representations of real physical systems. 
To ease building idealised systems, 'frameworks' have been developed to provide pre-compiled equations, algorithms, solver interfaces and/or input/output features. A framework becomes a 'model' only when data is added that describes real physical systems \cite{Pfenninger2018OpeningLearned}.
In this view, PyPSA is a framework and PyPSA\=/Eur and PyPSA\=/Earth are models for the European and Earth energy system, respectively. 
Nowadays, the open-source community is rich in energy system modelling frameworks that can provide similar functionalities. Table \ref{tab:model comparison} compares some available functionalities across selected widely-adopted energy system modelling frameworks \cite{Brown2018PyPSA:Analysis, Victoria2020EarlyOff, Neumann2022, Poncelet2020UnitFlexibility}. Undoubtedly, each developer team might be capable of filling missing features, but the functionality of the frameworks is only one important part of models, the other one, often even more relevant, is the integration of data.

\begin{table*}[!ht]
 \caption{Comparison of selected features for energy system modelling frameworks that are applied in Africa.}
\label{tab:model comparison}
\begin{tabularx}{\textwidth}{@{}l*{10}{C}c@{}}
Software & 
\rotatebox[origin=c]{90}{Version}  &
\rotatebox[origin=c]{90}{Citation}  &
\rotatebox[origin=c]{90}{Language}  &
\rotatebox[origin=c]{90}{Free and Open}  &
\rotatebox[origin=c]{90}{Power Flow}  &
\rotatebox[origin=c]{90}{Transport Model}  &
\rotatebox[origin=c]{90}{LOPF}  &
\rotatebox[origin=c]{90}{SCOPF}  &
\rotatebox[origin=c]{90}{Unit Commitment}  &
\rotatebox[origin=c]{90}{Sector-Coupling}  &
\rotatebox[origin=c]{90}{Pathway Optimization}
\\ \midrule
Calliope & v0.6.8 & \cite{Pfenninger2018}  & Python & \checkmark &   & \checkmark  &  &   & \checkmark & \checkmark \\ 
Dispa-SET & v2.4  & \cite{Dispa-SET}  & GAMS & \checkmark  &   & \checkmark &  &  &  \checkmark &  & \\
GridPath & v0.14.1 & \cite{gridpath} & Python & \checkmark &   & \checkmark & \checkmark &  &  \checkmark &  &  \checkmark \\
LEAP & 2020.1.63 & \cite{LEAPmodel} & NA$^b$ &  &   &  &  & &  & \checkmark &  \\
NEMO & v1.7 & \cite{NEMOmodel} & Julia & \checkmark & \checkmark & \checkmark & \checkmark  & & \checkmark \\
OSeMOSYS & 2022 & \cite{HOWELLS20115850} & GNU$^a$ & \checkmark &  & \checkmark &  & & & \checkmark & \checkmark \\
PLEXOS & 9 & \cite{PLEXOSmodel} & NA$^b$ &  &  & \checkmark & \checkmark & \checkmark & \checkmark & \checkmark & \checkmark \\
\textbf{PyPSA} & v0.20.0 & \cite{Brown2018PyPSA:Analysis} & Python & \checkmark & \checkmark & \checkmark & \checkmark & \checkmark &  \checkmark & \checkmark & \checkmark \\
SPLAT-MESSAGE & 2022 & \cite{SPLAT-MESSAGEmodel} & GAMS & & & \checkmark &  &  & & \\
TIMES & 2022 & \cite{TIMESmodel} & GAMS & & & \checkmark & \checkmark & & \checkmark & \checkmark & \checkmark\\
\bottomrule%
\multicolumn{12}{l}{$^a$ Is available in GNU Mathprog, Python and GAMS.}\\
\multicolumn{12}{l}{$^b$ NA = no information available.}
\end{tabularx}
\end{table*}

Existing models are often designed to implement data with limited geographical coverage, such as a specific province, country or continent \cite{Ringkjb2018ARenewables}. 
Continental models with implemented high-resolution data have proven to be the most maintained and active, possibly by covering many regions of interest and giving the user options for the aggregation level \cite{Horsch2018PyPSA-Eur:System, PyPSA-Eur2021GithubPyPSA-Eur}. In contrast, there are several examples where single-country models have soon become outdated, poorly documented or inactive \cite{pypsa-za, Schlott2020, Liu2018, Calliope-Kenya}. While global energy system models exist, they currently have several shortcomings. 'GlobalEnergyGIS' \cite{Mattsson2021AnRegions}, an open source model which can create energy system model data for any arbitrary region, is used in the 'Supergrid' model \cite{Reichenberg2022}, but misses network data or a  workflow management system that are important for flexible and reproducible data processing \cite{Kiviluoma2022}. Similarly, the GENeSYS-MOD model is a global open-source model, however, it is written in GAMS preventing free use and offers no data processing workflows \cite{genesys-mod-public}. Another promising candidate is the recently released OSeMOSYS-global model, which includes a workflow management system but misses network topology data as well as unit-commitment and power flow constraints that have been shown to strongly affect model results \cite{Barnes2022, Neumann2022}.

Similarly, existing PyPSA models are geographically limited. While PyPSA as a framework is adopted worldwide by many companies, non-profit organisations and universities (see example studies in \cite{PyPSAdocumenation}), there is no global model solution available yet. Providing a global energy system model ecosystem solution has the potential to unlock collaboration potentials that accelerate and improve the energy transition planning.

\subsection{Contributions}

In this paper, we present PyPSA\=/Earth, the first open-source global energy system model with data in high spatial and temporal resolution. Users can flexibly model the world or any subset of it with high spatial and temporal resolved data, including a GIS-tagged network representation based on OpenStreetMap. Using an automated workflow procedure it can (a) generate energy system model data and (b) perform energy planning studies. The novel contributions of PyPSA\=/Earth are detailed as follows:




\begin{enumerate}
\item New model creates arbitrary high spatial and temporal resolution representation of energy systems around the world
\item Automated workflow generates national, regional, continental or global model-ready data for energy planning studies based on open or optionally closed data
\item Integration and linking of multiple data sources and open-source tools to process raw data from multiple sources, e.g. OpenStreetMap
\item Provision of new spatial clustering strategies to simplify the high-resolution model
\item Data and model validation for the African continent and Nigeria
\item Development of 2060 net-zero energy planning study for Nigeria
\end{enumerate}

Following the open-source spirit, the model is not built from scratch but derived from the popular European transmission system model -- PyPSA\=/Eur \cite{Horsch2018PyPSA-Eur:System}. Novel additions that have not been part of PyPSA-Eur include the use and integration of global data by default (e.g. demand, generators, transmission grid) and building a meshed electricity network from OpenStreetMap. New features in PyPSA-Earth that can be optionally activated are the use of administrative zones instead of network-derived zones, consideration of DC meshed grids, and the automatic creation of line expansion options. 

In this paper, the PyPSA\=/Earth model is presented and discussed with a focus on the African continent. This includes the data and model validation for the African continent and Nigeria. Finally, a 2060 net-zero energy planning study for Nigeria's electricity sector demonstrates the optimization capabilities.

All code and validation scripts are shared open-source under GPL 3.0 license. The data, often extracted by python script activation, is available under multiple open licenses. For a detailed license listing, see \cite{pypsaafrica}. 


\subsection{Organization of the paper}

The rest of the paper is organized as follows. Section \ref{sec:methodology} introduces the novel PyPSA\=/Earth model that is able to perform large-scale energy system modelling studies. The data processing novelties are described in detail in Section \ref{sec:data_and_methods}. Data validation for the African continent is performed in Section \ref{sec: validation}, and a quantitative case study on Nigeria is discussed in Section \ref{sec: Nigeria case study}. Finally, the limitations of the model are discussed, and the conclusions are drawn.

\section{PyPSA\=/Earth model}\label{sec:methodology}

This section describes the scope of the PyPSA\=/Earth model, its features as well as the role of the initiative that is facilitating the model developments. 

\subsection{Scope}

The PyPSA\=/Earth model is a novel open-source data management and optimization tool that aims to provide policymakers, companies and researchers with a shared platform for a wide range of macro-energy system analyses needed to achieve the energy transition together. The option to create a tailored country, continental or global model under a unique code repository maximises synergies and wider user-benefits. For instance, one user in Africa can implement new features and data, improve the documentation or implement bug fixes that immediately benefit all other users around the world. 


Studies that were already demonstrated in the PyPSA ecosystem \cite{PyPSAdocumenation}, and are now globally available include:

\begin{enumerate}
    \item energy system transition studies
    \item power system studies
    \item technology evaluation studies (e.g. energy storage, synthetic fuels and hydrogen pipelines)
    \item technology phase-out plans (e.g. coal and nuclear)
    \item supply diversification studies
    \item electricity market simulations
\end{enumerate}

The PyPSA\=/Earth model currently focuses on the power system and does not yet provide capabilities for sector coupling studies such as co-optimization of e.g. power, heat, transport and industry sectors. However, these features are in active development building on top of PyPSA\=/Eur\=/Sec~\cite{Brown2018SynergiesSystem}.

\subsection{Features}
\label{sec: pypsa earth features}




The following features are implemented in PyPSA\=/Earth:
\begin{enumerate}
    \item flexible model scope: from Earth to any subregion
    \item high temporal and spatial resolution
    \item model-ready data creation
    \item co-optimization of investment and operation
    \item single or multi-year optimization
    \item flexible addition of arbitrary optimization constraints, e.g. socio-economic, technical, or economic
\end{enumerate}


Moreover, the PyPSA\=/Earth model has been developed with the following non-functional requirements:
\begin{enumerate}
    \item easy to use and learn
    \item highly customizable and flexible
    \item modular to include new features and data
    \item fully reproducible
\end{enumerate}

The proposed features of PyPSA\=/Earth are a novelty as compared to the literature in Section \ref{sec:introduction}. Furthermore, new features can be created in or adopted from other PyPSA-based models that share a similar backbone. Examples are the work on endogenous learning with pathway optimization and multiple investment periods \cite{ZeyenPathways2022}, dynamic line rating constraints based on spatially differing environmental conditions \cite{Glaum2022}, the implementation of generic constraint settings that enable equity constraints such as applied in \cite{Neumann2020CostsSystem} and uncertainty analyses by input parameter sweeps or by exploring the near-optimal solution space \cite{Neumann2021BroadNear-Optimality}.

The data and methods Section \ref{sec:data_and_methods} presents more details on the presented features.

\begin{figure}[h!]
\centering
\includegraphics[width=1\textwidth]{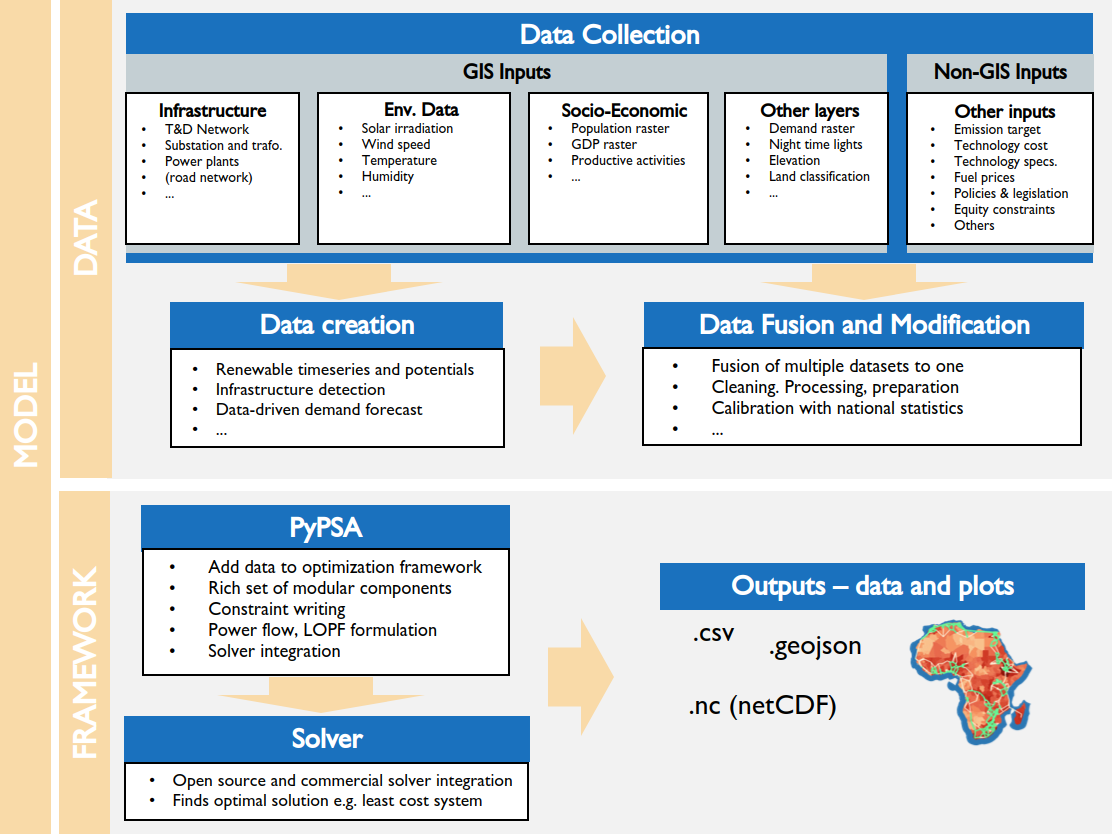}
\caption{PyPSA\=/Earth model design. After providing the configuration parameters and countries of interest, data is collected and processed to be then fed into the PyPSA model framework which enables to perform the desired optimization studies such as least-cost system transition scenarios.}
\label{fig:high level model design}
\end{figure}

\subsection{PyPSA meets Earth initiative}\label{sec:initiative}



The PyPSA meets Earth initiative is an independent research initiative that aims to improve energy system planning with open solutions. It supports, builds and maintains the PyPSA\=/Earth model and is therefore briefly introduced. The initiative's vision is to support transparent and debatable decision-making on the energy matter that cannot be achieved with the status quo ruled by commercial inscrutable closed-source \enquote{black-box} tools. Current research activities in the initiative can be categorised into three distinct groups:



\begin{itemize}
    \item open data
    \item open energy system model and
    \item open source solver 
\end{itemize}

First, the open data activities focus on open data creation, collection, fusion, modification, prediction and validation for energy system models. These data activities are not limited to aggregated country information but prioritise work on high spatial and temporal resolution data, which is fundamental for scalable and accurate mini-grid and macro-energy system model solutions. Second, the open energy system modelling activities focus on implementing new functions and data streams into the model, such as building a sector-coupled model with multi-horizon optimization that is useful across the globe. Third, open source solver-related activities deal with benchmarks and efficient interfaces that help to adopt and develop open source solvers.
For instance, we created a benchmark that became a successful public funding proposal and attracted sufficient funding for the open-source solver HiGHS \cite{Parzen2022OptimizationSolvers}
. This activity pushes breakthroughs in large-scale optimization performance required for energy system models, which were until now reserved only for people that can afford commercial proprietary solvers.


In order to assure a continuous inflow of people that maintain, improve and use the software, as needed by open-source software \cite{Steinmacher2019}, the initiative supports a free and open community where anyone can contribute. The initiative adopts:
\begin{itemize}
    \item \textit{GitHub} to publicly record issues, requests, solutions or source code-based discussions
    \item \textit{Discord} as a voice channel and messaging social platform for regular public meetings and exchanges
    \item \textit{Google Drive} to publicly store files and meeting notes
\end{itemize}
Together, these tools provide the backbone of the open community supporting the initiative goals and activities (data, model, solver).





\section{Data and methods}\label{sec:data_and_methods}

In this section, the PyPSA\=/Earth methodology sketched in Figure \ref{fig:high level model design} is described in detail. First, we introduce the workflow management tool that supports the model user experience. Then, data creation and processing approaches are discussed considering the main data blocks used by the PyPSA\=/Earth model: power grid topology and spatial shapes, electricity demand, renewable potential and power plant locations. Further, we describe some advanced pre-processing techniques such as clustering and line augmentation used to introduce data into the model in a robust and efficient way. Finally, we describe the energy system modelling and optimization framework with its solver interfaces. 

\subsection{Workflow management tool}\label{subsec: workflow}

\begin{figure}[h!]
\centering
\includegraphics[trim={0cm 0cm 0cm 0cm}, clip, width=1\textwidth]{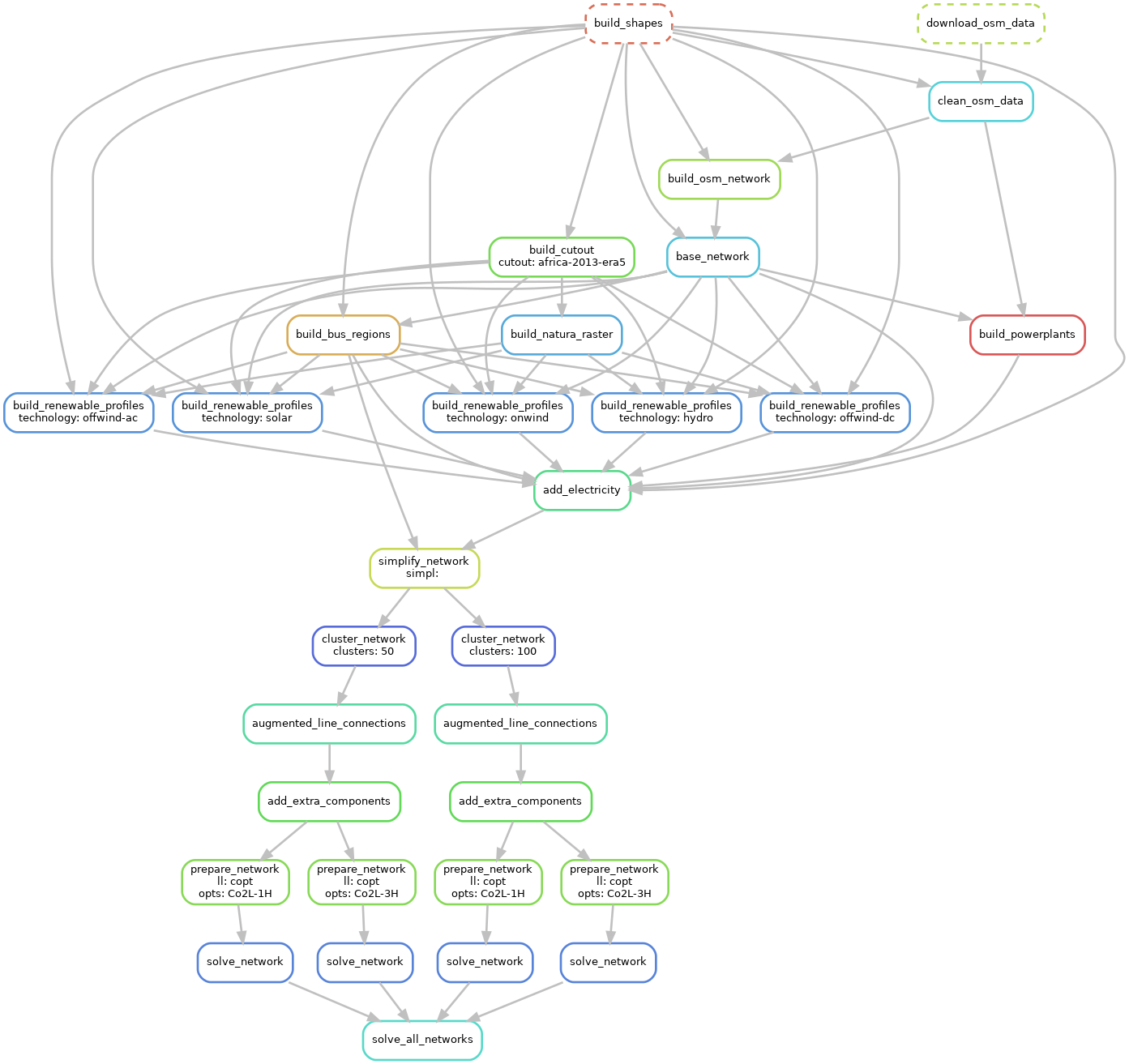}
\caption{Directed acyclic graph of a PyPSA\=/Earth example workflow. Each box represents a rule that modifies or creates the input, resulting in outputs. Already computed rules are shown as boxes with dashed borders, solid bordered rules indicate rules yet to be computed during the workflow execution. The workflow creates here four least-cost scenarios with varying spatial and temporal resolution and is triggered by a single line of code. Wildcards define options for the spatial resolution with 50 and 100 clusters and the temporal resolution with 1 and 3 hours (Co2L-1H and Co2l-3H).}
\label{fig: workflow}
\end{figure}

First of all, inherited from PyPSA\=/Eur \cite{Horsch2018PyPSA-Eur:System}, PyPSA\=/Earth relies on the 'Snakemake' workflow management \cite{Koster2012} that decomposes a large software process into a set of subtasks, or 'rules', that are automatically chained to obtain the desired output. Accordingly, 'Snakemake' helps sustainable software design that enables reproducible, adaptable and transparent science, as described in \cite{Moelder2021}.
For example, Figure \ref{fig: workflow} represents a workflow of PyPSA\=/Earth automatically created by 'Snakemake' for which the user can execute any part of the workflow with a single line of code. 
The implications of such design on the energy system model are described in the following.

Starting with essential usability features, the implemented Snakemake procedure enables the user to flexibly execute the entire workflow with various options without writing a single line of code. For instance, the user can model the world energy system or any subset of countries only using the required data. Wildcards, which are special generic keys that can assume multiple values depending on the configuration options, help to execute large workflows with parameter sweeps and various options. Examples are shown in Figure \ref{fig: workflow}, where the function 'build\_renewable\_profile' is executed with the wildcards 'offwind-ac', 'solar', 'onwind', 'hydro' and 'offwind-dc' to estimate renewable potentials and create synthetic time series for renewable feed-in availability by technology. Lastly, workflows allow easy scaling by parallel execution. A few more benefits also exist from the developer's perspective. For instance, the modular design of rules allows parallel code development and deployment with continuous integration tests of the complete workflow that guarantee the stability of PyPSA\=/Earth. 

\subsection{Network topology and model}

The electricity network topology is one of the main inputs needed to build an energy system model which accounts for realistic power flow approximations across regions. The most comprehensive and accurate data on power grids are curated by the transmission system operator. 
In practice, the availability of open power grid data is still relatively low for many parts of the world, with the situation in Africa being extremely sparse. 


A natural way to address the lack of power grid data provided by transmission system operators' is to utilize open geospatial datasets. Currently, a few open source packages have been published to extract and build networks from such datasets (e.g. Gridkit \cite{Medjroubi2014}, Transnet \cite{Rivera2018}, SciGrid \cite{Medjroubi2014}). However, each of these packages focuses on applications for a particular world region rather than on the global coverage and there is still no ready-to-use solution which could be implemented into a global model. To fill this gap, we have developed an original approach which reconstructs the network topology by relying solely on open globally-available data. The developed approach is based on the OpenStreetMap (OSM) datasets that are a crowd-sourced collection of geographic information, which is daily updated and includes geolocation references \cite{OpenStreetMap}.

\begin{figure}[h!]
    \centering
    \subfigure[]{\includegraphics[width=\textwidth]{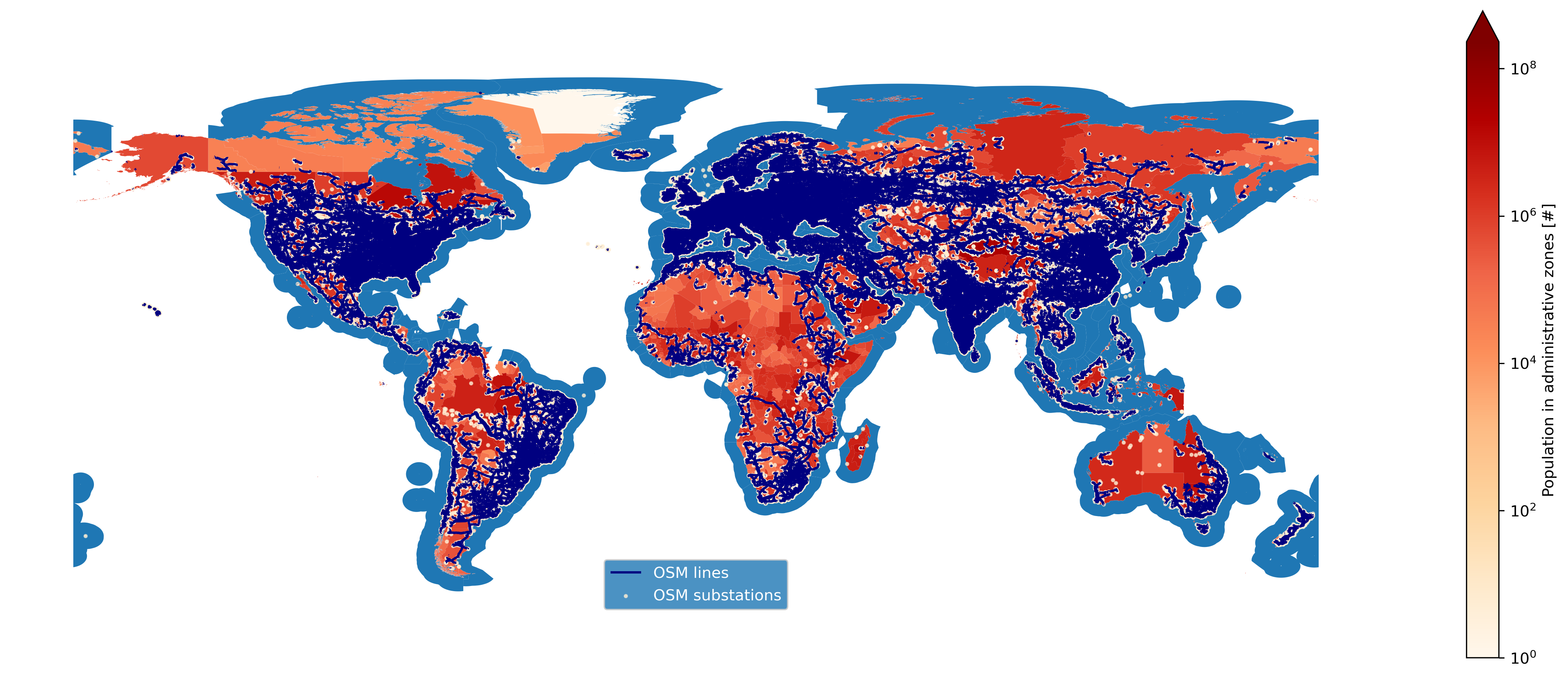}} 
    \subfigure[]{\includegraphics[width=0.48\textwidth, trim={0cm 0cm 3.4cm 0cm}, clip]{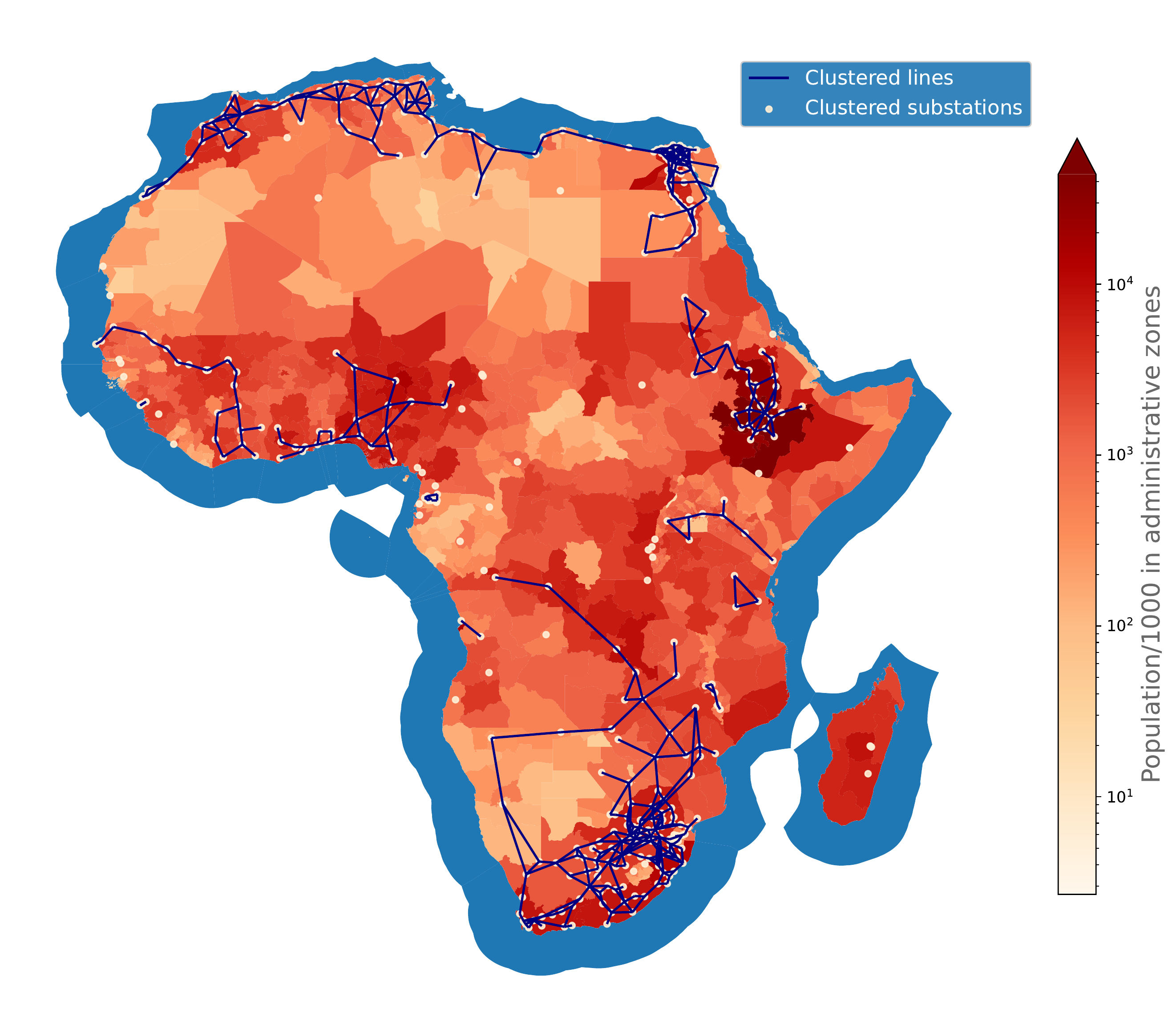}}
    \subfigure[]{\includegraphics[width=0.48\textwidth, trim={0cm 0cm 3.0cm 0cm}, clip]{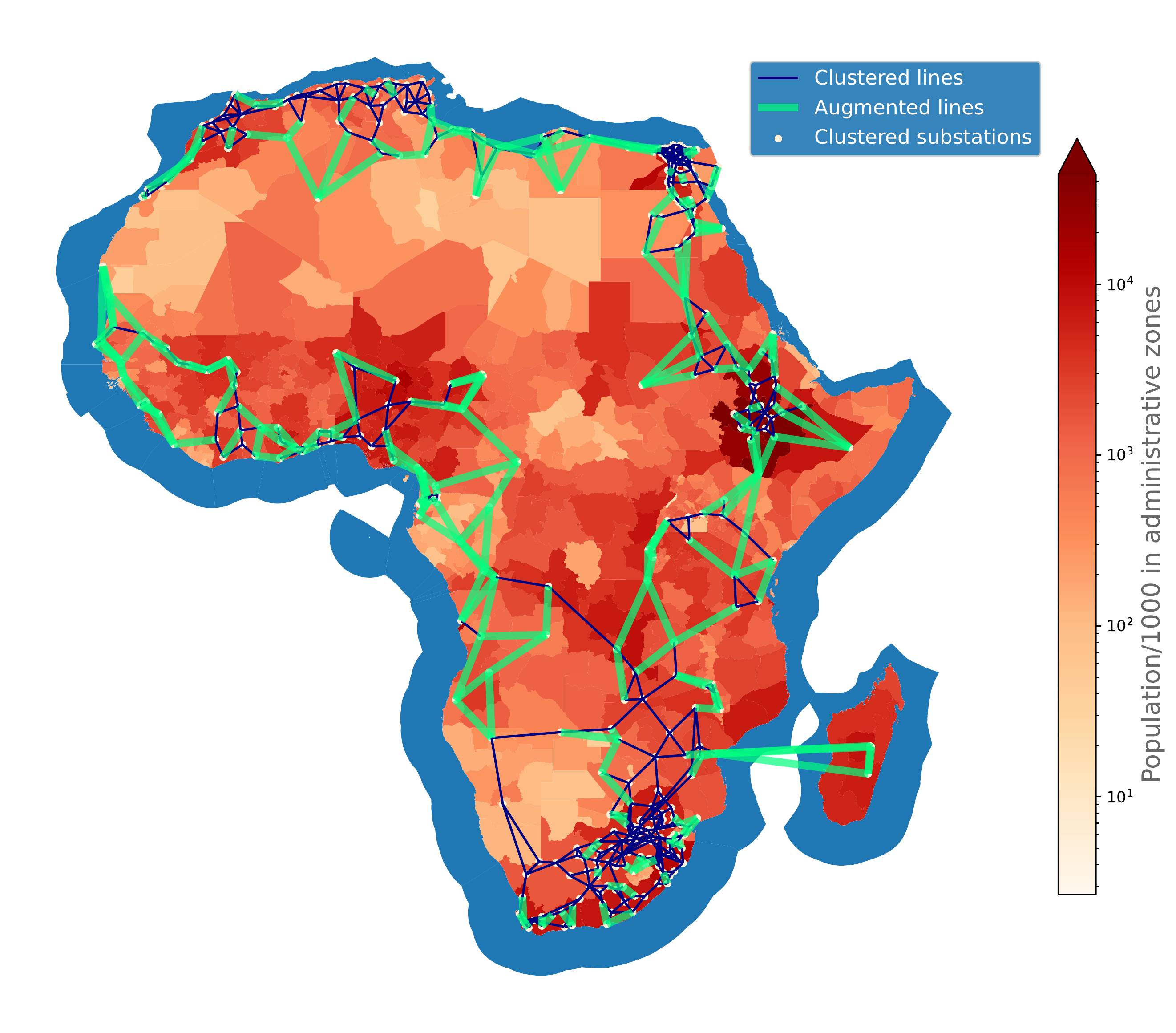}}
    \caption{Representation of transmission networks and shapes produced by PyPSA\=/Earth show: (a) a sample Open Street Map transmission network, (b) a clustered 420 node African transmission network and (c) its augmented version with additional line connections to test the benefits of additional interconnections. In case of c), the applied k-edge augmentation guarantees that every node has at least a certain number of connections, three in the case of the figure. The augmented lines are connected by a minimum spanning tree algorithm to the nearest neighbour.}
    \label{fig:clustered and augmented meshed network}
\end{figure}

The electricity network topology is created in three novel steps: i) downloading, ii) filtering and cleaning the data, and iii) building a meshed network dataset with transformer, substation, converter and high voltage alternating current (HVAC) as well as high voltage direct current (HVDC) components. Figure \ref{fig:clustered and augmented meshed network} shows sample raw and cleaned networks along with the options for clustering and line augmentation that are introduced in Section \ref{subsec: clustering} and \ref{subsec:augmented line connections}, respectively.

The following explains the most critical steps to building the electricity network topology and model.
First, the model implements an interface to download the OSM database. From this interface, substations and network data of any available voltage level are retrieved with essential operating details of the infrastructure. The interface thereby runs with the \textit{esy-osm} tool that allows fast retrieval of OSM data through multi-threaded processing \cite{esy-osm}.


To select only power-related data, the raw OSM database is filtered using the "power" key. By doing so, electrical substations, generators, high-voltage transmission lines and underground cables information can be obtained. A further data cleaning procedure is designed to restore the missed pieces of information and ensure that the overall data quality conforms with the requirements of the PyPSA modelling framework. In particular, the power infrastructure elements are being filtered by the validity of their geometry and the desired voltage level. Available data on the electrical overhead and underground cables are used to improve the representation of the high-voltage transmission lines, while the frequency values are applied to identify the HVDC transmission systems.

The cleaned dataset is pre-processed to build the network topology. The pre-processing includes:

\begin{enumerate}
    \item extracting additional data on electrical substations from the transmission line dataset,
    \item cleaning duplicated records,
    \item splitting substations by the voltage level to represent each voltage by a separate substation,
    \item deriving data on transformers and AC/DC converters from the information on the transmission lines.
\end{enumerate}

Beyond that, an approach has been developed to improve the quality of the OSM-extracted grid topology: substations within a given tolerance (few kilometres) from power transmission lines with the same voltage level are merged and connected to the power lines. Otherwise, the network may result in incomplete and unexpected distortions because of small missing information, such as short underground cables connecting the last mile of a transmission line with a transformer in a substation. 

\subsection{Fundamental shapes}\label{subsec: fundamental shape}

\begin{figure}[h!]
\centering
\subfigure[]{\includegraphics[width=0.32\textwidth]{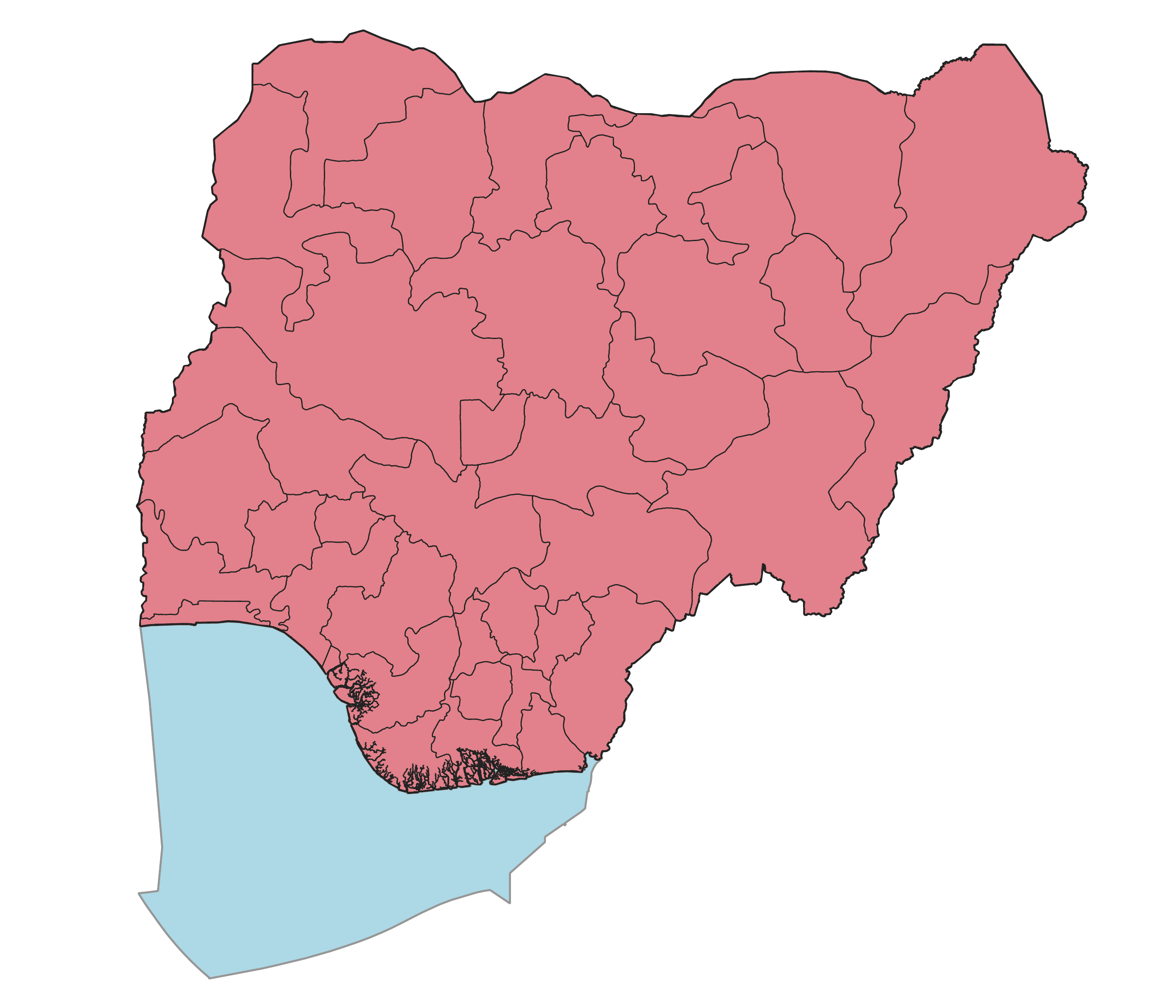}} 
\subfigure[]{\includegraphics[width=0.32\textwidth]{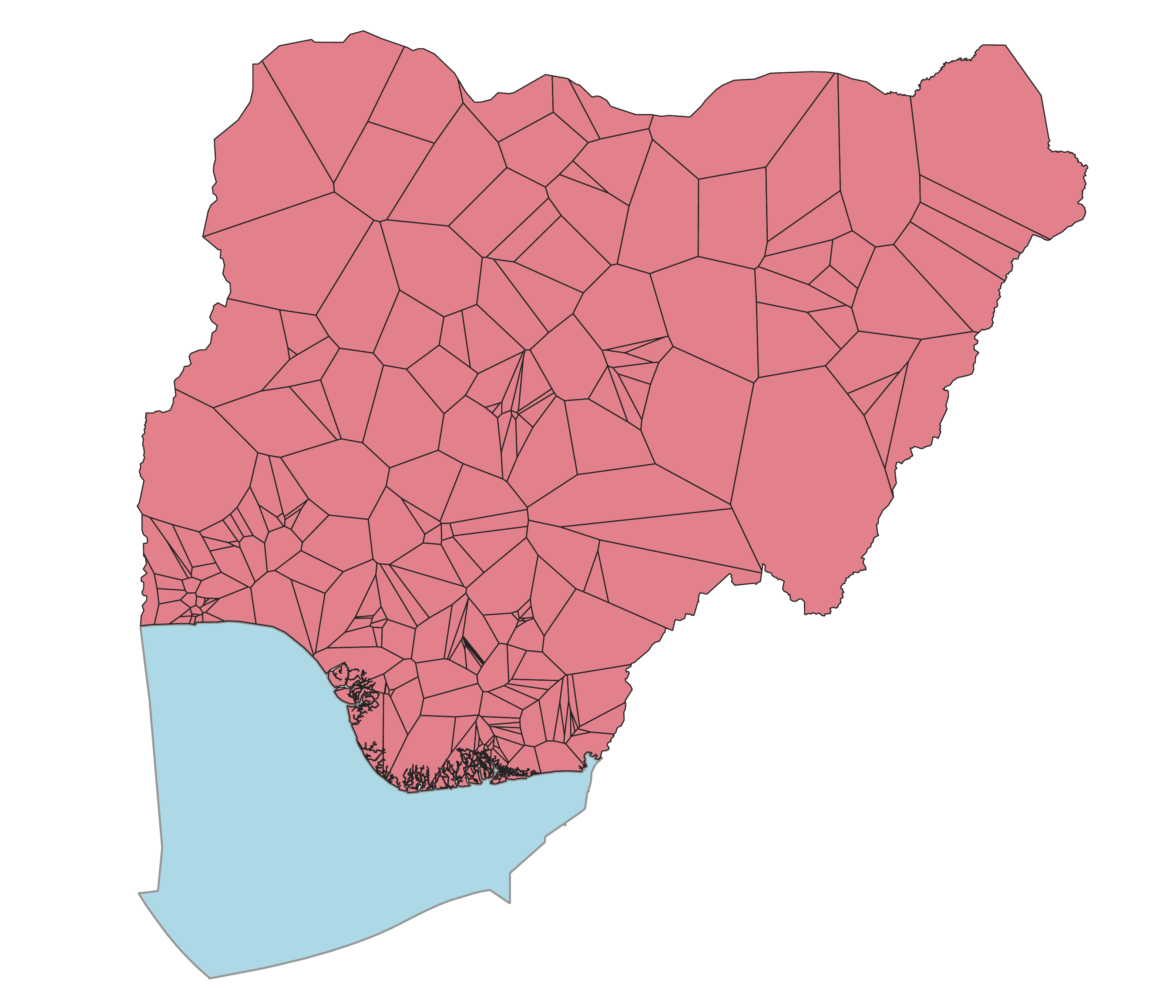}} 
\subfigure[]{\includegraphics[width=0.32\textwidth]{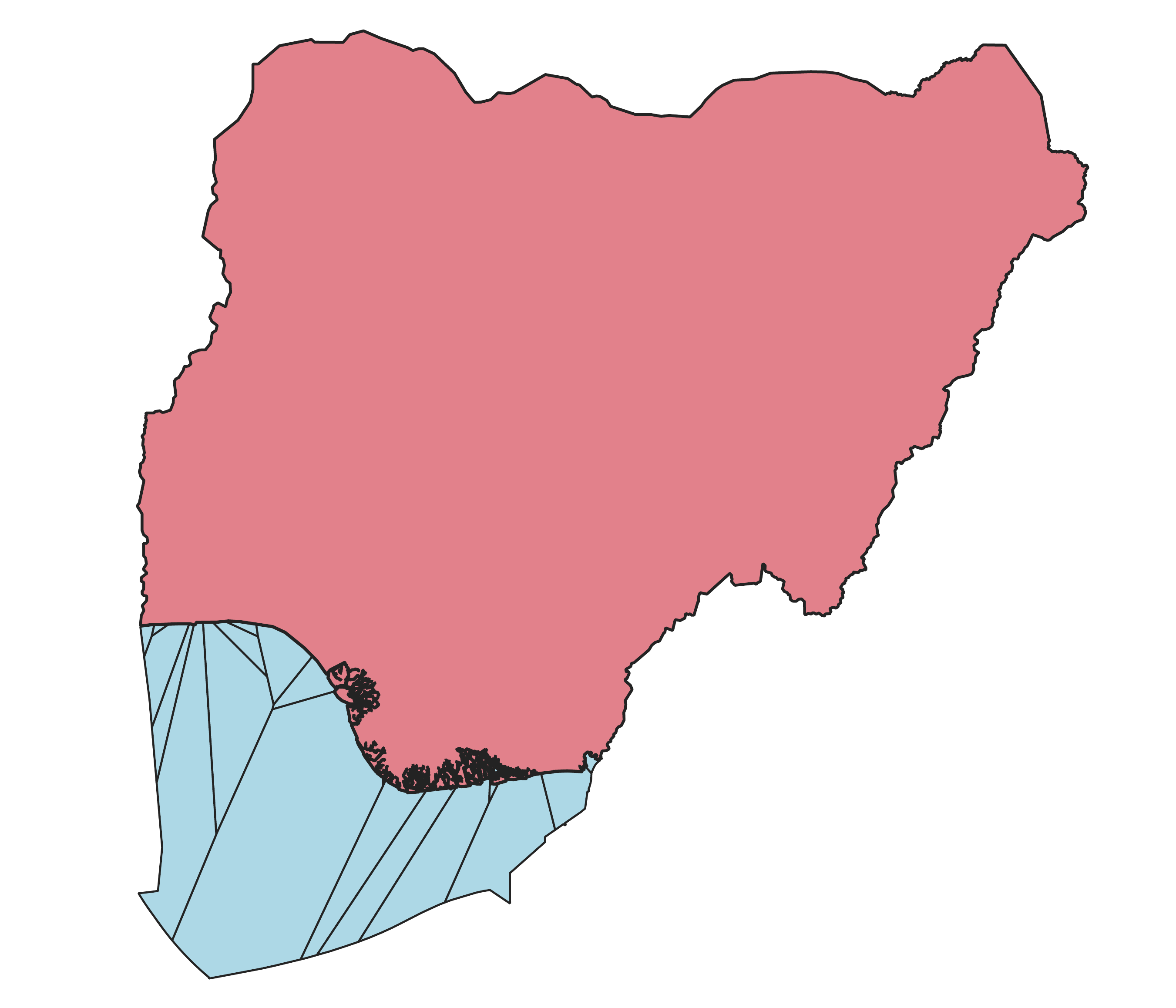}}
\caption{Fundamental shapes of Nigeria in PyPSA\=/Earth: (a) shows the onshore regions represented by the GADM zones at level 1, (b) shows the onshore regions represented by Voronoi cells that are derived from the network structure, and (c) shows the offshore regions also represented by Voronoi cells based on the closest onshore nodes.}
\label{fig:fundamental data shape}
\end{figure}

    
    Fundamental shapes represent the smallest defined regions that gather various data types to characterise the energy system (see Figure \ref{fig:fundamental data shape}). Before being ready for the model-framework execution, data is often provided in many different ways. For instance, it can be represented as geo-referenced point location for power plants, raster in many different resolutions such for population and GDP zones or lines for country boundaries. The data needs to be combined so that each fundamental shape represents a network node from which meshed energy systems can be built. These shapes are used multiple times across the workflow. For instance, one can apply network meshing strategies and clustering approaches, as illustrated in Figure \ref{fig:clustered and augmented meshed network}, to explore or reduce the complexity of optimization problems depending on the user requirements. Another example is to use the fundamental shapes to spatially distribute the national electricity demand or as catchment areas to aggregate renewable sources (see Figure \ref{fig:atlite demonstration}).
    
    
    
    For onshore regions, the model provides two ways to build fundamental data shapes. The first retrieves the so-called Global Administrative Areas (GADM) that represent administrative zones at various levels of detail (e.g. national, regional, province, municipality) \cite{gadm2022}. The second one uses the substation GIS location to create Voronoi partitioned areas for each substation, which boundary is defined as equidistant to the centroid of the nearest sites \cite{Frysztacki2021TheSolar}. The latter approach is beneficial to replicating the network accurately, while the former helps communicate results.
    
 
   For offshore regions, the model uses only Voronoi partitioned areas to create fundamental shapes. These Voronoi areas are built from high voltage onshore nodes and are limited to the offshore extent by the Maritime Boundaries and Exclusive Economic Zones (EEZ) data for each country \cite{marineregions2019}. The same Voronoi approach is applied to the offshore zones splitting them into separate regions and linking each to the closest onshore shape.
   


\subsection{Electricity consumption and prediction}\label{subsec: demand}


The model currently provides globally hourly demand predictions for 2030, 2040, 2050 and 2100. Each year includes demand predictions for the ERA5 reanalysis weather years of 2011, 2013 and 2018 which ERA5 provides in principle between 1959-2021. The demand time series is created using the new contributed \textit{synde} package~\cite{Synde2022} which implements a workflow management system to extract the demand data created with the open source Global-Energy GIS (GEGIS) package \cite{Mattsson2021AnRegions}.

In principle, GEGIS produces hourly demand time series by applying machine learning methods \cite{Mattsson2021AnRegions}. It uses information such as temperature profiles, population, GDP, and predicted values from the 'Shared Socioeconomic Pathways' \cite{SSP2017} to forecast the annual and hourly demand over the following decades. This approach is not new as it was already applied and tested in \cite{TOKTAROVA2019160}. 
The observed absolute error of GEGIS in the validation test is considered acceptable for energy studies as it is 8\% across 44 countries, yet with generally worse performance in low-income countries \cite{Mattsson2021AnRegions}. 

\begin{figure}[h!]
\centering
\includegraphics[trim={0cm 0cm 0cm 0cm}, clip, width=0.99\textwidth]{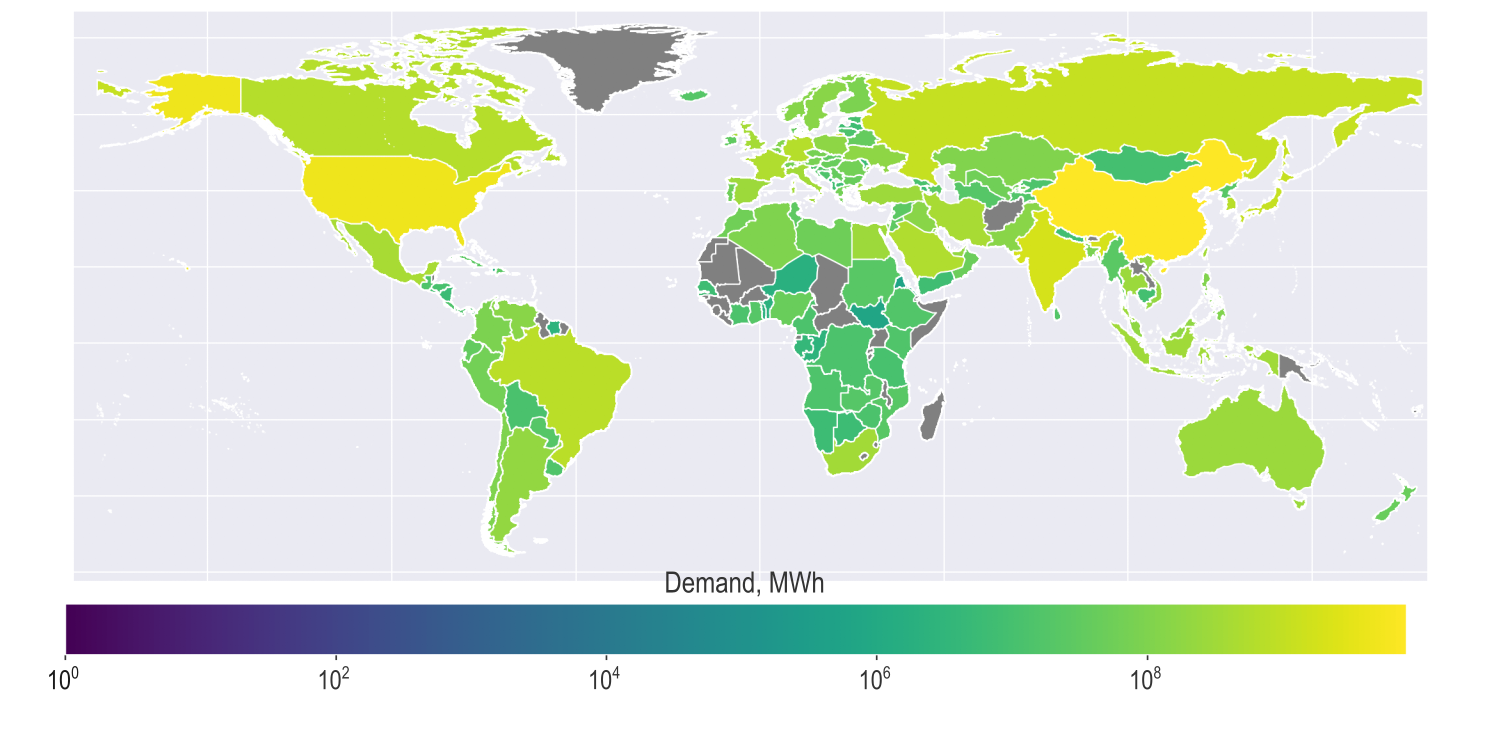}
\caption{
    Demand predictions created per country using the \textit{synde} workflow based on \textit{GEGIS}.
    For grey-coloured countries, synde does not provide data, however, a heuristic creates representative time series as described in Section \ref{subsec: demand}.
}
\label{fig:demand prediction coverage}
\end{figure}

The coverage of the \textit{synde} package is currently limited. Figure \ref{fig:demand prediction coverage} shows that there are no data outputs for especially low-demand countries. A heuristic creates data for the countries with missing data by scaling the Nigerian demand time series proportionally to population and GDP. We validate this approach in Section \ref{subsec: demand validation}.


\subsection{Renewable Energy Sources}\label{subsec: renewable}

\begin{figure}[!h]
    \centering
    \subfigure[]{\includegraphics[width=0.25\textwidth]{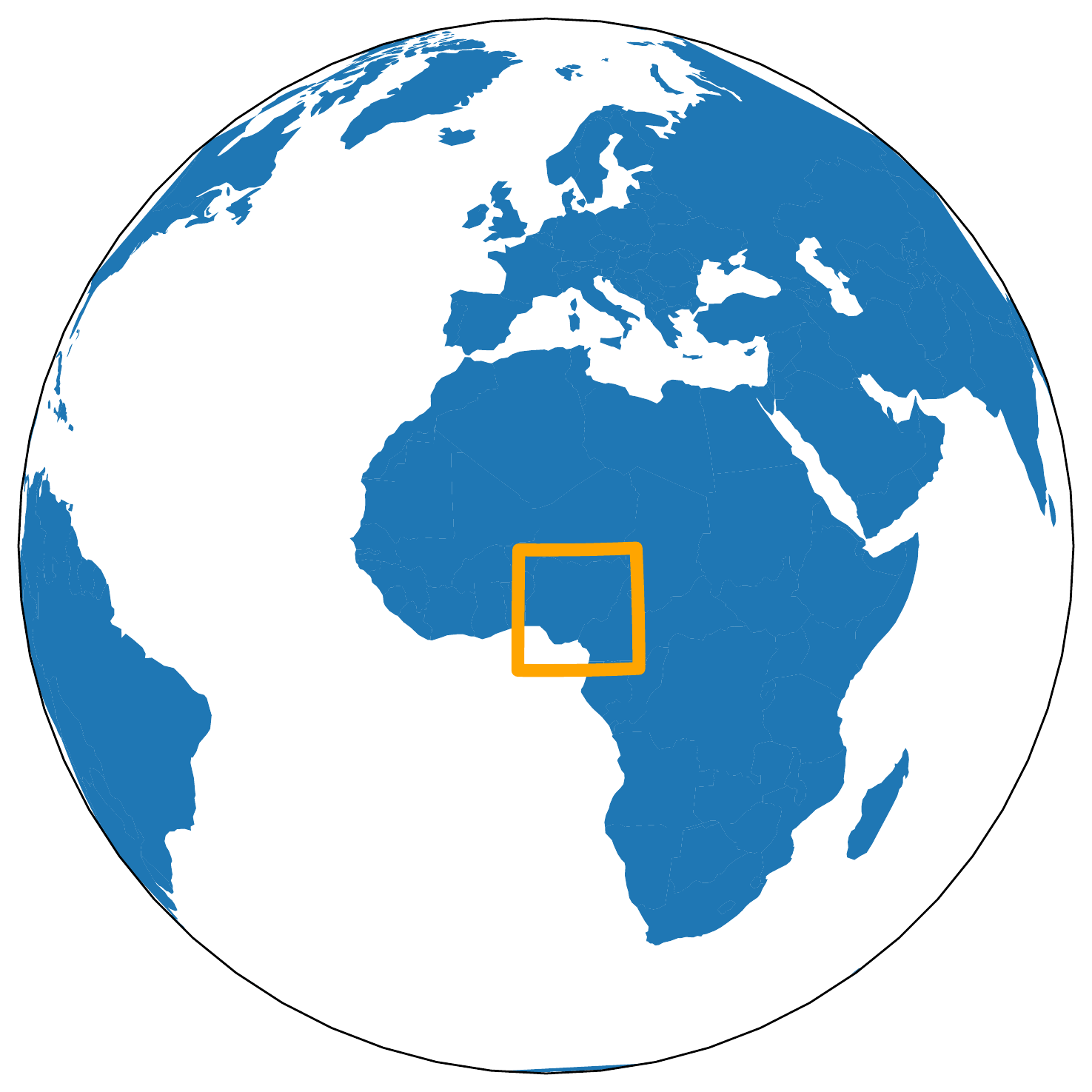}} 
    \subfigure[]{\includegraphics[width=0.39\textwidth]{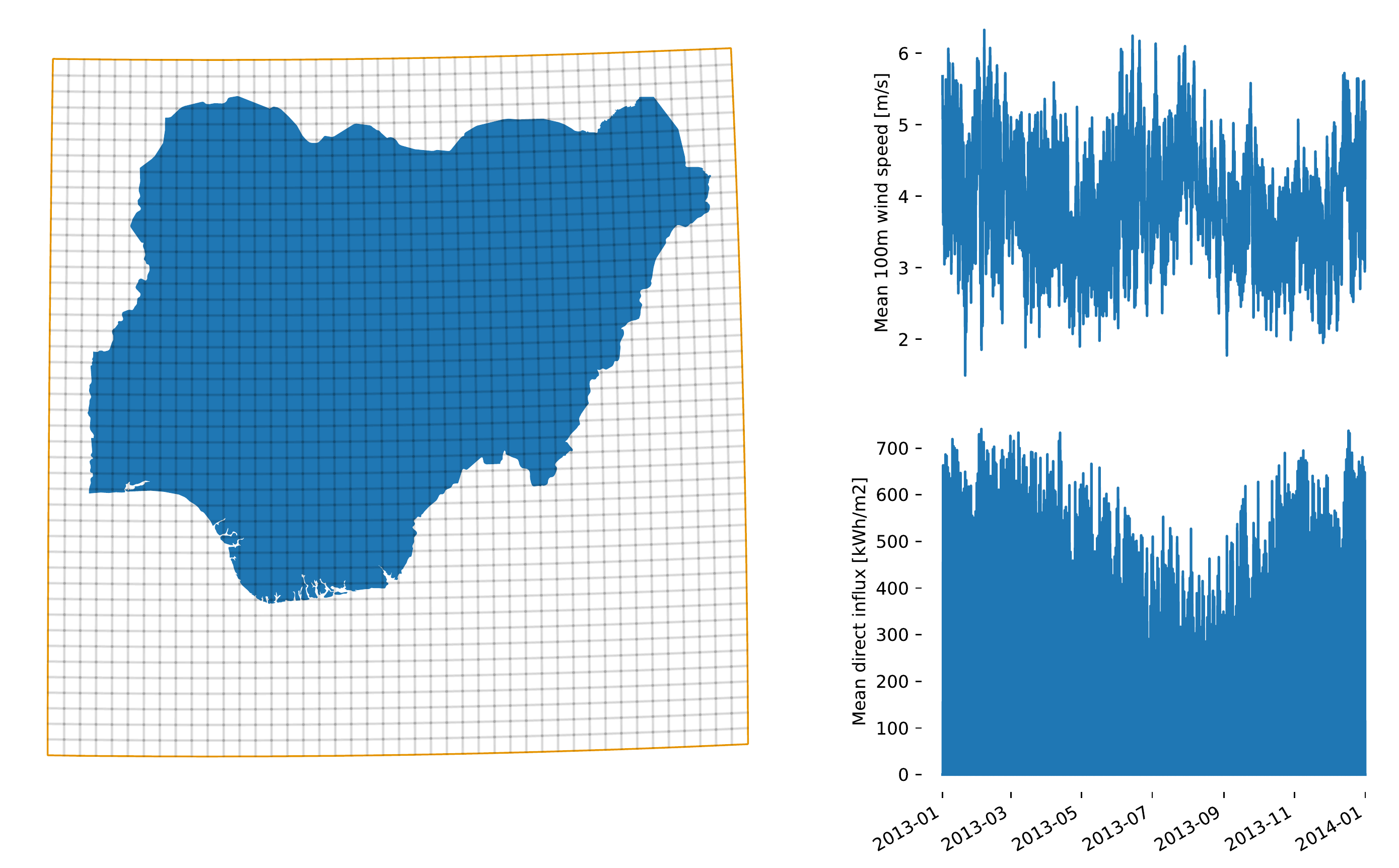}} 
    \subfigure[]{\includegraphics[width=0.32\textwidth]{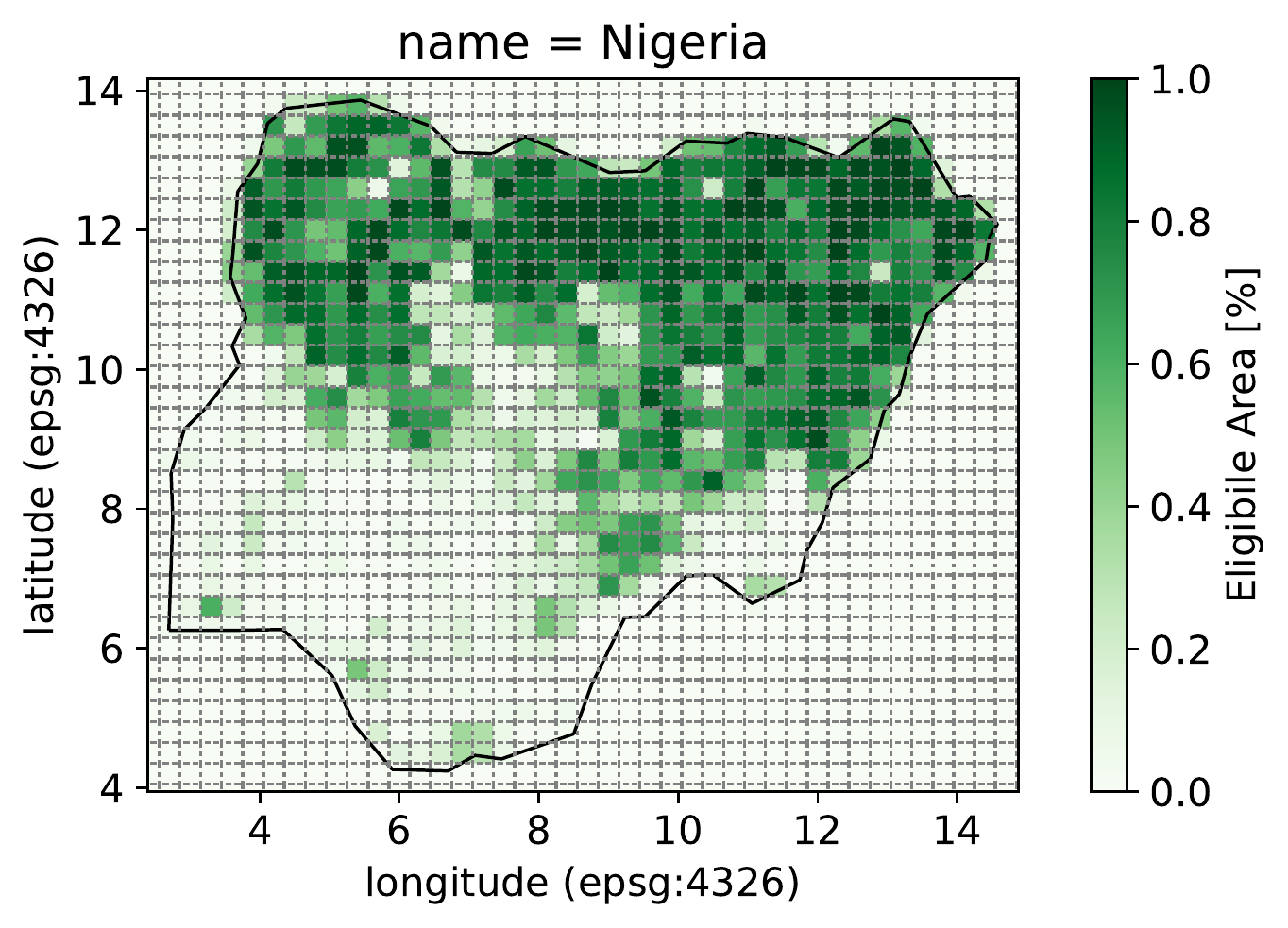}}
    \subfigure[]{\includegraphics[width=0.32\textwidth]{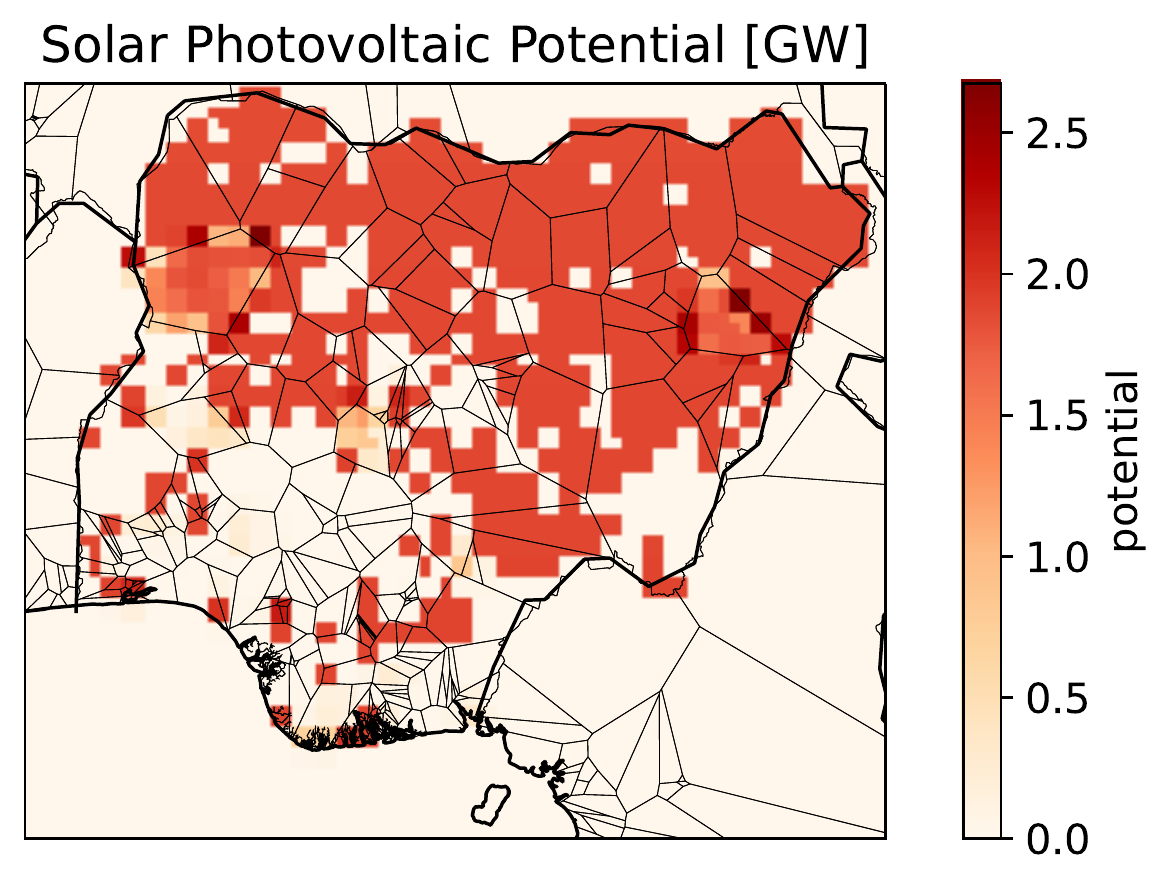}}
    \subfigure[]{\includegraphics[width=0.32\textwidth]{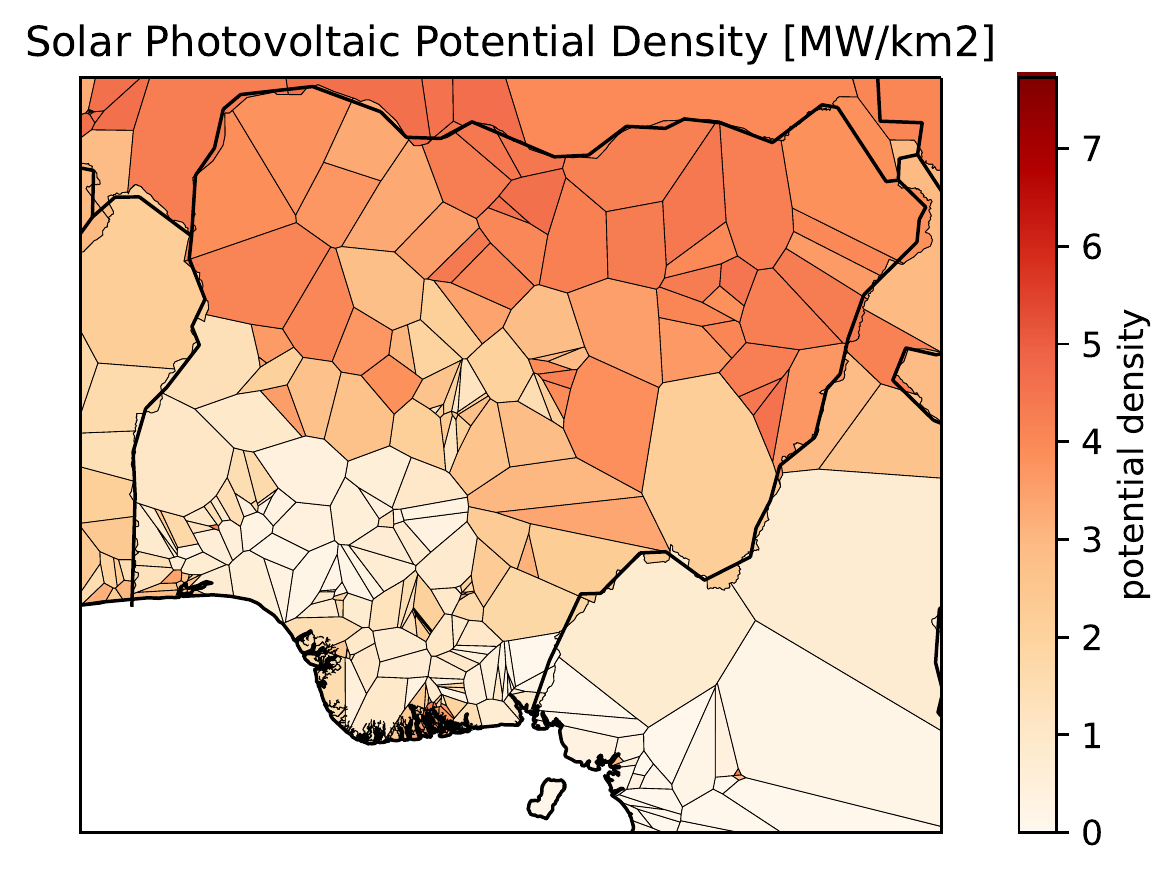}}
    \subfigure[]{\includegraphics[width=0.32\textwidth]{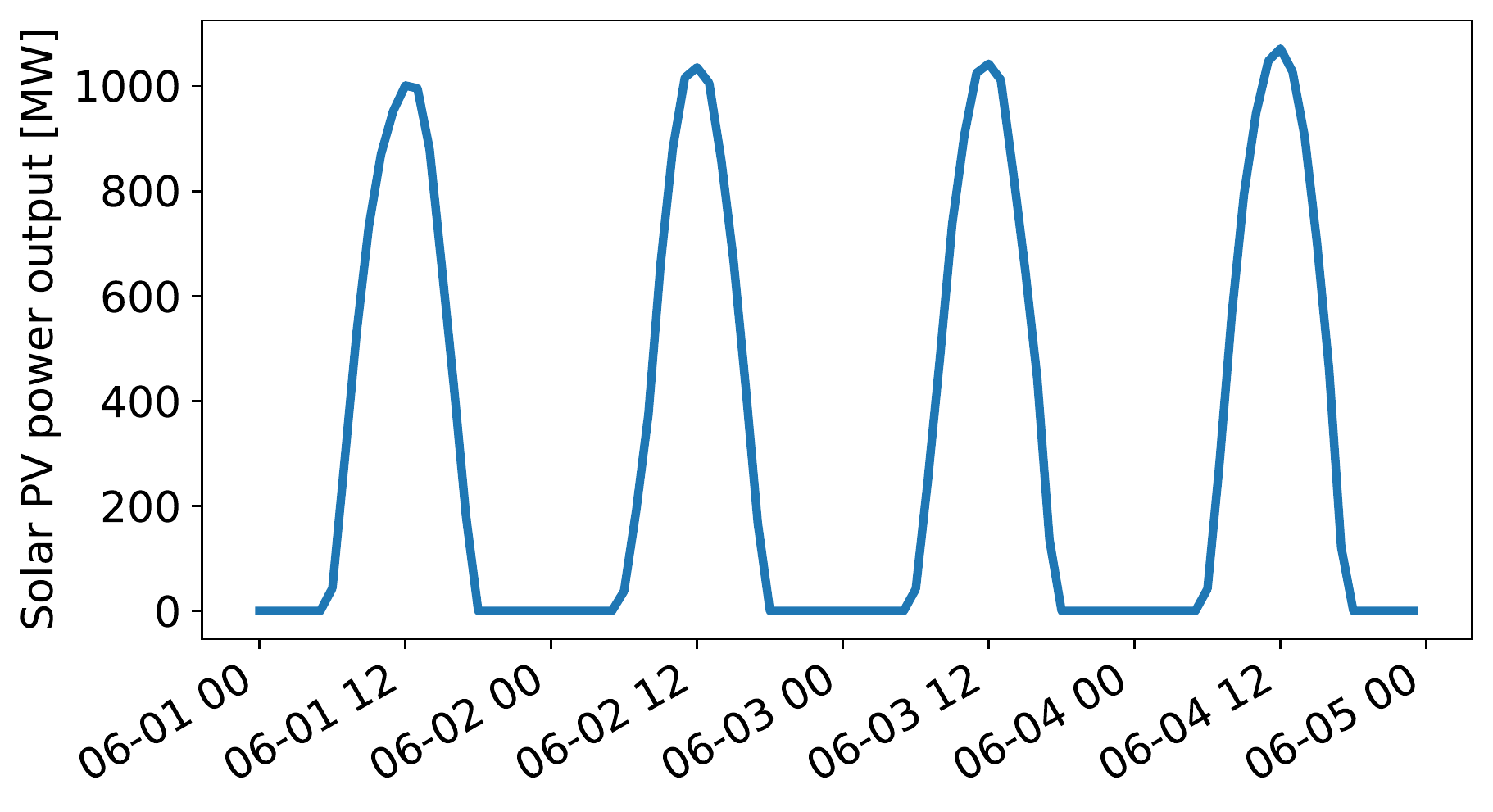}}
    \caption{A concept demonstration of Atlite for Nigeria. (a) Shows that environmental and weather data is extracted in a \textit{cutout} for the region of interest. (b) The cutout is split in a raster of $(0.25^{\circ})^2$ or roughly $(27.5km)^2$ (length varies along latitude), whereby each cell contains static or hourly time series data. The example wind speed and direct irradiation influx time series are shown for one cutout cell that contains an \textit{ERA5} extract of the Copernicus Data Store \cite{Hersbach2020}. (c) Shows the eligible area raster, which is built by excluding protected and reserved areas recorded in \textit{protectedplanet.net} and excluding specific land-cover types from \textit{Copernicus Global Land Service} whose eligibility can vary depending on the technology. (d) Illustrates the maximal installable power raster, which is calculated by the eligible area and the socio-technical power density of a technology e.g. $4.6MW/km^2$ for solar photovoltaic. (e) The raster is then downsampled to the region of interest or fundamental shape by averaging the proportion of the overlapping areas. (f) Finally, by applying a PV technology model to (b) and combining it with (e) we can define per region the upper expansion limit and the maximal hourly availability constraint for a given technology.
    }
    \label{fig:atlite demonstration}
\end{figure}

Renewable energy sources such as solar, wind and hydro time series are modelled with the open-source package \textit{Atlite} \cite{HofmannAtlite:Series}. \textit{Atlite} i) creates cutouts that define spatio-temporal boundaries, ii) prepares cutouts, which means that environmental and weather data is added to geospatial boundaries by matching various datasets (ERA5 reanalysis data \cite{Hersbach2020}, SARAH-2 satellite data \cite{Sarah2}, and GEBCO bathymetry \cite{GEBCO}), and finally, iii) applies conversion functions to produce technology-specific spatially resolved time series and potentials \cite{HofmannAtlite:Series}. Currently, the \textit{PyPSA\=/Earth} model framework implements solar photovoltaic, on- and offshore wind turbines, hydro-runoff, reservoir and dam power resources. In the case of hydro, the runoff time series are obtained by \textit{Atlite} for each powerplant location, as described in Section \ref{subsec: generators}. As our new contribution, the hydro power output is thereby proportionally rescaled to match the reported total energy production of existing plants as reported per country by the open US Energy Information Administration (EIA) platform \cite{EIA2022}.
At the time of writing, available in \textit{Atlite} but not yet implemented in PyPSA\=/Earth are potentials and time series for concentrated solar power, solar thermal collectors, heat demand and dynamic line rating with a wide range of technology options. For details on the model implementation for each technology, we refer the reader to the \textit{PyPSA\=/Eur} publication which the presented model mostly builds-upon \cite{ Horsch2018PyPSA-Eur:System}.
A brief concept demonstration of \textit{Atlite} is provided in the Figure \ref{fig:atlite demonstration}.

\subsection{Generators}\label{subsec: generators}

Given the limitation of reliable datasets for power plants for the African region, the existing powerplantmatching tool \cite{Gotzens2019PerformingDatabases} has been extended to include additional datasets, such as OpenStreetMap, to fine-tune the African model and validate the results with the final goal of maximizing accuracy and quality of the result.

\begin{figure}[h!]
\centering
\includegraphics[trim={0cm 0cm 0cm 0cm}, clip, width=.95\linewidth]{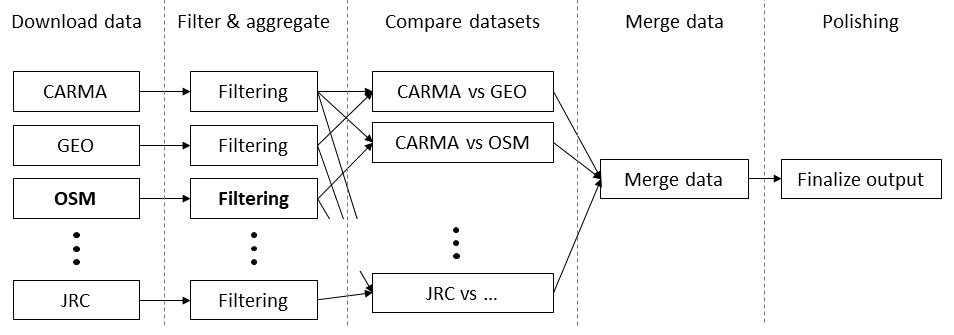}
\caption{Flowchart of the \textit{powerplantmatching} procedure, including the novel OSM input (in bold) which was developed for PyPSA\=/Earth.}
\label{fig: ppl_method}
\end{figure}

\textit{Powerplantmatching} has been successfully proposed to estimate the location and capacity of power plants in Europe. The validation performed with respect to the commercial World Electric Power Plants Database (WEPP) by Platts and the dataset by the Association of European Transmission System Operators (ENTSO\=/E) reaches an accuracy of around 90\% using only open data \cite{Gotzens2019PerformingDatabases}. By default various open data sources are included such as CARMA (discontinued)~\cite{CFGD2012}, ENTSO\=/E~\cite{ENTSOE2022}, ESE~\cite{Sandia2021}, GEO \cite{GEO2018}, OSPD~\cite{OSPD2020}, GPD (discontinued) \cite{WRI2021}, JRC database on hydro powerplants~\cite{JRC2019} and renewable statistics by IRENA~\cite{IRENA2022}.
The approach applied for \textit{powerplantmatching} is based on the procedure depicted in Figure \ref{fig: ppl_method}, where the raw datasets are first downloaded, then filtered to remove missing or damaged data, and aggregated. Once the refined data are obtained, the datasets are pairwise compared to identify duplicated entries. Finally, non-duplicated data are merged into a unique dataset and used as a source for PyPSA\=/Earth. Only a few of these datasets have global scope (GEO, GPD and IRENA) and have been validated for Africa. In particular for Africa, where data is lacking, including all available open data can be critical to maximizing the accuracy of the results. Therefore, inspired by future work suggested in \cite{Powerplantmatching2019}, we have extended the \textit{powerplantmatching} tool to optionally include and process OpenStreetMap data to improve the quality of outputs.

\subsection{Spatial clustering approach}\label{subsec: clustering}
In order to tackle the computational complexity of solving a co-optimisation problem of transmission and generation capacity expansion, the model offers state-of-the-art spatial clustering methods which are imported from the PyPSA package and PyPSA\=/Eur model \cite{Brown2018PyPSA:Analysis, Horsch2020PyPSA-Eur:Code} and adapted for the model to simulate future scenarios. Spatial clustering allows finding sets of nodes that are similar and aggregating them to a single node to represent the original set. This way, the network is reduced to a smaller number of nodes to manage the model's computational complexity.

The available clustering methods provide a focus on (i) conserving the representation of renewable potentials as well as the topology of the transmission grid, (ii) accurately representing the electrical parameters to improve estimates of electrical power flows in an aggregated model, (iii) aggregating spatially close nodes disregarding other a-priori information of the network, or (iv) according to their location with regards to the country's subdivisions facilitating results interpretation for policy recommendations. An analysis of suitable clustering methods that depend on the modelling application is provided in \cite{Frystacki2022comparison}. 

In summary, (i) the clustering approach that focuses on a better representation of variable sources or sinks of the model is inspired by \cite{SIALA201975}. It includes variable potentials, i.e. capacity factors or full load hours for solar and wind, or the variable electricity demand as a distance metric between nodes. This is combined with a hierarchical clustering approach, similar to the suggestions provided in \cite{Kueppers2020Data-DrivenOptimization}. However, we only allow nodes to be aggregated when a physical transmission line connects them instead of assuming a synthesised grid in contrast with \cite{Kueppers2020Data-DrivenOptimization}.
(ii) The clustering method that focuses on a better representation of the transmission grid was initially suggested by \cite{BIENER2020106349} to be applied for the case of electricity system modelling. It is a density-based hierarchical clustering operating on the line impedance.
(iii) The network can also be reduced using a weighted k-means algorithm on the locations of the network nodes as explained in detail in \cite{Frysztacki2021TheSolar}. (iv) Finally, using the GADM shapes allows aggregating all nodes in the same shape.

\begin{figure}[h!]
\centering
\includegraphics[trim={0cm 0cm 0cm 0cm}, clip, width=.95\linewidth]{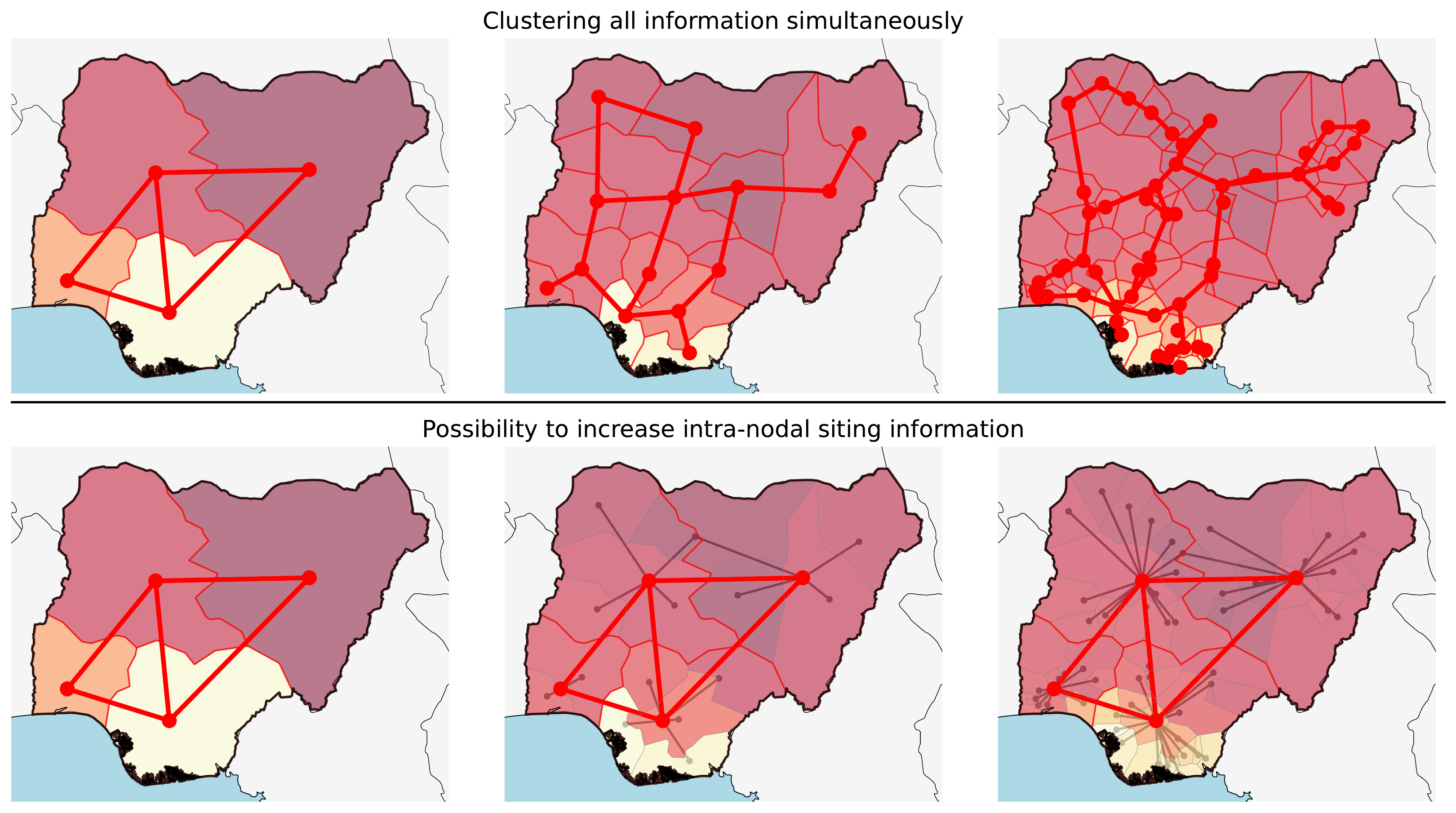}
\caption{Illustration of the clustering methodology applied for the transmission network (red nodes and edges) and resource resolution (grey nodes and edges). In the first row, we show how nodal data (i.e. generators, storage units, electrical loads etc.) is aggregated in tandem with the resolution of the transmission network. Three exemplary resolutions of the network for Nigeria are displayed here: 4, 14 and 54 nodes from left to right. The second row shows how the clustering also allows modelling the transmission grid at a different resolution than the resources. In this example, the transmission network contains 4 nodes connected by 4 lines (all in red) at every resolution, while 4, 14 and 54 generation sites become available (left to right). The background colour represents exemplary capacity factors in shades of red, for an arbitrary technology. The darker the colour, the higher the capacity factor.}
\label{fig: clustering}
\end{figure}

Any of these methods can be applied in a single or two distinct iterations, as displayed in Figure \ref{fig: clustering} for Nigeria. In each of these two iterations, a different method can be applied, choosing from (i)-(iv). In the first iteration, all nodes are clustered to a desired number of representative nodes, aggregating generators, flexibility options (electricity storage and transmission lines) and electrical demand. The second iteration is optional and allows the remaining nodes to be clustered again. However, now only the transmission network is effectively reduced such that the representation of renewable resources is fixed to the resolution of the previous iteration (compare the first row and second row of Figure \ref{fig: clustering}). The spatial resolution of the transmission network must always be larger or equal to the resource resolution, i.e. the clustering of the first iteration sets an upper bound.

\subsection{Augmented line connection}\label{subsec:augmented line connections}

The African network is often not well interconnected. This is due to isolated national planning data or the presence of isolated mini-grids that are popular electrification measures \cite{Fioriti2021CouplingMicrogrid}. Therefore, we propose an algorithm to mesh a given network and assess different grades of connectivity. To investigate the benefits of meshed networks, PyPSA\=/Earth can perform a k-edge augmentation algorithm that guarantees every node has a modifiable number of connections to other nodes. Only if nodes do not already fulfil the connectivity condition, the algorithm will create new lines to the nearest neighbour by a minimum spanning tree. The new 'augmented' lines can be set to an insignificant size (e.g. 1 MW) to create new options for line expansion in the investment optimization. For example, Figure \ref{fig:clustered and augmented meshed network} shows the comparison between the standard clustered network with 420 nodes and its augmented version. Only the model that includes augmented line connections can explore an interconnected continent.

\subsection{Model framework and solver interface}\label{subsec: pypsa and solver}

The PyPSA-Earth model integrates the PyPSA model framework with its solver interfaces to perform energy system planning studies. Using PyPSA has several benefits compared to other tools that are briefly introduced in the following. First, PyPSA enables large-scale optimization in Python.
Python is well known for being user-friendly \cite{PythonOliphant}, but when analysing the memory consumption and speed for building optimization problems it was considered non-competitive compared to tools based on the programming language `Julia` or `C++` \cite{Linopy} -- a bottleneck which also hinders large-scale optimization required for PyPSA\=/Earth. As reaction, developers in the PyPSA ecosystem built \textit{nomopyomo} overcoming the bottlenecks \cite{Groissbock2019AreUse}. More recently, the same group is working on a general package called \textit{Linopy} that promises a 4-6 runtime speed up and a 50\% improvement in memory consumption compared to the optimization problem formulator Pyomo, possibly making it also more memory efficient than the Julia alternative \textit{JuMP} as indicated in \cite{Linopy}. Another point making the PyPSA dependency attractive is that it is one of the most popular tools, as suggested by GitHub stars in the GPST benchmark \cite{GpstBenchmark}, possibly due to its standard component objects and the continuously maintained documentation \cite{PyPSAdocumenation}. Finally, the framework offers several solver interfaces (HiGHS, Cbc, GLPK, Gurobi, among others) providing flexibility in solving various optimization problems with open-source and proprietary solutions.
\section{Validation}\label{sec: validation}

The data validation section aims to assess the data quality with publicly available data: at a continental level in Africa and a country level in Nigeria.

\subsection{Network topology and length}\label{subsec: validation. network topology and length}


Validating the African power grid is challenging. Unlike in Europe, where ENTSO-E \cite{ENTSO-E2020ENTSO-EPlatform} provides reliable open data with continental scope, such a transparent data source is lacking in Africa, and only a few utilities release open data. The self-proclaimed most complete and up-to-date open map of Africa's electricity network is offered by the World Bank Group, which implements Open Street Map data, as well as indicative maps data from multiple sources \cite{worldbank-network}. However, the World Bank data should not be used as a single validation set, because it may report outdated data, given that it has not been updated after 2020, and is partially based on indicative maps rather than on geo-referenced data, making the post-validation time-consuming. 
Conversely, PyPSA\=/Earth builds its grid topology directly from daily updated Open Street Map data. Finally, the World Bank data also provides less detailed information than Open Street Map; for instance, it does not give any information on the frequency, circuit or cable number, limiting the information that can be used for validation. In the following, grid statistics and topology are compared on a Nigerian and African scale. This also includes nationally reported data from the Nigerian energy commission.

First, the transmission lines are validated by comparing the total circuit lengths at different alternate current (AC) voltage levels. Transmission lines can carry one or more 3-phase circuits, whereby each circuit has at least three cables. Instead of looking only at the line length, which is the distance between high voltage towers, it is common to report the total circuit length, which multiplies each line length, e.g. distance from tower to tower, with the number of circuits \cite{Horsch2018PyPSA-Eur:System}.
Table \ref{tab:AC line length validation} indicates that the Nigerian network length reported at the World Bank aligns approximately with the official transmission company statistics \cite{NigerianTSO}, suggesting that the World Bank data is either accurate in Nigeria or used as a reference by the transmission system operator. This official reported total circuit length is approximately 35\% longer than the original Open Street Map data or the modified and cleaned PyPSA\=/Earth derivative. Conversely, on a continental scale, Open Street Map provides approximately a 117\% longer total circuit length than the reported World Bank data. To summarise, while Open Street Map data is qualitatively less available in Nigeria by looking at the statistics, it offers significantly more data on a continental scale. 

\begin{table}[!h]
\centering
\caption{HVAC and HVDC circuit line lengths of Nigeria and Africa from different sources}
\label{tab:AC line length validation}
\resizebox{\columnwidth}{!}{%
\begin{tabular}{|l|ccc|ccc|c|}
\hline
\multicolumn{1}{|c|}{\multirow{2}{*}{Circuit lengths in 1000km}} &
  \multicolumn{3}{c|}{Nigeria} &
  \multicolumn{3}{c|}{Africa} &
  \multicolumn{1}{c|}{\multirow{2}{*}{Ref}} \\
\multicolumn{1}{|c|}{} &
  110-220kV &
  220-380kV &
  \textgreater{}380kV &
  110-220kV &
  220-380kV &
  \textgreater{}380kV &
  \multicolumn{1}{c|}{} \\ \hline
World Bank Group$^a$               & 9.3 & 12.1 & 0.0 &    59.4 & 63.5 & 41.0 &    \cite{worldbank-network}\\ \cline{1-1}
Open Street Map (OSM)           & 6.3 & 9.1 & 0.0 &     87.9 & 180.7 & 76.7 & \cite{OpenStreetMap}  \\ \cline{1-1}
Transmission Company of Nigeria & \multicolumn{3}{c|}{More than 20} &  - & - & - & \cite{NigerianTSO} \\ \cline{1-1}
PyPSA\=/Earth (cleaned OSM)       & 6.7 & 9.1 & 0.0 &   88.3 & 183.7 & 82.9 &  \\ \hline

\multicolumn{7}{l}{$^a$ Information about circuits is missing.}
\end{tabular}%
}
\end{table}

\begin{figure}[h!]
    \centering
    \subfigure[]{\includegraphics[width=0.32\textwidth]{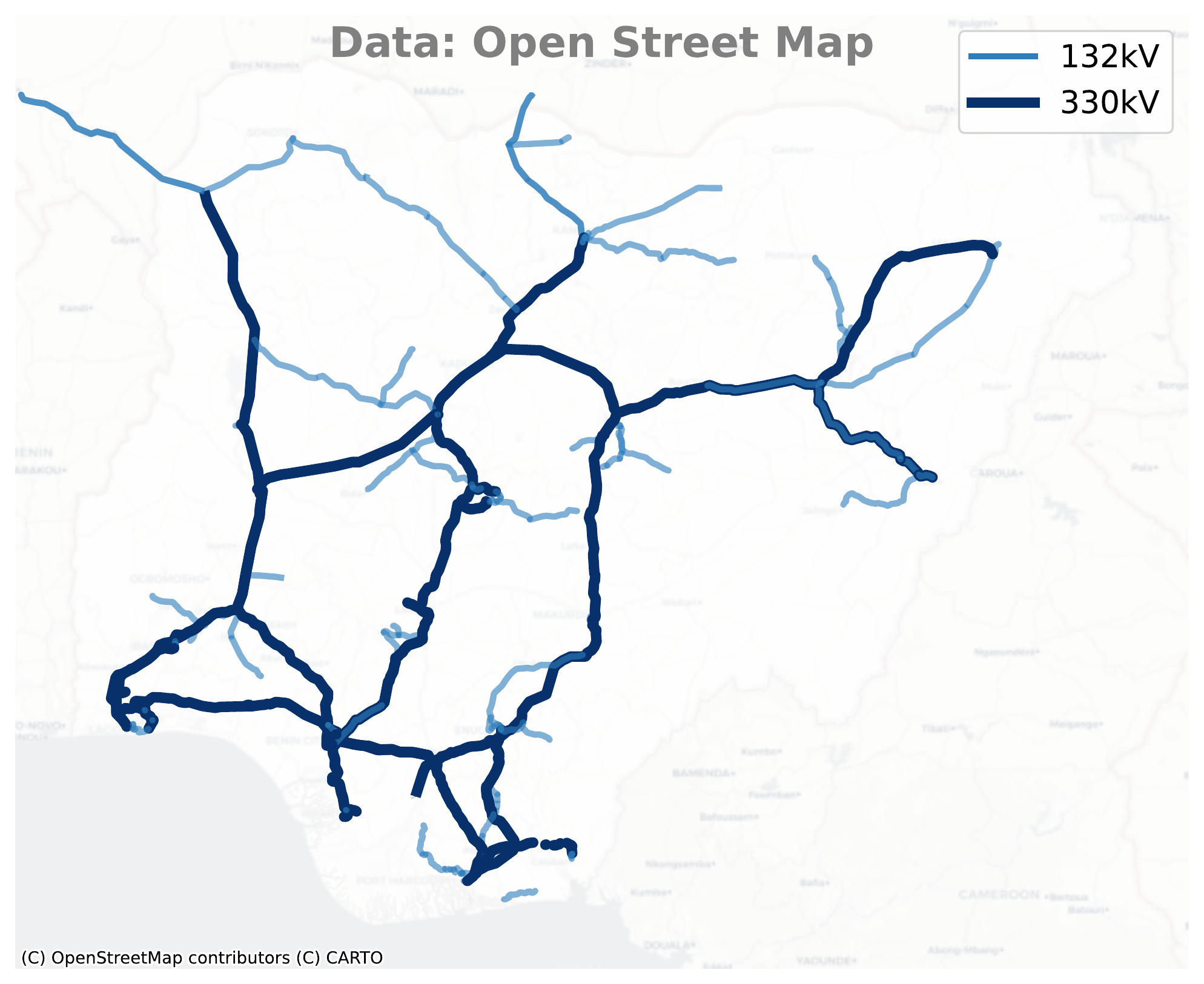}} 
    \subfigure[]{\includegraphics[width=0.32\textwidth]{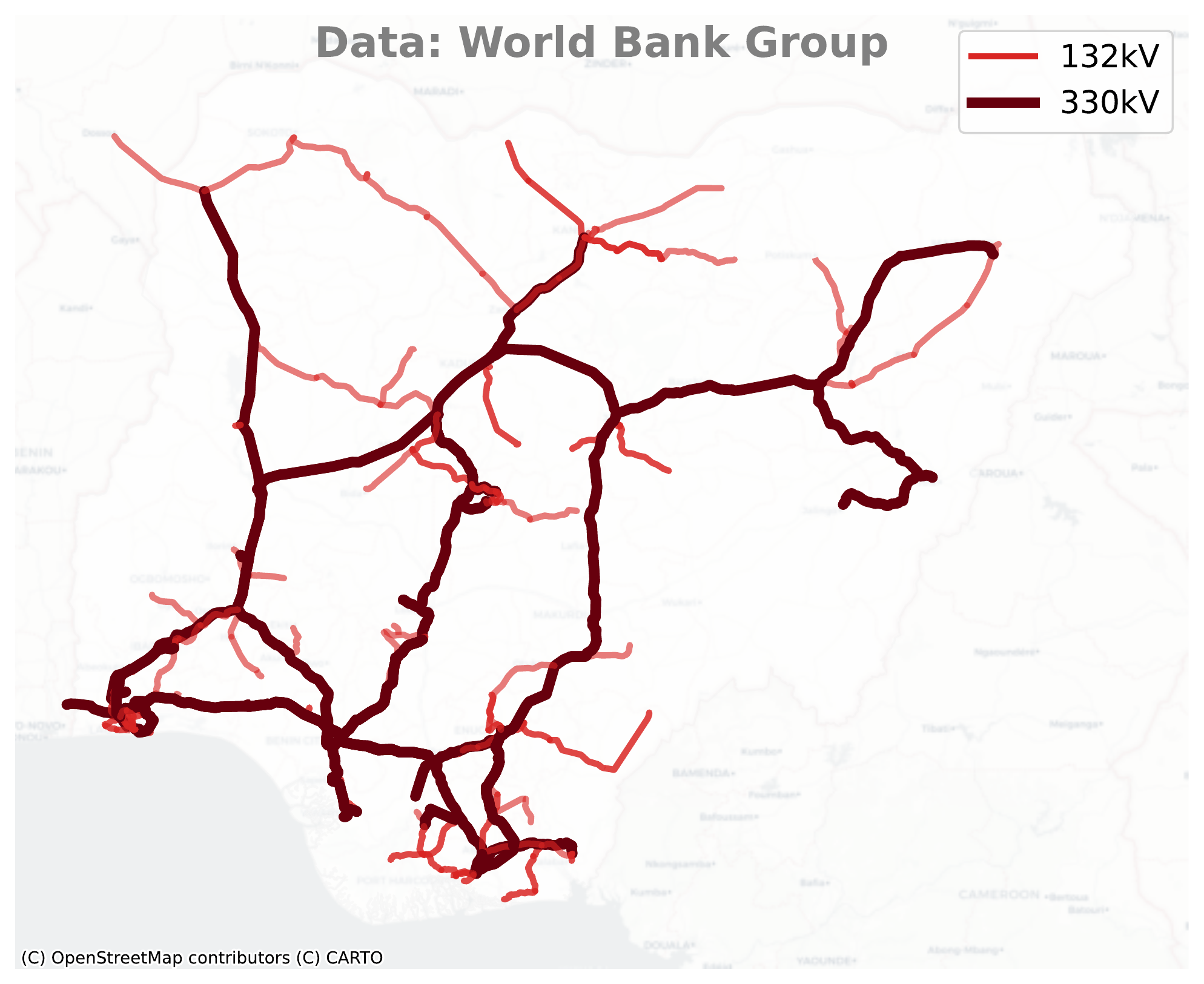}} 
    \subfigure[]{\includegraphics[width=0.32\textwidth]{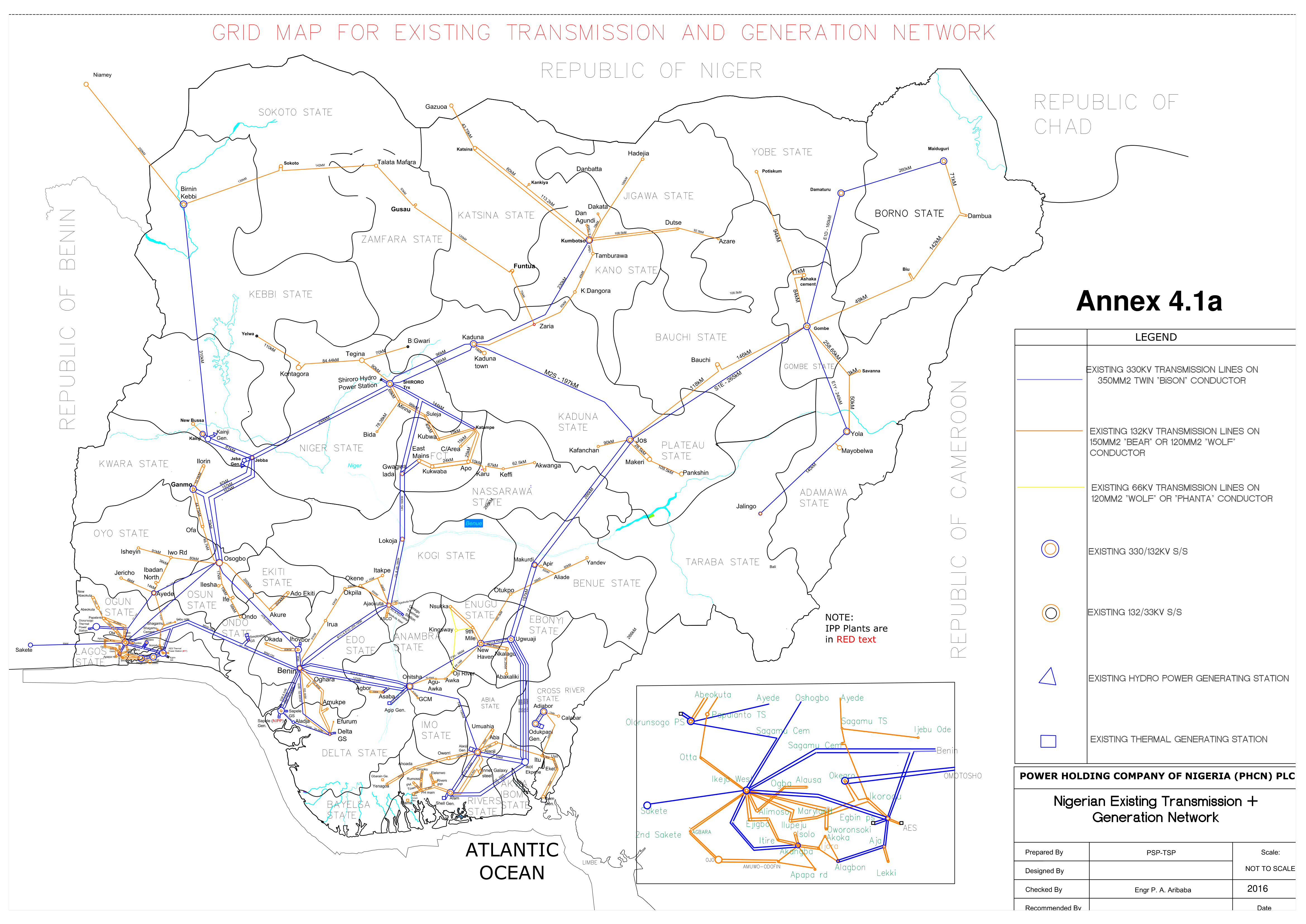}}
    \subfigure[]{\includegraphics[width=0.48\textwidth]{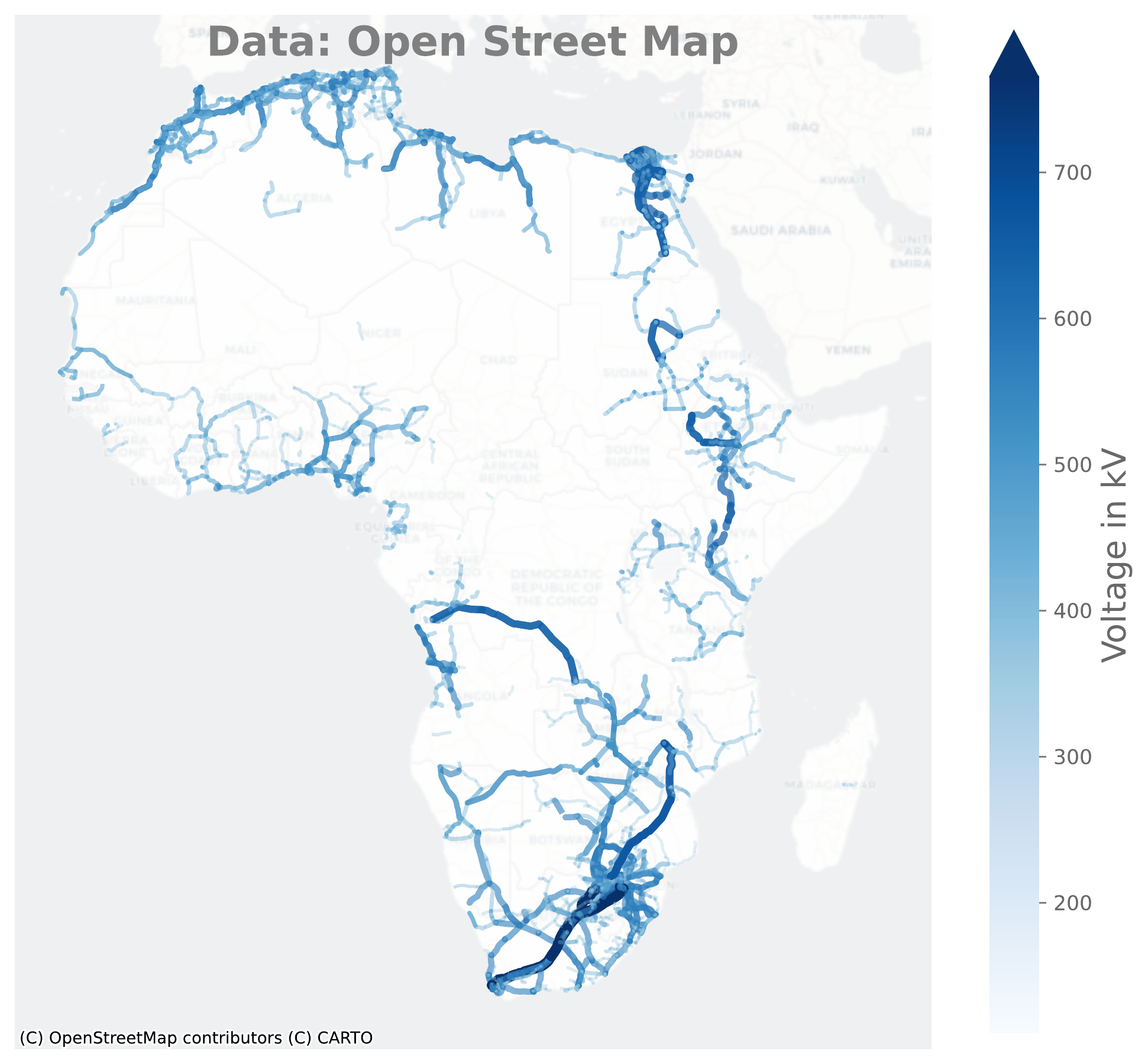}}
    \subfigure[]{\includegraphics[width=0.48\textwidth]{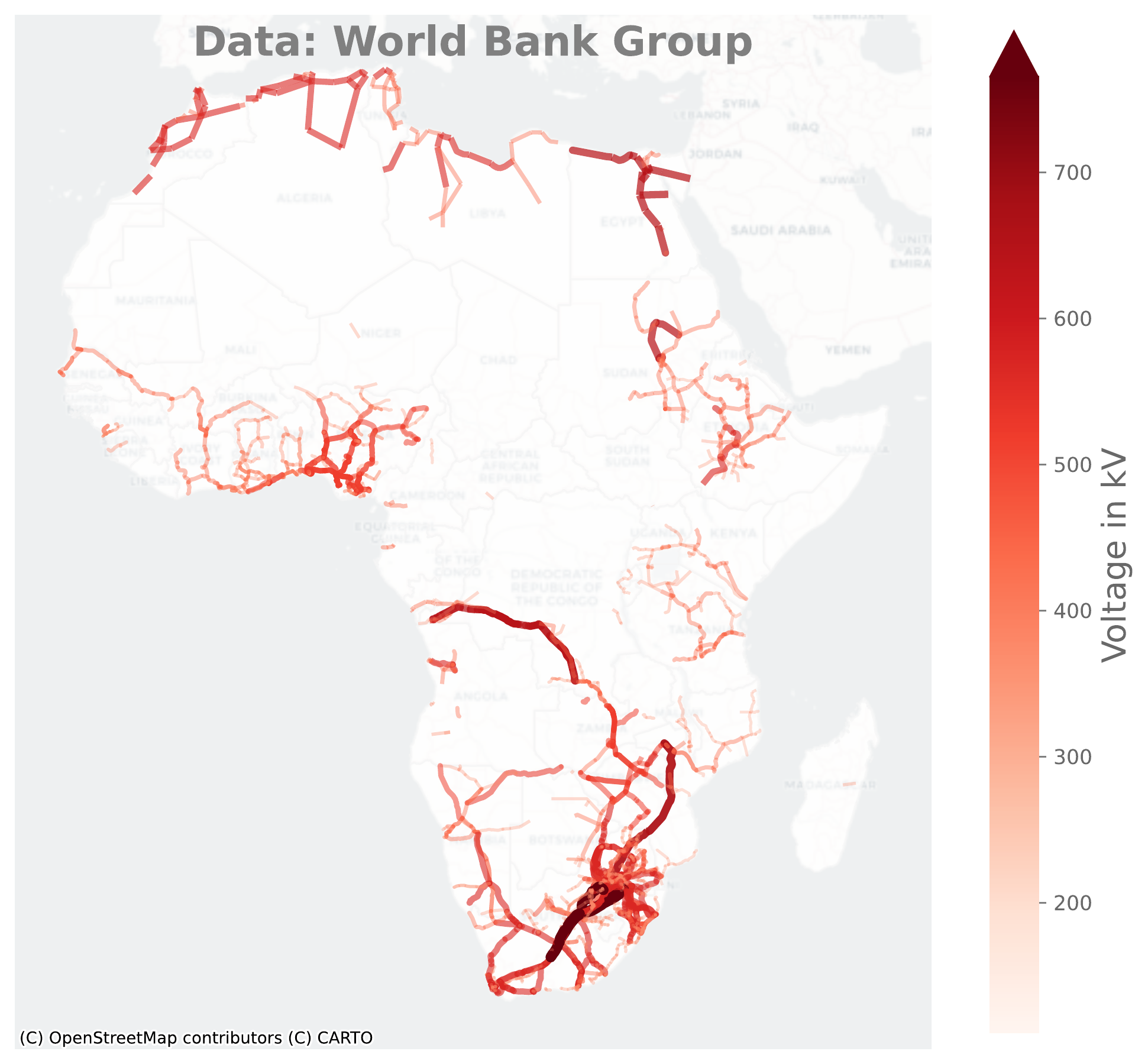}}
    \caption{Network topology of open available transmission network data (above 110kV) from (a) \& (d) Open Street Map, (b) \& (e) World Bank Group and (c) the Nigerian Transmission Company. On the African scale, the voltage ranges from 110-765 kV in both data sets. The line format varies with the voltage level and includes transparency, thickness and colour.}
    \label{fig:network topology}
\end{figure}

To further compare and validate the data, Figure \ref{fig:network topology} highlights good agreement between the network topology in Nigeria and Africa of Open Street Map and World Bank sources. However, in central and south Nigeria, the World Bank covers more power lines. On the African scale, the opposite is observed. Open Street Map covers more network structures in East and North Africa.

\subsection{Electricity consumption}\label{subsec: demand validation}

\begin{figure}[h!]
    \centering
    \includegraphics[width=1\textwidth]{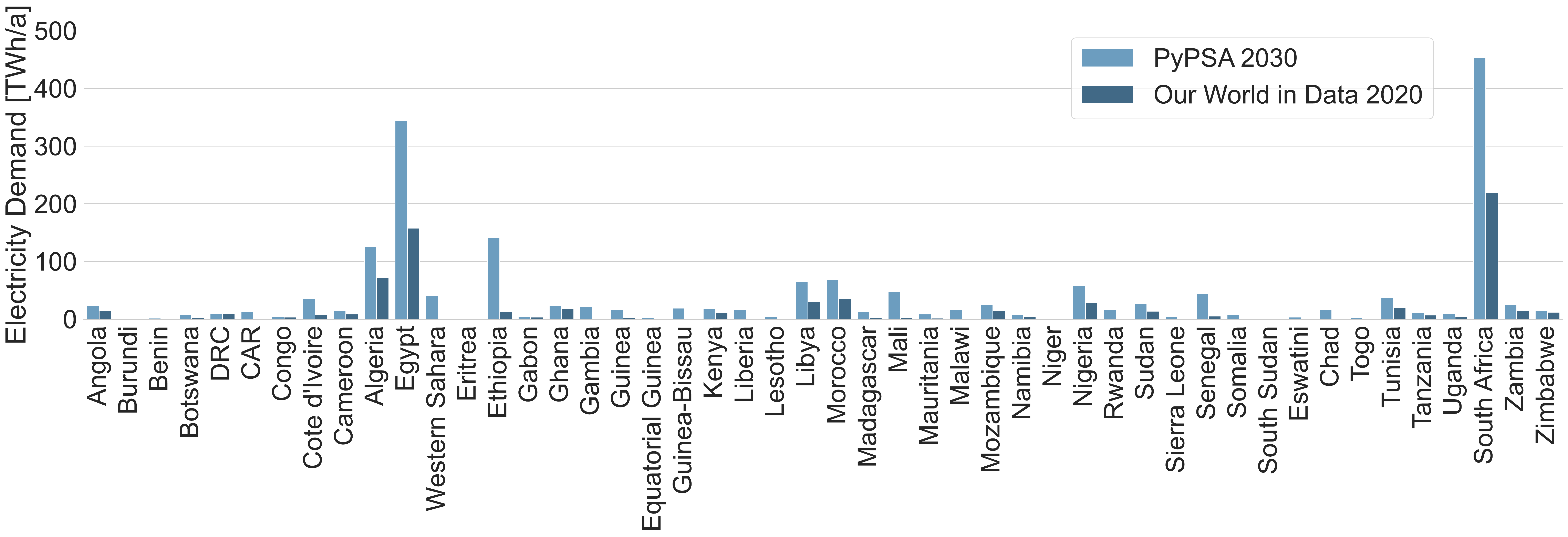} 
    \caption{Comparison of reported \cite{OurWorldInDataDemand} and predicted annual electricity consumption data across African countries indicate in every country demand growth. For 2030, the African total electricity consumption of PyPSA aligns with other predictions from \citet{IRENA} and \citet{Alova2021}}.
    \label{fig: annual electricity consumption}
\end{figure}

This subsection validates the demand prediction on the example year 2030 for every country in Africa by comparing the individual country consumption for 2020 and 2030 with official continental annual electricity consumption used in PyPSA\=/Earth. Figure \ref{fig: annual electricity consumption} shows 2020 reported electricity consumption data per country, published from \textit{Our World in Data} that is additionally refined by data from the global energy think-tank Ember and BP's statistical review of world energy \cite{OurWorldInDataDemand}. The used electricity demand data in PyPSA\=/Earth roughly doubles from 2020 to 2030, indicating demand growth. While national demand predictions are often not available, the demand prediction is further validated by comparing it to other - more common - continental demand predictions. In Africa, \textit{Our World in Data} reported an electricity consumption in 2020 of $782$~TWh/a. For 2030, \citet{IRENA} predicted $1924$TWh/a, \citet{Alova2021} $1877$~TWh/a and the PyPSA\=/Earth model data $1866$~TWh/a, predicting more than a doubling of Africa's electricity consumption by 2030. In summary, looking at the total African electricity consumption suggests that the data used in the global PyPSA model is in the range of others. 

\subsection{Solar and wind power potentials}\label{subsec: validation solar and wind}

The validation of solar and wind potential is performed by comparing statistics by international organizations, such as IRENA, with the outputs of the PyPSA\=/Earth model, both including total generation capabilities and the specific power densities per unit of available land.

Solar and wind potentials are well reported across the African continent. In 2021, the Global Wind Energy Council estimated for Africa a technical potential for wind generation of $180.000 TWh$ (PyPSA-Earth: $108.700 TWh$), which is sufficient to electrify the continent 250 times relative to the 2019 demand \cite{WindpotentialIFC}. Similar, the International Renewable Energy Agency estimated in 2014 that the technical potential in Africa is $660.000 TWh$ (PyPSA-Earth: $122.200 TWh$), which is sufficient to electrify the continent 916 times \cite{IRENA2014}. The discrepancy between the technical potentials observed in the PyPSA-Earth model and the institutional reports is due to the underlying assumptions. In fact, how many renewables can be installed in a region depends on two main assumptions: the excluded areas $[km^2]$ and the power density per technology $[MW/km^2]$, both discussed in the following. 

While we define the available areas in a data-driven way similar to \cite{IRENA2014} and \cite{Horsch2018PyPSA-Eur:System} (see details in Section \ref{subsec: renewable}), the remaining eligible area quantifies the technical potential per technology through the technical power density factor. However, this density applies only to land specifically and uniquely allocated to renewable production, yet this cannot easily be generalized to all non-protected land areas at the country level. In fact, land areas are also necessary for non-technical activities such as economic activities, industries, farming, well-being, and housing, among others. Accordingly, in PyPSA\=/Earth we considered a more conservative power density coefficient to account for such socio-economic considerations.

Focusing on solar photovoltaic power plants, we assessed the power density of the 41 largest installations in the world \cite{Bolinger2022, WikipediaSolar}: the average power density is $46.4~MW/km^2$, the minimum $10.41~MW/km^2$, and the maximum $150.0~MW/km^2$.
The type of solar module and the solar photovoltaic plant design are driving factors for this extensive range of values.
For instance, the Cestas Solar Park in France uses high-performing solar modules and additionally contains a compact east-west orientation solar field design leading to the $150.0~MW/km^2$ extreme. 
Similarly to \cite{Horsch2018PyPSA-Eur:System}, we reduced the technical power density to $10\%$ of the average power plant density to represent the socio-technical limit: $4.6~MW/km^2$ for solar photovoltaic. 

For onshore and offshore wind farm technologies, we verify power density assumptions by analysing seven existing utility-scale wind farms. The observed average technically feasible power density for onshore wind farms is 6.2~$MW/km^2$, and for offshore wind farms, 4~$MW/km^2$ \cite{GithubWindList}. Using the same approach as \cite{Horsch2018PyPSA-Eur:System}, we reduced these values from 6.2 to 3~$MW/km^2$ and from 4 to 2~$MW/km^2$ for onshore and offshore wind farms, respectively, to represent socio-technical power densities and give wind farms space to lower generation reducing wake-effects.

Currently, we apply the same socio-technical power density irrespective of the land cover type. However, roughly about $43\%$ of the continent is characterised as extreme deserts \cite{Reich2004}, giving the opportunity in these regions to be less conservative about the social-technical power density.

\subsection{Power plant database}\label{subsec: powerplant data validation} 

In this section, we compare the site-specific power plant database used in PyPSA-Earth to national statistics provided by IRENA and USAID.

Data on existing power plants is critical for accurate energy simulations as they affect long-term investments, dispatch, and stability of the energy systems. For validation purpose
relevant country statistics are provided by IRENA \cite{IRENAPxWeb2022} and USAID \cite{Usaid2022}. While the PyPSA\=/Earth data is geo-referenced, hence including the location, type and nominal capacities of each power plant, the other sources only provide country statistics. Therefore, data used in PyPSA\=/Earth is of higher quality, especially for energy system modelling in a high spatial resolution that would be impossible to perform with the IRENA and USAID sources only.

\begin{figure}[h!]
    \centering
    \subfigure[\label{fig: capacity_country_validation}]{\includegraphics[width=\textwidth]{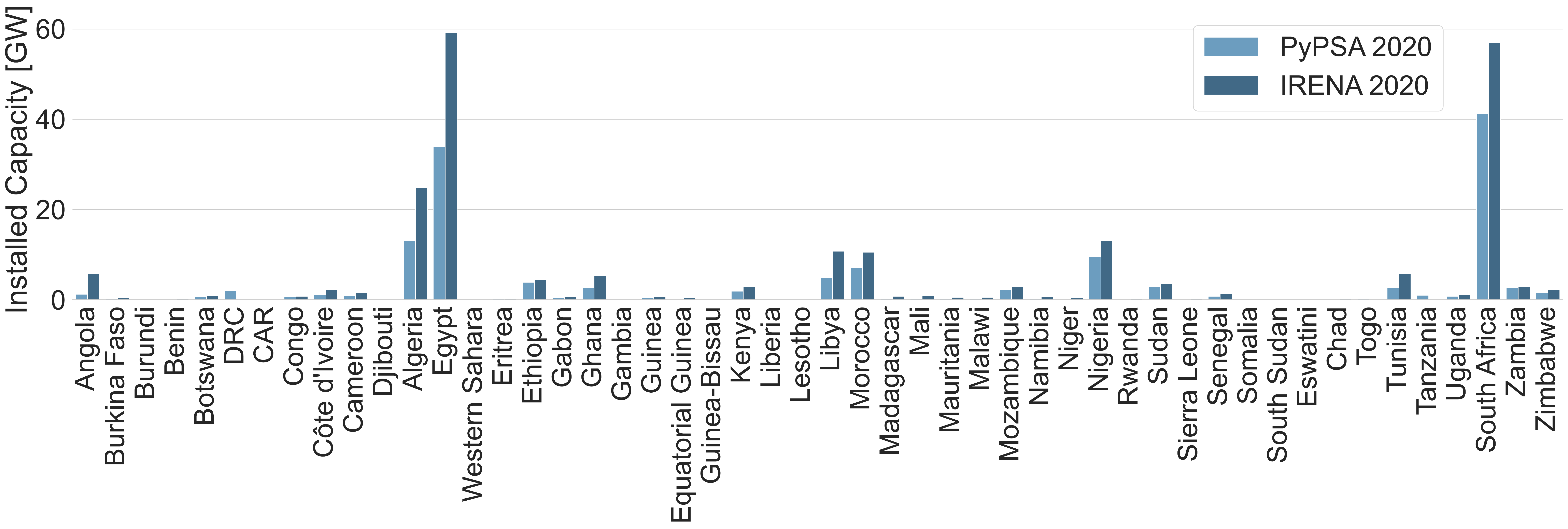}}
    \subfigure[\label{fig: capacity_technology_validation}]{\includegraphics[width=0.48\textwidth]{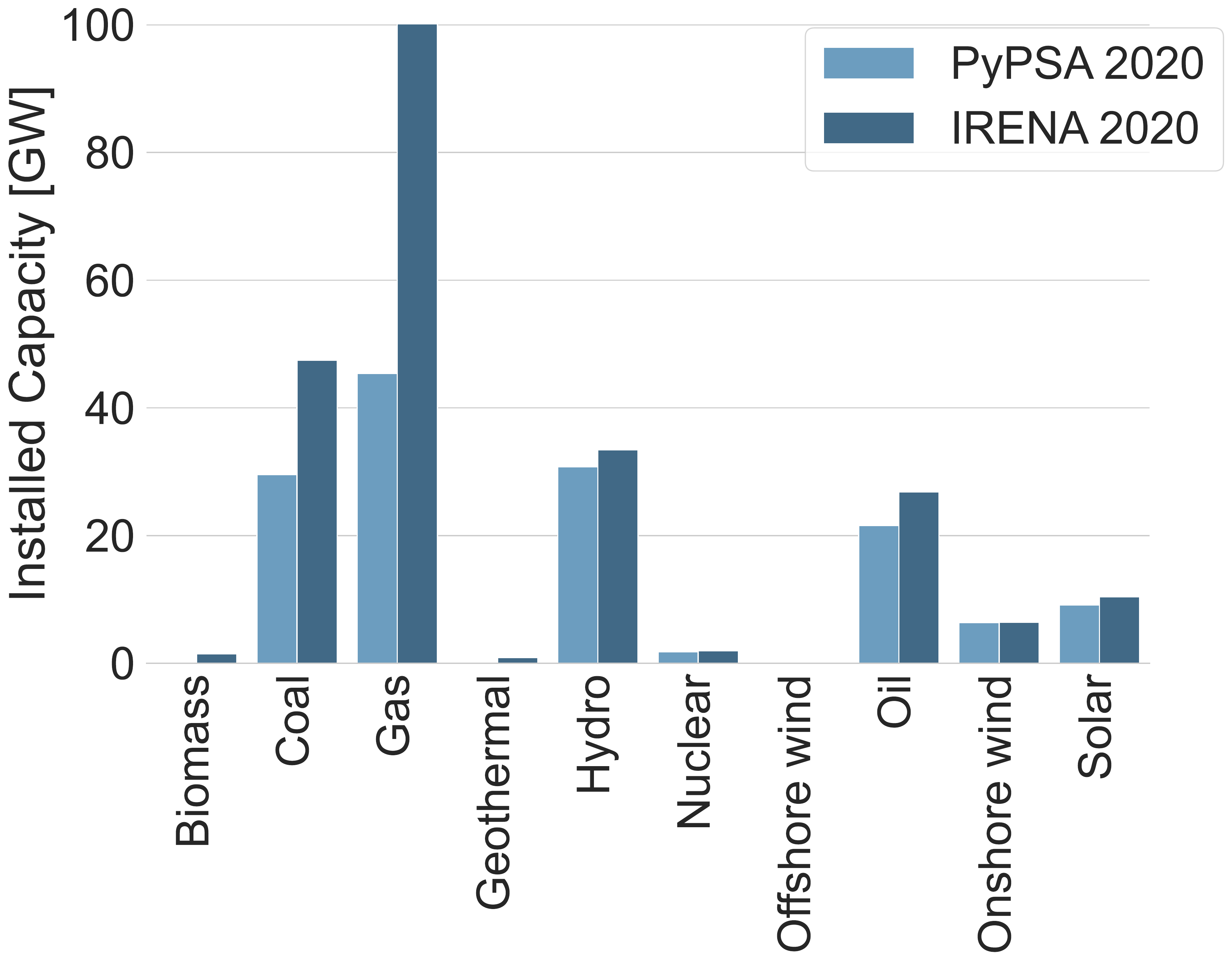}}\hfill
    \subfigure[\label{fig: capacity_technology_validation_ng}]{\includegraphics[width=0.48\textwidth]{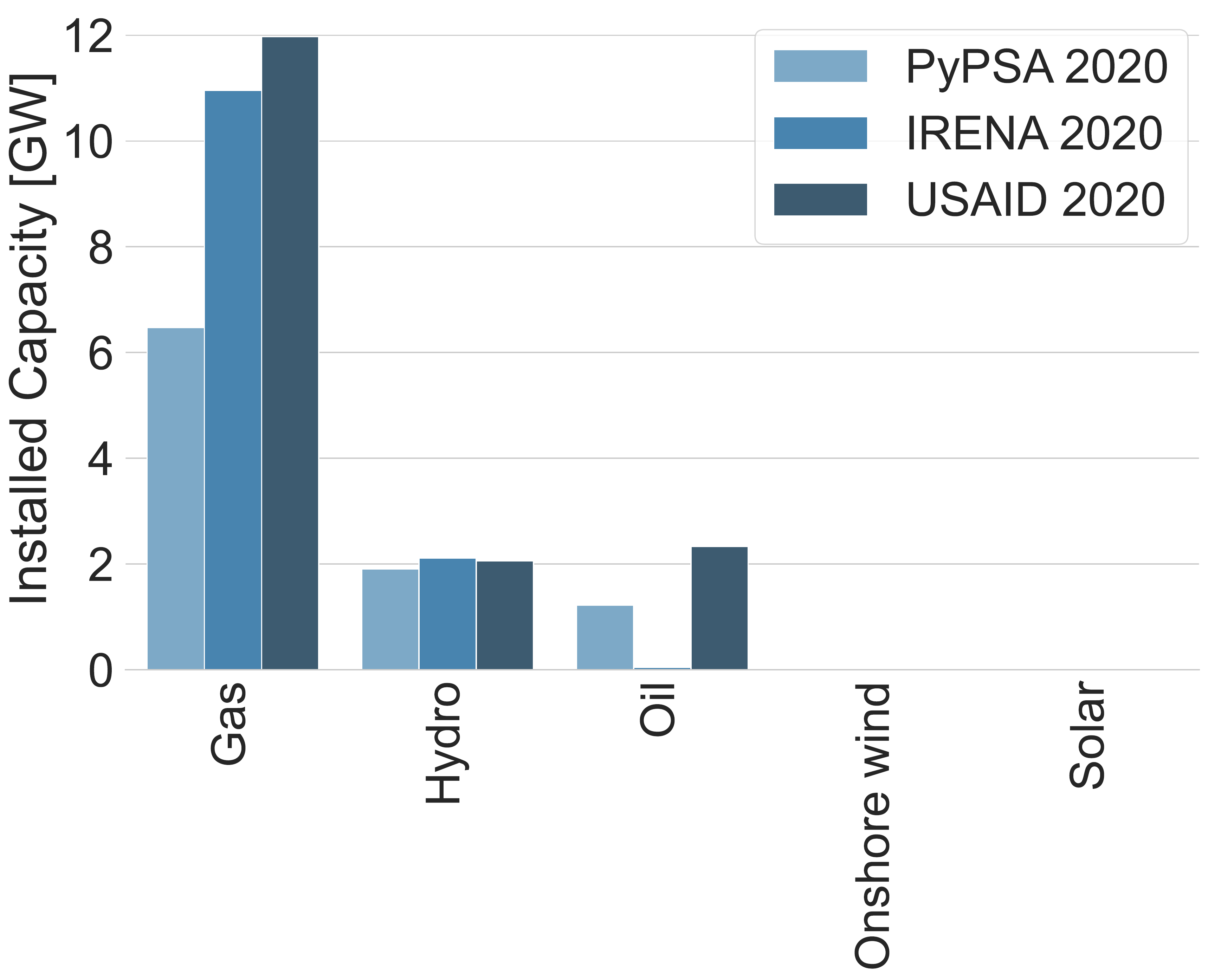}}\\
    \caption{Total installed generation capacity in Africa by (a) country and (b) technology, including a focus on (c) the installed capacities in Nigeria.}
    \label{fig: capacity_validation}
\end{figure}

Figure \ref{fig: capacity_technology_validation} shows that the PyPSA-Earth model matches the largest fraction of the installed capacity of existing databases, with 165~GW out of the 229~GW reported by IRENA. Most technologies are matched with adequate accuracy (2--15\% error), yet larger differences occur especially for coal and gas power plants, partially due to the recent installation of power plants over the last 3-4 years, whose data has not been updated by the sources described in Section \ref{subsec: generators}. In future work, adding more recent data sources may improve the data situation \cite{GlobalEnergyMonitor}. Furthermore, we note that, although the current PyPSA-Earth procedure does not include geothermal and CSP technologies, their capacity can be relevant for certain countries (e.g. Kenya, Morrocco and South Africa), but at an African scope, these technologies still represent a small fraction of the installed capacity. Therefore, the proposed validation is considered of good accuracy, supporting the appropriateness of the PyPSA\=/Earth model.

\section{Demonstration of optimization capabilities in Nigeria}\label{sec: Nigeria case study}


At COP26, Nigeria's president Buhari committed to net zero emissions by 2060 \cite{ClimateActionTracker}. To demonstrate that the presented model can be useful for Nigeria's energy planning activities, we showcase the optimization capabilities of PyPSA\=/Earth. In particular, this section covers two least-cost power system optimizations, one for 2020 to reproduce the historical system behaviour and one representing a decarbonised 2060 scenario (see Figure \ref{fig: nigeria 2020} and \ref{fig: nigeria 2060}).

\subsection{Nigeria 2020 - Dispatch validation}

The 2020 scenario applies a dispatch optimization with linear optimal power flow constraints to simulate and validate the optimization results for Nigeria. Accordingly, only the operation of existing infrastructure is optimized for the lowest system cost, excluding any infrastructure expansion e.g. generation or transmission line expansion (see Figure \ref{fig: nigeria 2020}). 

Starting with the scenario design. The power grid retrieved from OpenStreetMap is clustered to 54 nodes, representing the aggregation zones for the demand and supply. Since the existing network is more meshed than the OpenStreetMap based PyPSA-Earth network (see Figure \ref{fig:network topology}), a few augmented line connections with a negligible minimal capacity of 1~MW are added such that every node has at least two line connections, see (b) in Figure \ref{fig: nigeria 2020}, and overcome short missing network data.
A total demand of 29.5~TWh is considered for 2020 using the national demand profiles provided in PyPSA-Earth. The magnitude aligns with reports from \textit{Our World in Data} (28.2 TWh) \cite{OurWorldInDataDemand}. The demand profiles are distributed across all nodes proportional to GDP and population. With the available hourly electricity demand time series and the existing 2020 power plant fleet (validated in Section \ref{subsec: powerplant data validation}), the model calculates the optimal generator dispatch considering power flow constraints. 

The dispatch validation shown in Table \ref{tab:nigeria_2020_comparison} compares the generation shares of the PyPSA\=/Earth results to those reported at \textit{Our World in Data} \cite{OurWorldInDataGeneration}. The comparison highlights that PyPSA\=/Earth adequately represents the total electricity production shares by source in Nigeria with acceptable accuracy. Model results for the solar generation have a $100\%$ accuracy compared to data provided by \textit{Our World in Data}, gas generation is $2TWh$ ($10\%$) higher than the benchmark, while hydro generation is $0.3~TWh$ ($5\%$) lower. These deviations could be explained by the $1.3~TWh$ ($4\%$) higher assumption of total electricity demand and differences in the specific marginal costs of resources. Using the cost assumptions from \cite{PyPSA-Eur2021GithubPyPSA-Eur}, we derive an average marginal price for electricity of 59~\euro/MWh, which aligns with reported production costs in the range of 45-70~\euro/MWh \cite{nesg2017}.



\begin{table}[!htbp]
    \centering
    \caption{Nigeria 2020 dispatch comparison}
    \begin{tabular}{c|c|cccccc}
         & Total & Hydro & Coal & Gas & Wind & Solar \\\hline
        PyPSA\=/Earth [TWh] & 29.5 & 5.8 & - & 23.6 & 0 & 0.04 \\
        Our World in Data [TWh] & 28.2 & 6.1 & 0.6 & 21.4 & 0 & 0.04
    \end{tabular}
    \label{tab:nigeria_2020_comparison}
\end{table}

\begin{figure}[h!]
    \centering
    \includegraphics[width=0.7\textwidth]{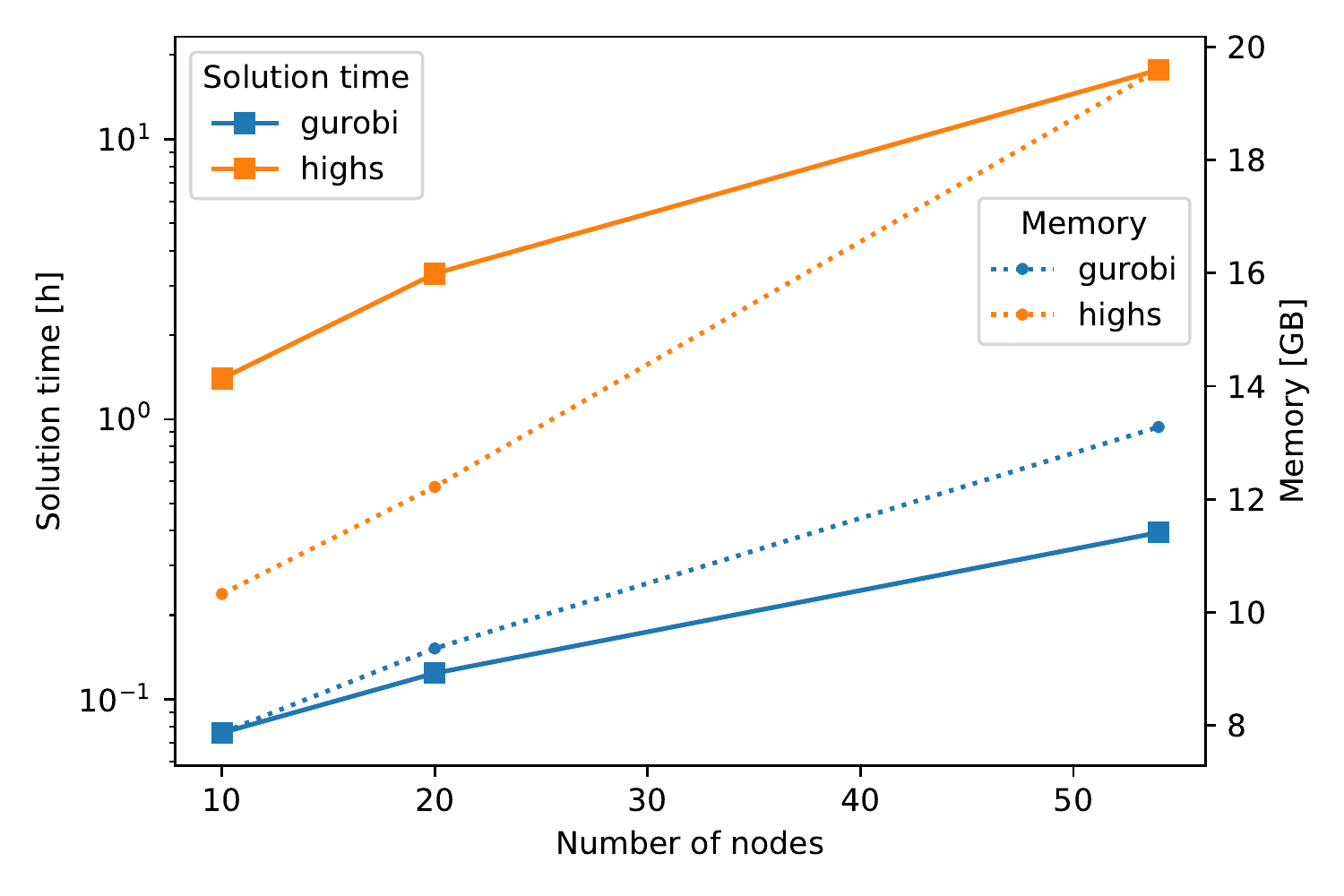}
    \caption{Solution time and memory requirements for the 2020 Nigeria dispatch optimization for HiGHS 1.2.1 and Gurobi 9.5.1 solver at different spatial resolution; solution time is weighted by threads.}
    \label{fig: solver_comparison}
\end{figure}

The computational needs for this scenario in terms of total solving time (computational time times the average load of the processors) and memory, are shown in Figure \ref{fig: solver_comparison}. We used 4 threads with Gurobi 9.5.1 solver while only a single-core with HiGHS 1.2.1, since its parallel solving capabilities are currently limited. While the commercial Gurobi solver is very efficient, the results in Figure \ref{fig: solver_comparison} confirm that the open-source HiGHS solver can also optimize the network below one day with memory requirements that are available for laptops. Given the expected improvements for open-source solvers, the computational requirements are likely to decrease significantly \cite{Parzen2022OptimizationSolvers}.

\subsection{Nigeria 2060 - Net-zero study}

\begin{figure}[h!]
    \centering
    \subfigure[]{\includegraphics[width=0.65\textwidth, trim={0cm 0cm 0cm 0cm}, clip,]{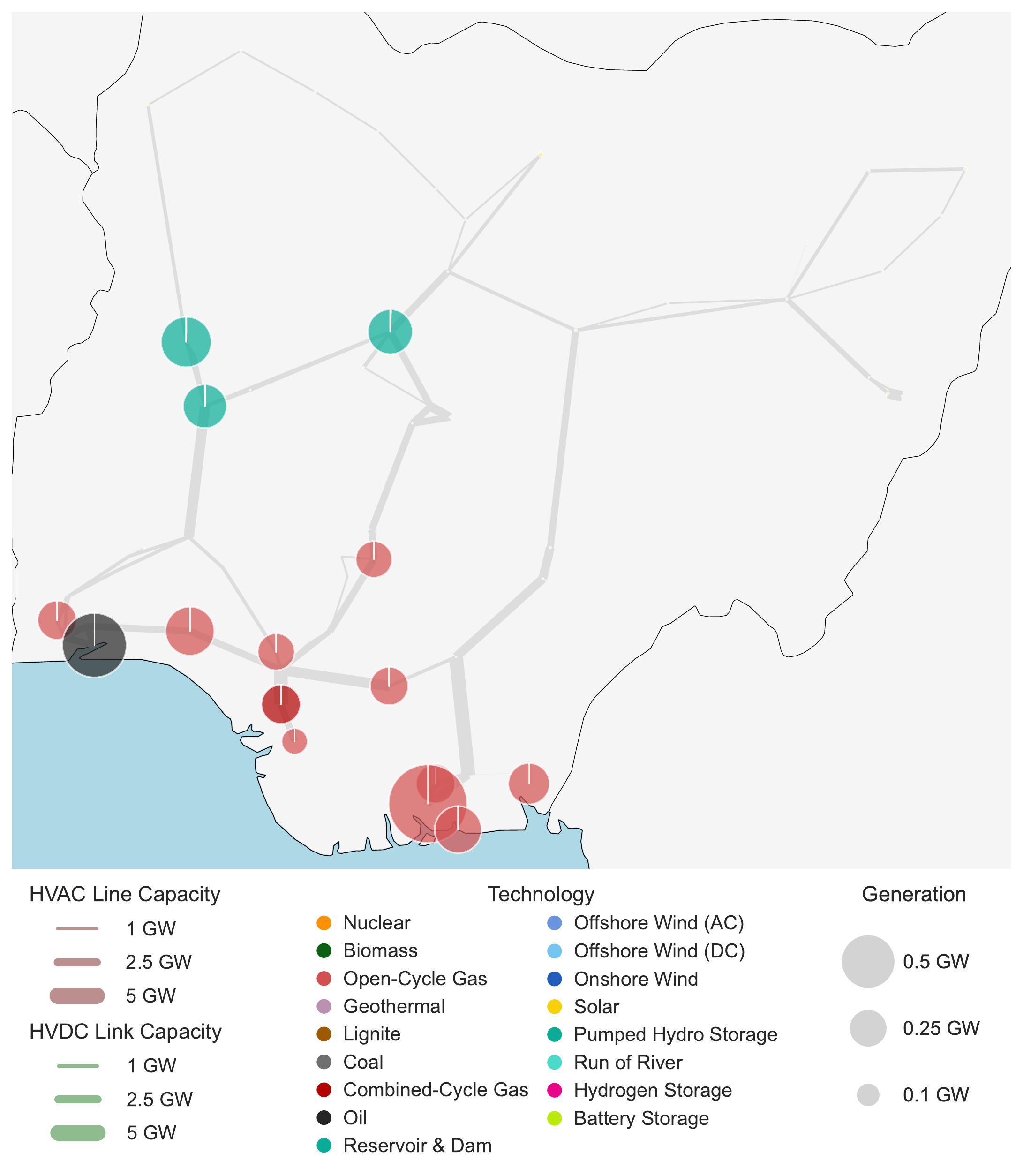}} 
    \subfigure[]{\includegraphics[width=0.3\textwidth, trim={0cm 0cm 0cm 0cm}, clip,]{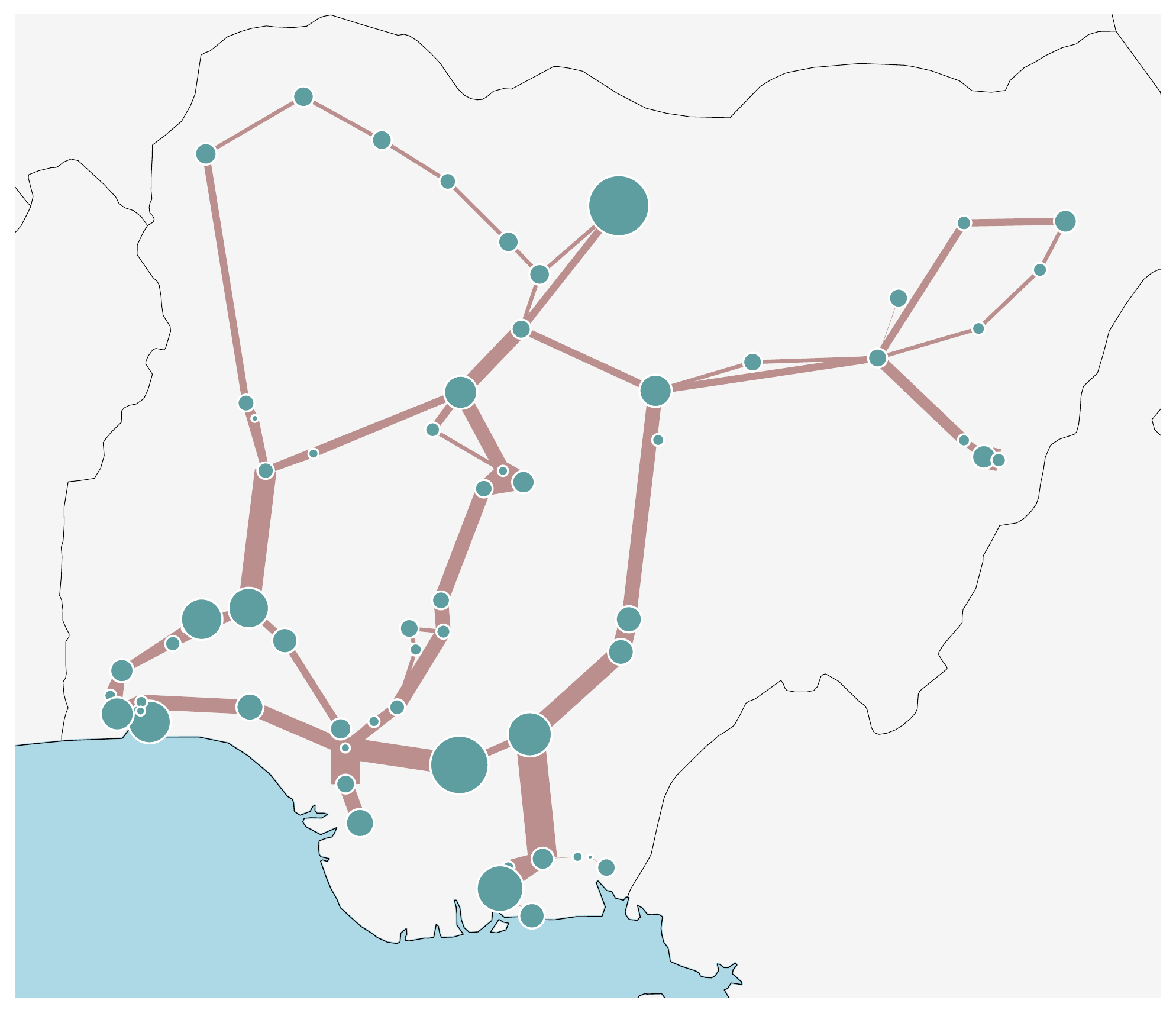}} 
    \caption{Optimization results of Nigeria's (a) 2020 power system. The coloured points represent installed capacities. (b) Shows all network options on a different scale as (a) with the total electricity consumption per node.}
    \label{fig: nigeria 2020}
\end{figure}

\begin{figure}[h!]
    \subfigure[]{\includegraphics[width=0.65\textwidth]{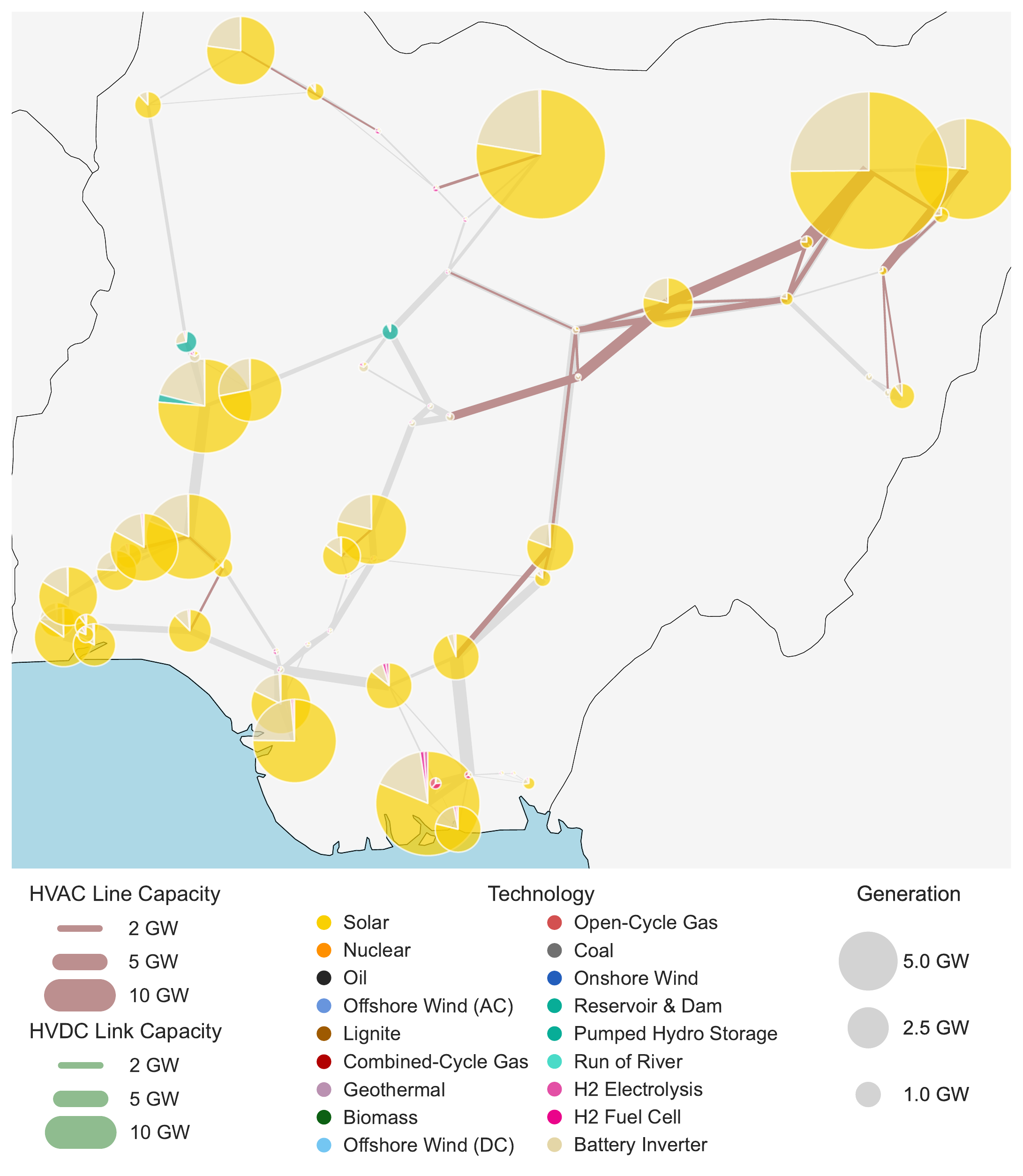}}
    \subfigure[]{\includegraphics[width=0.3\textwidth, trim={0cm 0cm 0cm 0cm}, clip,]{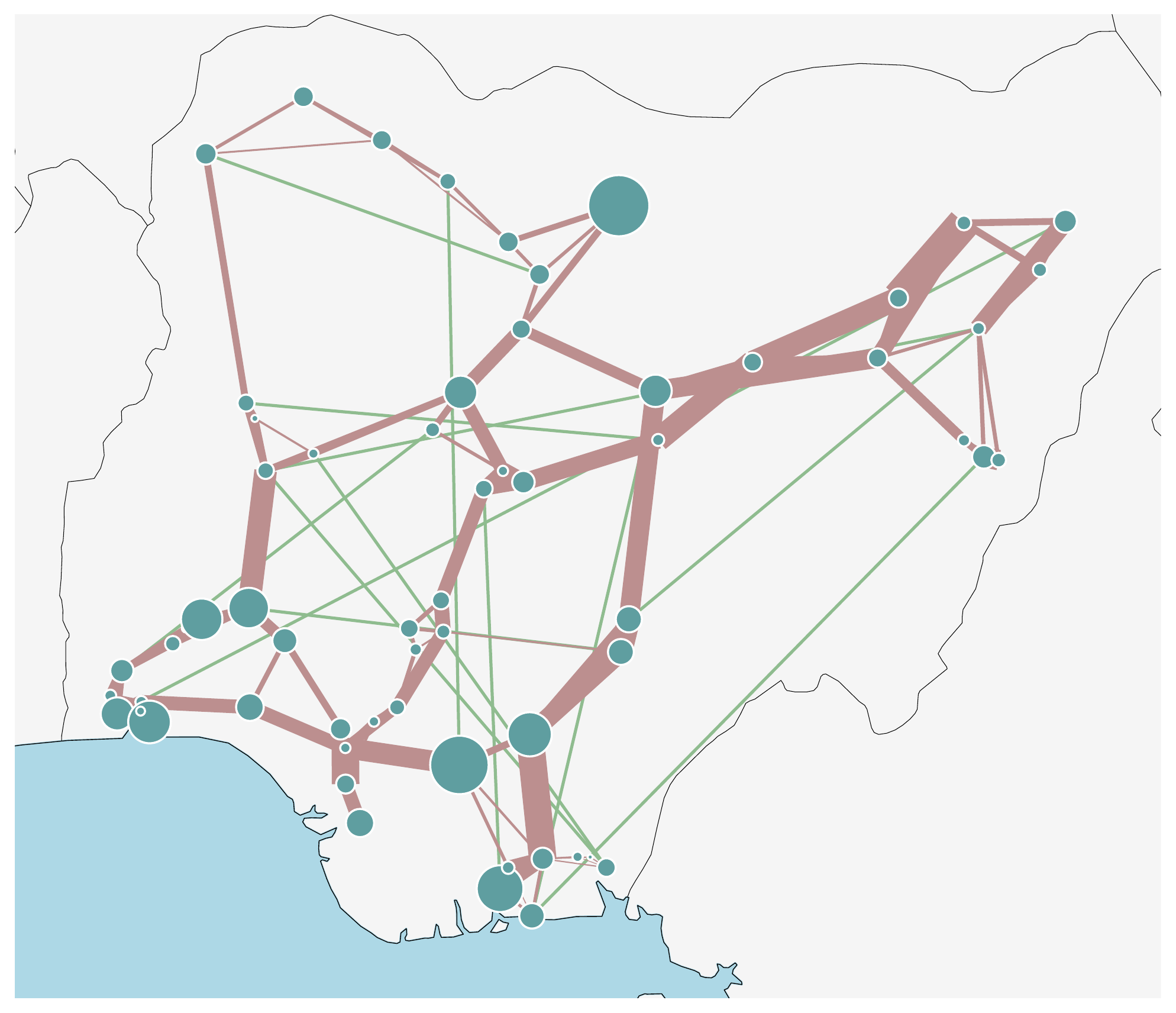}}
    \caption{Optimization result represent Nigeria's (a) 2060 power system. The coloured points represent installed capacities. Light grey and dark grey lines are existing and newly optimized transmission lines, respectively. (b) Shows all network options on a different scale as (a) with the total electricity consumption per node.}
    \label{fig: nigeria 2060}
\end{figure}

In the 2060 net-zero scenario, we perform a brownfield capacity expansion optimization. This means that new renewable energy and transmission capacity can be built on top of existing infrastructure. Simultaneously, a dispatch optimisation is performed subject to linear optimal power flow constraints. To explore new transmission grid structures, the meshing strategy is increased such that each node connects HVAC lines to at least three nearest neighbouring nodes and that a random selection of far distance nodes above $600km$ connects HVDC lines, see b) in Figure \ref{fig: nigeria 2060}). Using a random selection for long-distance HVDC can help identify valuable line connections before applying any heuristic that might not find these. Additional to the net-zero emission constraint, the 2060 total demand has been calibrated in agreement to \cite{Kim2021} to about 250 TWh by linear interpolation of the Stated Energy Policies of IEA for Nigeria \cite{IRENA2022}.
Observing the optimized infrastructure in Figure \ref{fig: nigeria 2020}, the overall optimal least-cost power system can be mostly supplied with solar energy and a mix of battery energy storage. Hydrogen energy storage with steel tanks is included as an expansion option. However, it is not significantly optimized, probably because we ignore fuel trade with other countries and the unique geo-location of the country. Nigeria lies close to the equator, where solar irradiation is homogenous across the year, requiring less seasonal energy storage. The battery storage consists of an inverter component $[\euro/MW]$ and a Li-Ion battery stack $[\euro/MWh]$ that can be independently scaled by the model such as applied in \cite{Parzen2021BeyondSystems}. The energy to power ratio (EP) indicating the sizing between these storage components is optimized in the range $4.6~h$ -- $15.1~h$ with an average of $7.4~h$. The optimal solar capacity distribution is spatially uneven. Most solar is expanded in the country's north, where the solar potential is significantly higher \cite{Umar2021}. It is also cost-optimal to build new transmission routes in the north and east of Nigeria, enabling the spatial distribution of the electricity. The HVDC options are not used significantly, indicating it is not cost-optimal in the scenario. Notably, to be conservative, with cost assumptions for 2050 \cite{lisazeyen_2022_6885392}, the average marginal prices reach only 51~\euro/MWh, compared to 59~\euro/MWh in the 2020 scenario. As a result, the optimized renewable energy future for Nigeria can be more economical than the current energy system design.

\section{Limitations and future opportunities}\label{sec: limitations and future developments}

\subsection{Missing network topology data}
\label{subsec: satellite image detection}
Modelling can only be as good as underlying data -- the same applies for PyPSA\=/Earth. By relying on open sources to model energy systems, their data quality is a concern that we also acknowledge for the present paper. Yet, we also describe possible procedures based on image recognition to not only improve the data situation in PyPSA\=/Earth but potentially all energy models. This effort may complement the traditional effort by public institutions that disclose data of public relevance, such as installed network infrastructure, as performed by ENTSO\=/E \cite{ENTSOE2022}.

Compared to Europe or North America, institutions that provide infrastructure data with geolocation for modelling have no analogues in Africa. The missing network data situation is limiting the use of energy system models.
However, other types of data from which energy system components can be inferred exist on a much larger scale.
Satellite imagery is one such data type.
As part of the PyPSA meets Earth initiative, we are exploring opportunities to use neural network based object detection applied to satellite imagery to enrich the existing datasets on energy infrastructure.
In the past, such efforts have either been hard to apply on larger scales due to high requirements on manual input \cite{devseed2017} or are coarse approximations to the true grid structure \cite{Arderne2020PredictiveData}.
Under the umbrella of this initiative, we aim to develop precise and scalable methods and base our efforts on recent advances in the field \cite{huang2021gridtracer, talukdar2018transfer, deng2021unbiased, maxar}.



\subsection{Missing demand time series and prediction biases}\label{subsec: missing demand}

Demand is a significant uncertainty factor in Africa due to the growth in magnitudes over the following decades that has implications on results created by PyPSA\=/Earth. Therefore, improving demand predictions is essential. We acknowledge this limitation in the data fed into our model and highlight research opportunities to address this challenge.
PyPSA-Earths demand data is limited, as indicated in Section \ref{subsec: demand}, by poor prediction performance for low-income countries due to input data biases and by missing machine learning output data for some low-demand countries due to software bugs. Additionally, while the open-source \textit{synde} package~\cite{Synde2022} used in PyPSA-Earth extended the original GEGIS package for demand prediction by a workflow, there are opportunities to create a package focusing only on demand prediction substituting the GEGIS design that provides all energy model data in one package \cite{Mattsson2021AnRegions}.

Developing a package focusing on electricity demand forecast for macro-energy system modelling worldwide is an opportunity to improve the status quo of existing tools. Its scope can also be generalised to other energy carriers. Adopting the findings from \cite{Synde2022}, \cite{Mattsson2021AnRegions} and \cite{TOKTAROVA2019160}, the output of such package can be static and time series information for the demand of any year beyond 2000, depending on the available features, for arbitrary regions on a sub-national and national scale. Creating such a package is feasible but needs considerable work and a sustainably funded group of people who enhance the software.


\subsection{Imprecise global data}\label{subsec: linkers}


PyPSA\=/Earth relies on open data with global scope. This means that sometimes data is used that approximate country-specific details needed for national energy planning studies. 
While improving the global open data situation is one opportunity \cite{Ritchie2021}, another is to enable the integration of national and regional more precise data that can also be used as a source for validation.  
Therefore, PyPSA\=/Earth does not only use global data as default but also allows the integration of national or regional more precise data. Accordingly, specialized functions here coined as "linkers", can enable the fetching of these region-specific to replace the default global data. 
These linkers allow, for instance, local transmission system operators to execute models with national data. PyPSA\=/Earth does not include this national potentially more precise data by default but workflow contributions to make these optional are welcome and can help maximise benefits for the society by better planning studies.


\subsection{Additional technologies}
\label{sec: additional techs}

PyPSA\=/Earth includes the major transmission, generation and storage technologies, however, some are not yet included. Examples of not implemented generation technologies are Concentrating Solar Power (CSP), location-based geothermal, and other secondary technologies such as wave/tidal energy harvesting. While at a global scale, these technologies represent a minor fraction, for country-specific analyses, they may have substantial implications, such as in the case of Kenya for geothermal or Morocco for CSP. Moreover, while currently only lithium-ion batteries and hydrogen energy storage are considered, additional technologies may be considered and tested, such as the well-known Redox Flow batteries, Compressed-Air Energy Storage (CAES), Liquified-Air Energy Storage (LAES), that can have a large market in the future. Moreover, the dynamic calculation of the transmission capacity as a function of weather conditions \cite{Horsch2018PyPSA-Eur:System}, also known as Dynamic Line Rating (DLR) \cite{Poli2019}, is not yet included.

These limitations, at the time of writing, represent future opportunities to improve the model and capture relevant technologies to perform detailed energy studies for all countries. 


\section{Conclusions}\label{sec: conclusion}


This paper presents the PyPSA\=/Earth model, which is the first open-source global energy system model in high spatial and temporal resolution. It is making high-resolution modelling accessible to countries which so far had not detailed energy planning scenarios developed.
Using a novel comprehensive workflow procedure PyPSA\=/Earth automatically downloads open data, provides model-ready data and integrates optimization features to address large scale energy system planning.
In agreement with the open-source spirit, the model is not built from scratch but derived from the European-focused PyPSA\=/Eur model adding global data as well as several new features.

The methodology is confirmed to be flexible and accommodate a high temporal and spatial resolution energy model for national and regional energy planning with global scope. The validation performed for the African continent highlighted that PyPSA\=/Earth successfully provides power network and installed generation data that match trustworthy third-party national data with adequate accuracy, hence suggesting PyPSA\=/Earth to be a reliable model for energy planning. The 2020 and 2060 planning studies for Nigeria have further confirmed that net-zero emission scenarios for the electricity sector can be performed using PyPSA\=/Earth, leading to realistic results comparable with similar studies but in higher spatial detail. That further stresses the robustness of the approach and the flexibility of the methodology to be used in practical projects.

Given the need for reliable tools to foster the energy transition and the need for the efficient use of resources, PyPSA\=/Earth can successfully support policymakers, utilities, and scholars in providing reliable, transparent, and efficient decision-making on energy studies. While several open source projects are developed but discontinued, the authors of this paper and PyPSA\=/Earth developers aim to foster collaborative energy system modelling on the same code-base to provide a well-maintained and robust tool, rather than disperse resources across multiple models that get easily outdated. Given the flexibility of the approach, additional improvements can be integrated, and scholars interested in contributing are invited to contact the PyPSA\=/Earth team to join forces. Accordingly, this paper and the proposed tool can serve as a backbone for further research and business activities built on top of PyPSA\=/Earth, to meet various energy transition planning needs that must be cheap and fast to develop for every nation and community on Earth.

Further studies, may address the sector-coupled version of PyPSA\=/Earth, to account for sectors beyond power (e.g. industry and/or transport), the interface of energy modelling with economics modelling for better energy policy decisions, the improvement of the demand forecasts also in alignment to climate change scenarios, the improvement of imprecise global network data using object detection on satellite images or the validation of the model in other regions. 

\section*{Code and Data availability}\label{sec:codedata}
Code and data to reproduce results and illustrations are available by using PyPSA-Earth v0.1 \cite{pypsaafrica}. Further, instruction and configurations to reproduce the scenarios and plots are provided here: {\color{blue}\href{ https://github.com/pz-max/pypsa-earth-paper}{https://github.com/pz-max/pypsa-earth-paper}}.

\section*{Credit authorship contribution statement} 


M.P, D.F., F.N., conceptualised the study; M.P., D.F administrated the project; M.P., D.F., H.A., E.F., M.M., M.F., J.H., L.S. contributed to the software development, validation and figure production,
A.K. acquired the funding; all authors contributed to writing and revising the manuscript.



\section*{Declaration of Competing Interest}

All authors declare that there are no competing interests.

\section*{Acknowledgements} 
This research was supported by UK Engineering and Physical Sciences Research Council (EPSRC) grant EP/P007805/1 for the Centre for Advanced Materials for Renewable Energy Generation (CAMREG) and EPSRC grant EP/V042955/1 DISPATCH. Maximilian Parzen would like to thank Tom Brown, Aminu Haruna Isa, Dahunsi Okekunle, Matija Pavičević, Michael Dioha and every one of our continuous supporters for their helpful comments and inspiring discussion.


\section{Appendix}

\bibliographystyle{elsarticle-num-names}
\bibliography{references.bib, manual_references.bib}

\begin{thebibliography}{107}
\expandafter\ifx\csname natexlab\endcsname\relax\def\natexlab#1{#1}\fi
\providecommand{\url}[1]{\texttt{#1}}
\providecommand{\href}[2]{#2}
\providecommand{\path}[1]{#1}
\providecommand{\DOIprefix}{doi:}
\providecommand{\ArXivprefix}{arXiv:}
\providecommand{\URLprefix}{URL: }
\providecommand{\Pubmedprefix}{pmid:}
\providecommand{\doi}[1]{\href{http://dx.doi.org/#1}{\path{#1}}}
\providecommand{\Pubmed}[1]{\href{pmid:#1}{\path{#1}}}
\providecommand{\bibinfo}[2]{#2}
\ifx\xfnm\relax \def\xfnm[#1]{\unskip,\space#1}\fi
\bibitem[{Pfenninger(2017)}]{Pfenninger2017EnergyWorkings}
\bibinfo{author}{S.~Pfenninger},
\newblock \bibinfo{title}{Energy scientists must show their workings},
\newblock \bibinfo{journal}{Nature} \bibinfo{volume}{542}
  (\bibinfo{year}{2017}) \bibinfo{pages}{393--393}.
  \DOIprefix\doi{10.1038/542393a}.
\bibitem[{Pfenninger et~al.(2017)Pfenninger, DeCarolis, Hirth, Quoilin, and
  Staffell}]{Pfenninger2017TheBehind}
\bibinfo{author}{S.~Pfenninger}, \bibinfo{author}{J.~DeCarolis},
  \bibinfo{author}{L.~Hirth}, \bibinfo{author}{S.~Quoilin},
  \bibinfo{author}{I.~Staffell},
\newblock \bibinfo{title}{{The importance of open data and software: Is energy
  research lagging behind?}},
\newblock \bibinfo{journal}{Energy Policy} \bibinfo{volume}{101}
  (\bibinfo{year}{2017}) \bibinfo{pages}{211--215}.
  \DOIprefix\doi{10.1016/j.enpol.2016.11.046}.
\bibitem[{{International Renewabe Energy Agency}(2021)}]{IRENA2021}
\bibinfo{author}{{International Renewabe Energy Agency}},
  \bibinfo{title}{{IRENA and IAEA to Help African Union Develop Continental
  Power Master Plan with EU support}}, \bibinfo{year}{2021}. \URLprefix
  \url{https://irena.org/newsroom/articles/2021/Sep/IRENA-and-IAEA-Selected-to-Help-African-Union-Develop-Continental-Power-Master-Plan-with-EU-support}.
\bibitem[{Brown et~al.(2018)Brown, H\"orsch, and
  Schlachtberger}]{Brown2018PyPSA:Analysis}
\bibinfo{author}{T.~Brown}, \bibinfo{author}{J.~H\"orsch},
  \bibinfo{author}{D.~Schlachtberger},
\newblock \bibinfo{title}{{PyPSA: Python for Power System Analysis}},
\newblock \bibinfo{journal}{Journal of Open Research Software}
  \bibinfo{volume}{6} (\bibinfo{year}{2018}). \URLprefix
  \url{https://doi.org/10.5334/jors.188}. \DOIprefix\doi{10.5334/jors.188}.
  \href{http://arxiv.org/abs/1707.09913}{{\tt arXiv:1707.09913}}.
\bibitem[{{European Commission}(2020)}]{EC2020a}
\bibinfo{author}{{European Commission}}, \bibinfo{title}{{OPEN SOURCE SOFTWARE
  STRATEGY 2020 – 2023 - Think Open}}, \bibinfo{type}{Technical Report},
  \bibinfo{year}{2020}. \URLprefix
  \url{https://ec.europa.eu/info/sites/default/files/en_ec_open_source_strategy_2020-2023.pdf}.
\bibitem[{Groissb{\"{o}}ck(2019)}]{Groissbock2019AreUse}
\bibinfo{author}{M.~Groissb{\"{o}}ck},
\newblock \bibinfo{title}{{Are open source energy system optimization tools
  mature enough for serious use?}},
\newblock \bibinfo{journal}{Renewable and Sustainable Energy Reviews}
  \bibinfo{volume}{102} (\bibinfo{year}{2019}) \bibinfo{pages}{234--248}.
  \DOIprefix\doi{10.1016/j.rser.2018.11.020}.
\bibitem[{Kiviluoma(2022)}]{Juha2022}
\bibinfo{author}{J.~Kiviluoma}, \bibinfo{title}{Trends of open-sourcing models
  in the energy space}, \bibinfo{year}{2022}. \URLprefix
  \url{https://www.esig.energy/trends-of-open-sourcing-models-in-the-energy-space/}.
\bibitem[{Pfenninger et~al.(2018)Pfenninger, Hirth, Schlecht, Schmid, Wiese,
  Brown, Davis, Gidden, Heinrichs, Heuberger, Hilpert, Krien, Matke, Nebel,
  Morrison, M{\"{u}}ller, Ple{\ss}mann, Reeg, Richstein, Shivakumar, Staffell,
  Tr{\"{o}}ndle, and Wingenbach}]{Pfenninger2018OpeningLearned}
\bibinfo{author}{S.~Pfenninger}, \bibinfo{author}{L.~Hirth},
  \bibinfo{author}{I.~Schlecht}, \bibinfo{author}{E.~Schmid},
  \bibinfo{author}{F.~Wiese}, \bibinfo{author}{T.~Brown},
  \bibinfo{author}{C.~Davis}, \bibinfo{author}{M.~Gidden},
  \bibinfo{author}{H.~Heinrichs}, \bibinfo{author}{C.~Heuberger},
  \bibinfo{author}{S.~Hilpert}, \bibinfo{author}{U.~Krien},
  \bibinfo{author}{C.~Matke}, \bibinfo{author}{A.~Nebel},
  \bibinfo{author}{R.~Morrison}, \bibinfo{author}{B.~M{\"{u}}ller},
  \bibinfo{author}{G.~Ple{\ss}mann}, \bibinfo{author}{M.~Reeg},
  \bibinfo{author}{J.~C. Richstein}, \bibinfo{author}{A.~Shivakumar},
  \bibinfo{author}{I.~Staffell}, \bibinfo{author}{T.~Tr{\"{o}}ndle},
  \bibinfo{author}{C.~Wingenbach},
\newblock \bibinfo{title}{{Opening the black box of energy modelling:
  Strategies and lessons learned}},
\newblock \bibinfo{journal}{Energy Strategy Reviews} \bibinfo{volume}{19}
  (\bibinfo{year}{2018}) \bibinfo{pages}{63--71}.
  \DOIprefix\doi{10.1016/j.esr.2017.12.002}.
\bibitem[{Victoria et~al.(2020)Victoria, Zhu, Brown, Andresen, and
  Greiner}]{Victoria2020EarlyOff}
\bibinfo{author}{M.~Victoria}, \bibinfo{author}{K.~Zhu},
  \bibinfo{author}{T.~Brown}, \bibinfo{author}{G.~B. Andresen},
  \bibinfo{author}{M.~Greiner},
\newblock \bibinfo{title}{{Early decarbonisation of the European energy system
  pays off}},
\newblock \bibinfo{journal}{Nature Communications} \bibinfo{volume}{11}
  (\bibinfo{year}{2020}) \bibinfo{pages}{6223}. \URLprefix
  \url{http://www.nature.com/articles/s41467-020-20015-4}.
  \DOIprefix\doi{10.1038/s41467-020-20015-4}.
\bibitem[{Neumann et~al.(2022)Neumann, Hagenmeyer, and Brown}]{Neumann2022}
\bibinfo{author}{F.~Neumann}, \bibinfo{author}{V.~Hagenmeyer},
  \bibinfo{author}{T.~Brown},
\newblock \bibinfo{title}{Assessments of linear power flow and transmission
  loss approximations in coordinated capacity expansion problems},
\newblock \bibinfo{journal}{Applied Energy} \bibinfo{volume}{314}
  (\bibinfo{year}{2022}). \DOIprefix\doi{10.1016/j.apenergy.2022.118859}.
\bibitem[{Poncelet et~al.(2020)Poncelet, Delarue, and
  D'haeseleer}]{Poncelet2020UnitFlexibility}
\bibinfo{author}{K.~Poncelet}, \bibinfo{author}{E.~Delarue},
  \bibinfo{author}{W.~D'haeseleer},
\newblock \bibinfo{title}{{Unit commitment constraints in long-term planning
  models: Relevance, pitfalls and the role of assumptions on flexibility}},
\newblock \bibinfo{journal}{Applied Energy} \bibinfo{volume}{258}
  (\bibinfo{year}{2020}). \DOIprefix\doi{10.1016/j.apenergy.2019.113843}.
\bibitem[{Pfenninger and Pickering(2018)}]{Pfenninger2018}
\bibinfo{author}{S.~Pfenninger}, \bibinfo{author}{B.~Pickering},
\newblock \bibinfo{title}{Calliope: a multi-scale energy systems modelling
  framework},
\newblock \bibinfo{journal}{Journal of Open Source Software}
  \bibinfo{volume}{3} (\bibinfo{year}{2018}) \bibinfo{pages}{825}. \URLprefix
  \url{https://doi.org/10.21105/joss.00825}.
  \DOIprefix\doi{10.21105/joss.00825}.
\bibitem[{{Dispa-SET contributors}(2022)}]{Dispa-SET}
\bibinfo{author}{{Dispa-SET contributors}}, \bibinfo{title}{{Dispa-SET}},
  \bibinfo{year}{2022}. \URLprefix
  \url{https://github.com/energy-modelling-toolkit/Dispa-SET},
  \bibinfo{note}{accessed: 17/05/2022}.
\bibitem[{{Blue Marble Analytics}(2022)}]{gridpath}
\bibinfo{author}{{Blue Marble Analytics}}, \bibinfo{title}{Gridpath: Github
  repository}, \bibinfo{year}{2022}. \URLprefix
  \url{https://github.com/blue-marble/gridpath}, \bibinfo{note}{accessed:
  30/07/2022}.
\bibitem[{Heaps(2022)}]{LEAPmodel}
\bibinfo{author}{C.~Heaps}, \bibinfo{title}{{LEAP: The Low Emissions Analysis
  Platform}}, \bibinfo{year}{2022}. \URLprefix \url{https://leap.sei.org},
  \bibinfo{note}{accessed: 17/05/2022}.
\bibitem[{Institute(2022)}]{NEMOmodel}
\bibinfo{author}{S.~E. Institute}, \bibinfo{title}{{NEMO: Next Energy Modeling
  system for Optimization}}, \bibinfo{year}{2022}. \URLprefix
  \url{https://sei-international.github.io/NemoMod.jl/stable/},
  \bibinfo{note}{accessed: 17/05/2022}.
\bibitem[{Howells et~al.(2011)Howells, Rogner, Strachan, Heaps, Huntington,
  Kypreos, Hughes, Silveira, DeCarolis, Bazillian, and
  Roehrl}]{HOWELLS20115850}
\bibinfo{author}{M.~Howells}, \bibinfo{author}{H.~Rogner},
  \bibinfo{author}{N.~Strachan}, \bibinfo{author}{C.~Heaps},
  \bibinfo{author}{H.~Huntington}, \bibinfo{author}{S.~Kypreos},
  \bibinfo{author}{A.~Hughes}, \bibinfo{author}{S.~Silveira},
  \bibinfo{author}{J.~DeCarolis}, \bibinfo{author}{M.~Bazillian},
  \bibinfo{author}{A.~Roehrl},
\newblock \bibinfo{title}{Osemosys: The open source energy modeling system: An
  introduction to its ethos, structure and development},
\newblock \bibinfo{journal}{Energy Policy} \bibinfo{volume}{39}
  (\bibinfo{year}{2011}) \bibinfo{pages}{5850--5870}. \URLprefix
  \url{https://www.sciencedirect.com/science/article/pii/S0301421511004897}.
  \DOIprefix\doi{https://doi.org/10.1016/j.enpol.2011.06.033},
  \bibinfo{note}{sustainability of biofuels}.
\bibitem[{{Energy Exemplar}(2022)}]{PLEXOSmodel}
\bibinfo{author}{{Energy Exemplar}}, \bibinfo{title}{{PLEXOS}},
  \bibinfo{year}{2022}. \URLprefix \url{https://www.energyexemplar.com/plexos},
  \bibinfo{note}{accessed: 18/05/2022}.
\bibitem[{{International Renewable Energy Agency}(2022)}]{SPLAT-MESSAGEmodel}
\bibinfo{author}{{International Renewable Energy Agency}},
  \bibinfo{title}{{SPLAT-MESSAGE}}, \bibinfo{year}{2022}. \URLprefix
  \url{https://irena.org/energytransition/Energy-System-Models-and-Data/System-Planning-Test-Model},
  \bibinfo{note}{accessed: 17/05/2022}.
\bibitem[{Loulou et~al.(2021)Loulou, Goldstein, Kanudia, Lettila, and
  Remme}]{TIMESmodel}
\bibinfo{author}{R.~Loulou}, \bibinfo{author}{G.~Goldstein},
  \bibinfo{author}{A.~Kanudia}, \bibinfo{author}{A.~Lettila},
  \bibinfo{author}{U.~Remme}, \bibinfo{title}{{Documentation for the TIMES
  Model - PART I 1–78}}, \bibinfo{type}{Technical Report}, ETSAP,
  \bibinfo{year}{2021}. \URLprefix
  \url{https://iea-etsap.org/index.php/documentation}.
\bibitem[{Ringkj{\o}b et~al.(2018)Ringkj{\o}b, Haugan, and
  Solbrekke}]{Ringkjb2018ARenewables}
\bibinfo{author}{H.~K. Ringkj{\o}b}, \bibinfo{author}{P.~M. Haugan},
  \bibinfo{author}{I.~M. Solbrekke},
\newblock \bibinfo{title}{{A review of modelling tools for energy and
  electricity systems with large shares of variable renewables}},
\newblock \bibinfo{journal}{Renewable and Sustainable Energy Reviews}
  \bibinfo{volume}{96} (\bibinfo{year}{2018}) \bibinfo{pages}{440--459}.
  \DOIprefix\doi{10.1016/j.rser.2018.08.002}.
\bibitem[{H{\"{o}}rsch et~al.(2018)H{\"{o}}rsch, Hofmann, Schlachtberger, and
  Brown}]{Horsch2018PyPSA-Eur:System}
\bibinfo{author}{J.~H{\"{o}}rsch}, \bibinfo{author}{F.~Hofmann},
  \bibinfo{author}{D.~Schlachtberger}, \bibinfo{author}{T.~Brown},
\newblock \bibinfo{title}{{PyPSA-Eur: An open optimisation model of the
  European transmission system}},
\newblock \bibinfo{journal}{Energy Strategy Reviews} \bibinfo{volume}{22}
  (\bibinfo{year}{2018}) \bibinfo{pages}{207--215}.
  \DOIprefix\doi{10.1016/j.esr.2018.08.012}.
\bibitem[{{PyPSA-Eur}(2021)}]{PyPSA-Eur2021GithubPyPSA-Eur}
\bibinfo{author}{{PyPSA-Eur}}, \bibinfo{title}{{Github repository of
  PyPSA-Eur}}, \bibinfo{year}{2021}. \bibinfo{note}{Accessed: 15/08/2022}.
\bibitem[{Hörsch and Calitz(2017)}]{pypsa-za}
\bibinfo{author}{J.~Hörsch}, \bibinfo{author}{J.~Calitz},
\newblock \bibinfo{title}{{PyPSA-ZA: Investment and operation co-optimization
  of integrating wind and solar in South Africa at high spatial and temporal
  detail}}  (\bibinfo{year}{2017}). \URLprefix
  \url{http://arxiv.org/abs/1710.11199}.
\bibitem[{Schlott et~al.(2020)Schlott, Schyska, Viet, Phuong, Quan, Khanh,
  Hofmann, Bremen, Heinemann, and Kies}]{Schlott2020}
\bibinfo{author}{M.~Schlott}, \bibinfo{author}{B.~Schyska},
  \bibinfo{author}{D.~T. Viet}, \bibinfo{author}{V.~V. Phuong},
  \bibinfo{author}{D.~M. Quan}, \bibinfo{author}{M.~P. Khanh},
  \bibinfo{author}{F.~Hofmann}, \bibinfo{author}{L.~V. Bremen},
  \bibinfo{author}{D.~Heinemann}, \bibinfo{author}{A.~Kies},
\newblock \bibinfo{title}{Pypsa-vn: An open model of the vietnamese electricity
  system},
\newblock \bibinfo{publisher}{Institute of Electrical and Electronics Engineers
  Inc.}, \bibinfo{year}{2020}, pp. \bibinfo{pages}{253--258}.
  \DOIprefix\doi{10.1109/GTSD50082.2020.9303096}.
\bibitem[{Liu et~al.(2018)Liu, Brown, Andresen, Schlachtberger, and
  Greiner}]{Liu2018}
\bibinfo{author}{H.~Liu}, \bibinfo{author}{T.~Brown}, \bibinfo{author}{G.~B.
  Andresen}, \bibinfo{author}{D.~P. Schlachtberger},
  \bibinfo{author}{M.~Greiner},
\newblock \bibinfo{title}{The role of hydro power, storage and transmission in
  the decarbonization of the chinese power system}  (\bibinfo{year}{2018}).
  \URLprefix \url{http://arxiv.org/abs/1810.10347
  http://dx.doi.org/10.1016/j.apenergy.2019.02.009}.
  \DOIprefix\doi{10.1016/j.apenergy.2019.02.009}.
\bibitem[{{Calliope-Kenya}(2019)}]{Calliope-Kenya}
\bibinfo{author}{{Calliope-Kenya}}, \bibinfo{title}{{Github repository of
  Calliope-Kenya}}, \bibinfo{year}{2019}. \bibinfo{note}{Accessed: 15/08/2022}.
\bibitem[{Mattsson et~al.(2021)Mattsson, Verendel, Hedenus, and
  Reichenberg}]{Mattsson2021AnRegions}
\bibinfo{author}{N.~Mattsson}, \bibinfo{author}{V.~Verendel},
  \bibinfo{author}{F.~Hedenus}, \bibinfo{author}{L.~Reichenberg},
\newblock \bibinfo{title}{{An autopilot for energy models – Automatic
  generation of renewable supply curves, hourly capacity factors and hourly
  synthetic electricity demand for arbitrary world regions}},
\newblock \bibinfo{journal}{Energy Strategy Reviews} \bibinfo{volume}{33}
  (\bibinfo{year}{2021}). \DOIprefix\doi{10.1016/j.esr.2020.100606}.
\bibitem[{Reichenberg et~al.(2022)Reichenberg, Hedenus, Mattsson, and
  Verendel}]{Reichenberg2022}
\bibinfo{author}{L.~Reichenberg}, \bibinfo{author}{F.~Hedenus},
  \bibinfo{author}{N.~Mattsson}, \bibinfo{author}{V.~Verendel},
\newblock \bibinfo{title}{Deep decarbonization and the supergrid – prospects
  for electricity transmission between europe and china},
\newblock \bibinfo{journal}{Energy} \bibinfo{volume}{239}
  (\bibinfo{year}{2022}). \DOIprefix\doi{10.1016/j.energy.2021.122335}.
\bibitem[{Kiviluoma et~al.(2022)Kiviluoma, Pallonetto, Marin, Savolainen,
  Soininen, Vennström, Rinne, Huang, Kouveliotis-Lysikatos, Ihlemann, Delarue,
  O'Dwyer, O'Donnel, Amelin, Söder, and Dillon}]{Kiviluoma2022}
\bibinfo{author}{J.~Kiviluoma}, \bibinfo{author}{F.~Pallonetto},
  \bibinfo{author}{M.~Marin}, \bibinfo{author}{P.~T. Savolainen},
  \bibinfo{author}{A.~Soininen}, \bibinfo{author}{P.~Vennström},
  \bibinfo{author}{E.~Rinne}, \bibinfo{author}{J.~Huang},
  \bibinfo{author}{I.~Kouveliotis-Lysikatos}, \bibinfo{author}{M.~Ihlemann},
  \bibinfo{author}{E.~Delarue}, \bibinfo{author}{C.~O'Dwyer},
  \bibinfo{author}{T.~O'Donnel}, \bibinfo{author}{M.~Amelin},
  \bibinfo{author}{L.~Söder}, \bibinfo{author}{J.~Dillon},
\newblock \bibinfo{title}{Spine toolbox: A flexible open-source workflow
  management system with scenario and data management},
\newblock \bibinfo{journal}{SoftwareX} \bibinfo{volume}{17}
  (\bibinfo{year}{2022}). \DOIprefix\doi{10.1016/j.softx.2021.100967}.
\bibitem[{{GENeSYS-MOD}(2022)}]{genesys-mod-public}
\bibinfo{author}{{GENeSYS-MOD}}, \bibinfo{title}{{Github repository of
  GENeSYS-MOD}}, \bibinfo{year}{2022}. \URLprefix
  \url{https://git.tu-berlin.de/genesysmod/genesys-mod-public},
  \bibinfo{note}{accessed: 06/05/2022}.
\bibitem[{Barnes et~al.(2022)Barnes, Shivakumar, Brinkerink, and
  Niet}]{Barnes2022}
\bibinfo{author}{T.~Barnes}, \bibinfo{author}{A.~Shivakumar},
  \bibinfo{author}{M.~Brinkerink}, \bibinfo{author}{T.~Niet},
\newblock \bibinfo{title}{Osemosys global: An open-source, open data global
  electricity system model generator},
\newblock \bibinfo{journal}{Research Square}  (\bibinfo{year}{2022}).
\bibitem[{PyPSA(2022)}]{PyPSAdocumenation}
\bibinfo{author}{PyPSA}, \bibinfo{title}{Documentation: Users of pypsa with
  case studies}, \bibinfo{year}{2022}. \URLprefix
  \url{https://pypsa.readthedocs.io/en/latest/users.html},
  \bibinfo{note}{accessed: 01/07/2022}.
\bibitem[{PyPSA-Earth(2022)}]{pypsaafrica}
\bibinfo{author}{PyPSA-Earth}, \bibinfo{title}{Pypsa meets africa and pypsa
  meets earth}, \bibinfo{year}{2022}. \URLprefix
  \url{https://github.com/pypsa-meets-africa/pypsa-africa},
  \bibinfo{note}{accessed: 30/07/2022}.
\bibitem[{Brown et~al.(2018)Brown, Schlachtberger, Kies, Schramm, and
  Greiner}]{Brown2018SynergiesSystem}
\bibinfo{author}{T.~Brown}, \bibinfo{author}{D.~Schlachtberger},
  \bibinfo{author}{A.~Kies}, \bibinfo{author}{S.~Schramm},
  \bibinfo{author}{M.~Greiner},
\newblock \bibinfo{title}{{Synergies of sector coupling and transmission
  reinforcement in a cost-optimised, highly renewable European energy system}},
\newblock \bibinfo{journal}{Energy} \bibinfo{volume}{160}
  (\bibinfo{year}{2018}) \bibinfo{pages}{720--739}.
  \DOIprefix\doi{10.1016/j.energy.2018.06.222}.
\bibitem[{Zeyen et~al.(2022)Zeyen, Victoria, and Brown}]{ZeyenPathways2022}
\bibinfo{author}{E.~Zeyen}, \bibinfo{author}{M.~Victoria},
  \bibinfo{author}{T.~Brown}, \bibinfo{title}{Endogenous learning for green
  hydrogen in a sector-coupled energy model for europe}, \bibinfo{year}{2022}.
  \URLprefix \url{https://arxiv.org/abs/2205.11901}.
  \DOIprefix\doi{10.48550/ARXIV.2205.11901}.
\bibitem[{Glaum and Hofmann(2022)}]{Glaum2022}
\bibinfo{author}{P.~Glaum}, \bibinfo{author}{F.~Hofmann},
  \bibinfo{title}{Enhancing the german transmission grid through dynamic line
  rating}, \bibinfo{year}{2022}. \URLprefix
  \url{https://arxiv.org/abs/2208.04716}.
  \DOIprefix\doi{10.48550/ARXIV.2208.04716}.
\bibitem[{Neumann(2021)}]{Neumann2020CostsSystem}
\bibinfo{author}{F.~Neumann},
\newblock \bibinfo{title}{Costs of regional equity and autarky in a renewable
  european power system},
\newblock \bibinfo{journal}{Energy Strategy Reviews} \bibinfo{volume}{35}
  (\bibinfo{year}{2021}) \bibinfo{pages}{100652}. \URLprefix
  \url{https://www.sciencedirect.com/science/article/pii/S2211467X21000389}.
  \DOIprefix\doi{https://doi.org/10.1016/j.esr.2021.100652}.
\bibitem[{Neumann and Brown(2021)}]{Neumann2021BroadNear-Optimality}
\bibinfo{author}{F.~Neumann}, \bibinfo{author}{T.~Brown},
  \bibinfo{title}{{Broad Ranges of Investment Conngurations for Renewable Power
  Systems, Robust to Cost Uncertainty and Near-Optimality}},
  \bibinfo{type}{Technical Report}, \bibinfo{year}{2021}.
\bibitem[{Parzen et~al.(2022)Parzen, Hall, Jenkins, and
  Brown}]{Parzen2022OptimizationSolvers}
\bibinfo{author}{M.~Parzen}, \bibinfo{author}{J.~Hall},
  \bibinfo{author}{J.~Jenkins}, \bibinfo{author}{T.~Brown},
  \bibinfo{title}{Optimization solvers: the missing link for a fully
  open-source energy system modelling ecosystem.}, \bibinfo{type}{Technical
  Report}, \bibinfo{year}{2022}. \URLprefix
  \url{https://zenodo.org/record/6409433#.YlbOUVzMKEB}.
  \DOIprefix\doi{https://doi.org/10.5281/zenodo.6409433}.
\bibitem[{Steinmacher et~al.(2019)Steinmacher, Treude, and
  Gerosa}]{Steinmacher2019}
\bibinfo{author}{I.~Steinmacher}, \bibinfo{author}{C.~Treude},
  \bibinfo{author}{M.~A. Gerosa},
\newblock \bibinfo{title}{Let me in: Guidelines for the successful onboarding
  of newcomers to open source projects},
\newblock \bibinfo{journal}{IEEE Software} \bibinfo{volume}{36}
  (\bibinfo{year}{2019}) \bibinfo{pages}{41--49}.
  \DOIprefix\doi{10.1109/MS.2018.110162131}.
\bibitem[{Koster and Rahmann(2012)}]{Koster2012}
\bibinfo{author}{J.~Koster}, \bibinfo{author}{S.~Rahmann},
\newblock \bibinfo{title}{Snakemake--a scalable bioinformatics workflow
  engine},
\newblock \bibinfo{journal}{Bioinformatics} \bibinfo{volume}{28}
  (\bibinfo{year}{2012}) \bibinfo{pages}{2520--2522}.
  \DOIprefix\doi{10.1093/bioinformatics/bts480}.
\bibitem[{Mölder et~al.(2021)Mölder, Jablonski, Letcher, Hall, Tomkins-Tinch,
  Sochat, Forster, Lee, Twardziok, Kanitz, Wilm, Holtgrewe, Rahmann, Nahnsen,
  and Köster}]{Moelder2021}
\bibinfo{author}{F.~Mölder}, \bibinfo{author}{K.~P. Jablonski},
  \bibinfo{author}{B.~Letcher}, \bibinfo{author}{M.~B. Hall},
  \bibinfo{author}{C.~H. Tomkins-Tinch}, \bibinfo{author}{V.~Sochat},
  \bibinfo{author}{J.~Forster}, \bibinfo{author}{S.~Lee},
  \bibinfo{author}{S.~O. Twardziok}, \bibinfo{author}{A.~Kanitz},
  \bibinfo{author}{A.~Wilm}, \bibinfo{author}{M.~Holtgrewe},
  \bibinfo{author}{S.~Rahmann}, \bibinfo{author}{S.~Nahnsen},
  \bibinfo{author}{J.~Köster},
\newblock \bibinfo{title}{Sustainable data analysis with snakemake},
\newblock \bibinfo{journal}{F1000Research} \bibinfo{volume}{10}
  (\bibinfo{year}{2021}) \bibinfo{pages}{33}.
  \DOIprefix\doi{10.12688/f1000research.29032.2}.
\bibitem[{Medjroubi et~al.(2017)Medjroubi, Müller, Scharf, Matke, and
  Kleinhans}]{Medjroubi2014}
\bibinfo{author}{W.~Medjroubi}, \bibinfo{author}{U.~P. Müller},
  \bibinfo{author}{M.~Scharf}, \bibinfo{author}{C.~Matke},
  \bibinfo{author}{D.~Kleinhans},
\newblock \bibinfo{title}{Open data in power grid modelling: New approaches
  towards transparent grid models},
\newblock \bibinfo{journal}{Energy Reports} \bibinfo{volume}{3}
  (\bibinfo{year}{2017}) \bibinfo{pages}{14--21}.
  \DOIprefix\doi{https://doi.org/10.1016/j.egyr.2016.12.001}.
\bibitem[{Rivera et~al.(2018)Rivera, Nasirifard, Leimhofer, and
  Jacobsen}]{Rivera2018}
\bibinfo{author}{J.~Rivera}, \bibinfo{author}{P.~Nasirifard},
  \bibinfo{author}{J.~Leimhofer}, \bibinfo{author}{H.-A. Jacobsen},
\newblock \bibinfo{title}{Automatic generation of real power transmission grid
  models from crowdsourced data},
\newblock \bibinfo{journal}{IEEE Transactions on Smart Grid}
  \bibinfo{volume}{10} (\bibinfo{year}{2018}) \bibinfo{pages}{5436--5448}.
  \DOIprefix\doi{https://doi.org/10.1109/TSG.2018.2882840}.
\bibitem[{{OpenStreetMap contributors}(2022)}]{OpenStreetMap}
\bibinfo{author}{{OpenStreetMap contributors}}, \bibinfo{title}{{Planet dump
  retrieved}}, \bibinfo{year}{2022}. \URLprefix
  \url{https://www.openstreetmap.org}, \bibinfo{note}{accessed: 06/05/2022}.
\bibitem[{Pluta and Lünsdorf(2020)}]{esy-osm}
\bibinfo{author}{A.~Pluta}, \bibinfo{author}{O.~Lünsdorf},
\newblock \bibinfo{title}{esy-osmfilter – a python library to efficiently
  extract openstreetmap data},
\newblock \bibinfo{journal}{Journal of Open Research Software}
  \bibinfo{volume}{8} (\bibinfo{year}{2020}). \DOIprefix\doi{10.5334/jors.317}.
\bibitem[{{GADM}(2022)}]{gadm2022}
\bibinfo{author}{{GADM}}, \bibinfo{title}{Database of global administrative
  areas}, \bibinfo{year}{2022}. \URLprefix \url{https://gadm.org/},
  \bibinfo{note}{accessed: 15/08/2022}.
\bibitem[{Frysztacki et~al.(2021)Frysztacki, H{\"{o}}rsch, Hagenmeyer, and
  Brown}]{Frysztacki2021TheSolar}
\bibinfo{author}{M.~M. Frysztacki}, \bibinfo{author}{J.~H{\"{o}}rsch},
  \bibinfo{author}{V.~Hagenmeyer}, \bibinfo{author}{T.~Brown},
\newblock \bibinfo{title}{{The strong effect of network resolution on
  electricity system models with high shares of wind and solar}},
\newblock \bibinfo{journal}{Applied Energy} \bibinfo{volume}{291}
  (\bibinfo{year}{2021}). \DOIprefix\doi{10.1016/j.apenergy.2021.116726}.
\bibitem[{{GADM}(2019)}]{marineregions2019}
\bibinfo{author}{{GADM}}, \bibinfo{title}{Maritime boundaries v11},
  \bibinfo{year}{2019}. \URLprefix
  \url{https://www.marineregions.org/downloads.php}, \bibinfo{note}{accessed:
  15/08/2022}.
\bibitem[{Hampp(2022)}]{Synde2022}
\bibinfo{author}{J.~Hampp}, \bibinfo{title}{{Synde. An open source package to
  create synthetic demand time-series.}}, \bibinfo{year}{2022}. \URLprefix
  \url{https://github.com/euronion/synde}, \bibinfo{note}{accessed:
  07/06/2022}.
\bibitem[{Riahi et~al.(2017)Riahi, {van Vuuren}, Kriegler, Edmonds, O’Neill,
  Fujimori, Bauer, Calvin, Dellink, Fricko, Lutz, Popp, Cuaresma, KC, Leimbach,
  Jiang, Kram, Rao, Emmerling, Ebi, Hasegawa, Havlik, Humpenöder, {Da Silva},
  Smith, Stehfest, Bosetti, Eom, Gernaat, Masui, Rogelj, Strefler, Drouet,
  Krey, Luderer, Harmsen, Takahashi, Baumstark, Doelman, Kainuma, Klimont,
  Marangoni, Lotze-Campen, Obersteiner, Tabeau, and Tavoni}]{SSP2017}
\bibinfo{author}{K.~Riahi}, \bibinfo{author}{D.~P. {van Vuuren}},
  \bibinfo{author}{E.~Kriegler}, \bibinfo{author}{J.~Edmonds},
  \bibinfo{author}{B.~C. O’Neill}, \bibinfo{author}{S.~Fujimori},
  \bibinfo{author}{N.~Bauer}, \bibinfo{author}{K.~Calvin},
  \bibinfo{author}{R.~Dellink}, \bibinfo{author}{O.~Fricko},
  \bibinfo{author}{W.~Lutz}, \bibinfo{author}{A.~Popp}, \bibinfo{author}{J.~C.
  Cuaresma}, \bibinfo{author}{S.~KC}, \bibinfo{author}{M.~Leimbach},
  \bibinfo{author}{L.~Jiang}, \bibinfo{author}{T.~Kram},
  \bibinfo{author}{S.~Rao}, \bibinfo{author}{J.~Emmerling},
  \bibinfo{author}{K.~Ebi}, \bibinfo{author}{T.~Hasegawa},
  \bibinfo{author}{P.~Havlik}, \bibinfo{author}{F.~Humpenöder},
  \bibinfo{author}{L.~A. {Da Silva}}, \bibinfo{author}{S.~Smith},
  \bibinfo{author}{E.~Stehfest}, \bibinfo{author}{V.~Bosetti},
  \bibinfo{author}{J.~Eom}, \bibinfo{author}{D.~Gernaat},
  \bibinfo{author}{T.~Masui}, \bibinfo{author}{J.~Rogelj},
  \bibinfo{author}{J.~Strefler}, \bibinfo{author}{L.~Drouet},
  \bibinfo{author}{V.~Krey}, \bibinfo{author}{G.~Luderer},
  \bibinfo{author}{M.~Harmsen}, \bibinfo{author}{K.~Takahashi},
  \bibinfo{author}{L.~Baumstark}, \bibinfo{author}{J.~C. Doelman},
  \bibinfo{author}{M.~Kainuma}, \bibinfo{author}{Z.~Klimont},
  \bibinfo{author}{G.~Marangoni}, \bibinfo{author}{H.~Lotze-Campen},
  \bibinfo{author}{M.~Obersteiner}, \bibinfo{author}{A.~Tabeau},
  \bibinfo{author}{M.~Tavoni},
\newblock \bibinfo{title}{The shared socioeconomic pathways and their energy,
  land use, and greenhouse gas emissions implications: An overview},
\newblock \bibinfo{journal}{Global Environmental Change} \bibinfo{volume}{42}
  (\bibinfo{year}{2017}) \bibinfo{pages}{153--168}. \URLprefix
  \url{https://www.sciencedirect.com/science/article/pii/S0959378016300681}.
  \DOIprefix\doi{https://doi.org/10.1016/j.gloenvcha.2016.05.009}.
\bibitem[{Toktarova et~al.(2019)Toktarova, Gruber, Hlusiak, Bogdanov, and
  Breyer}]{TOKTAROVA2019160}
\bibinfo{author}{A.~Toktarova}, \bibinfo{author}{L.~Gruber},
  \bibinfo{author}{M.~Hlusiak}, \bibinfo{author}{D.~Bogdanov},
  \bibinfo{author}{C.~Breyer},
\newblock \bibinfo{title}{Long term load projection in high resolution for all
  countries globally},
\newblock \bibinfo{journal}{International Journal of Electrical Power and
  Energy Systems} \bibinfo{volume}{111} (\bibinfo{year}{2019})
  \bibinfo{pages}{160--181}. \URLprefix
  \url{https://www.sciencedirect.com/science/article/pii/S0142061518336196}.
  \DOIprefix\doi{https://doi.org/10.1016/j.ijepes.2019.03.055}.
\bibitem[{Hofmann et~al.(2022)Hofmann, Hampp, Neumann, and
  H{\"{o}}rsch}]{HofmannAtlite:Series}
\bibinfo{author}{F.~Hofmann}, \bibinfo{author}{J.~Hampp},
  \bibinfo{author}{F.~Neumann}, \bibinfo{author}{J.~H{\"{o}}rsch},
\newblock \bibinfo{title}{{atlite: A Lightweight Python Package for Calculating
  Renewable Power Potentials and Time Series}}  (\bibinfo{year}{2022}).
  \URLprefix \url{https://doi.org/10.21105/joss.03204}.
  \DOIprefix\doi{10.21105/joss.03204}.
\bibitem[{Hersbach et~al.(2020)Hersbach, Bell, Berrisford, Hirahara, Horányi,
  Muñoz‐Sabater, Nicolas, Peubey, Radu, Schepers, Simmons, Soci, Abdalla,
  Abellan, Balsamo, Bechtold, Biavati, Bidlot, Bonavita, Chiara, Dahlgren, Dee,
  Diamantakis, Dragani, Flemming, Forbes, Fuentes, Geer, Haimberger, Healy,
  Hogan, Hólm, Janisková, Keeley, Laloyaux, Lopez, Lupu, Radnoti, Rosnay,
  Rozum, Vamborg, Villaume, and Thépaut}]{Hersbach2020}
\bibinfo{author}{H.~Hersbach}, \bibinfo{author}{B.~Bell},
  \bibinfo{author}{P.~Berrisford}, \bibinfo{author}{S.~Hirahara},
  \bibinfo{author}{A.~Horányi}, \bibinfo{author}{J.~Muñoz‐Sabater},
  \bibinfo{author}{J.~Nicolas}, \bibinfo{author}{C.~Peubey},
  \bibinfo{author}{R.~Radu}, \bibinfo{author}{D.~Schepers},
  \bibinfo{author}{A.~Simmons}, \bibinfo{author}{C.~Soci},
  \bibinfo{author}{S.~Abdalla}, \bibinfo{author}{X.~Abellan},
  \bibinfo{author}{G.~Balsamo}, \bibinfo{author}{P.~Bechtold},
  \bibinfo{author}{G.~Biavati}, \bibinfo{author}{J.~Bidlot},
  \bibinfo{author}{M.~Bonavita}, \bibinfo{author}{G.~Chiara},
  \bibinfo{author}{P.~Dahlgren}, \bibinfo{author}{D.~Dee},
  \bibinfo{author}{M.~Diamantakis}, \bibinfo{author}{R.~Dragani},
  \bibinfo{author}{J.~Flemming}, \bibinfo{author}{R.~Forbes},
  \bibinfo{author}{M.~Fuentes}, \bibinfo{author}{A.~Geer},
  \bibinfo{author}{L.~Haimberger}, \bibinfo{author}{S.~Healy},
  \bibinfo{author}{R.~J. Hogan}, \bibinfo{author}{E.~Hólm},
  \bibinfo{author}{M.~Janisková}, \bibinfo{author}{S.~Keeley},
  \bibinfo{author}{P.~Laloyaux}, \bibinfo{author}{P.~Lopez},
  \bibinfo{author}{C.~Lupu}, \bibinfo{author}{G.~Radnoti},
  \bibinfo{author}{P.~Rosnay}, \bibinfo{author}{I.~Rozum},
  \bibinfo{author}{F.~Vamborg}, \bibinfo{author}{S.~Villaume},
  \bibinfo{author}{J.~Thépaut},
\newblock \bibinfo{title}{The era5 global reanalysis},
\newblock \bibinfo{journal}{Quarterly Journal of the Royal Meteorological
  Society} \bibinfo{volume}{146} (\bibinfo{year}{2020})
  \bibinfo{pages}{1999--2049}. \DOIprefix\doi{10.1002/qj.3803}.
\bibitem[{Pfeifroth et~al.(2017)Pfeifroth, Kothe, Müller, Trentmann, Hollmann,
  Fuchs, and Werscheck}]{Sarah2}
\bibinfo{author}{U.~Pfeifroth}, \bibinfo{author}{S.~Kothe},
  \bibinfo{author}{R.~Müller}, \bibinfo{author}{J.~Trentmann},
  \bibinfo{author}{R.~Hollmann}, \bibinfo{author}{P.~Fuchs},
  \bibinfo{author}{M.~Werscheck}, \bibinfo{title}{{Surface Radiation Data Set -
  Heliosat (SARAH) - Edition 2}}, \bibinfo{year}{2017}. \URLprefix
  \url{https://wui.cmsaf.eu/safira/action/viewDoiDetails?acronym=SARAH_V002}.
  \DOIprefix\doi{10.5676/EUM_SAF_CM/SARAH/V002}, \bibinfo{note}{accessed:
  27/07/2022}.
\bibitem[{{GEBCO Compilation Group}(2022)}]{GEBCO}
\bibinfo{author}{{GEBCO Compilation Group}}, \bibinfo{title}{{GEBCO 2022
  Grid}}, \bibinfo{year}{2022}. \URLprefix
  \url{https://www.gebco.net/data_and_products/gridded_bathymetry_data/}.
  \DOIprefix\doi{10.5285/e0f0bb80-ab44-2739-e053-6c86abc0289c},
  \bibinfo{note}{accessed: 27/07/2022}.
\bibitem[{{Energy Information Administration}(2022)}]{EIA2022}
\bibinfo{author}{{Energy Information Administration}},
  \bibinfo{title}{International - u.s. energy information administration
  (eia)}, \bibinfo{year}{2022}. \URLprefix
  \url{https://www.eia.gov/international/overview/world},
  \bibinfo{note}{accessed: 15/08/2022}.
\bibitem[{Gotzens et~al.(2019)Gotzens, Heinrichs, H{\"{o}}rsch, and
  Hofmann}]{Gotzens2019PerformingDatabases}
\bibinfo{author}{F.~Gotzens}, \bibinfo{author}{H.~Heinrichs},
  \bibinfo{author}{J.~H{\"{o}}rsch}, \bibinfo{author}{F.~Hofmann},
\newblock \bibinfo{title}{{Performing energy modelling exercises in a
  transparent way - The issue of data quality in power plant databases}},
\newblock \bibinfo{journal}{Energy Strategy Reviews} \bibinfo{volume}{23}
  (\bibinfo{year}{2019}) \bibinfo{pages}{1--12}.
  \DOIprefix\doi{10.1016/j.esr.2018.11.004}.
\bibitem[{{Center For Global Development}(2012)}]{CFGD2012}
\bibinfo{author}{{Center For Global Development}}, \bibinfo{title}{Carbon
  monitoring for action | center for global development | ideas to action},
  \bibinfo{year}{2012}. \URLprefix
  \url{https://www.cgdev.org/topics/carbon-monitoring-action},
  \bibinfo{note}{accessed: 15/08/2022}.
\bibitem[{ENTSO-E(2022)}]{ENTSOE2022}
\bibinfo{author}{ENTSO-E}, \bibinfo{title}{Entso-e - home},
  \bibinfo{year}{2022}. \URLprefix \url{https://www.entsoe.eu/}.
\bibitem[{{Sandia National Lab}(2021)}]{Sandia2021}
\bibinfo{author}{{Sandia National Lab}}, \bibinfo{title}{{DOE Global Energy
  Storage Database}}, \bibinfo{year}{2021}. \URLprefix
  \url{https://sandia.gov/ess-ssl/gesdb/public/}, \bibinfo{note}{accessed:
  15/08/2022}.
\bibitem[{Observatory(2018)}]{GEO2018}
\bibinfo{author}{G.~E. Observatory}, \bibinfo{title}{Power plants data},
  \bibinfo{year}{2018}. \URLprefix
  \url{http://globalenergyobservatory.org/docs/HelpGeoPower.php}.
\bibitem[{{Open Power System Data}(2020)}]{OSPD2020}
\bibinfo{author}{{Open Power System Data}}, \bibinfo{title}{Data package
  conventional power plants. version 2020-10-01.}, \bibinfo{year}{2020}.
  \URLprefix
  \url{https://data.open-power-system-data.org/conventional_power_plants}.
  \DOIprefix\doi{https://doi.org/10.25832/conventional_power_plants/2020-10-01},
  \bibinfo{note}{accessed: 15/08/2022}.
\bibitem[{{World Resources Institute}(2021)}]{WRI2021}
\bibinfo{author}{{World Resources Institute}}, \bibinfo{title}{{Global Power
  Plant Database - Datasets - Data}}, \bibinfo{year}{2021}. \URLprefix
  \url{https://datasets.wri.org/dataset/globalpowerplantdatabase},
  \bibinfo{note}{accessed: 15/08/2022}.
\bibitem[{{Joint Research Center}(2019)}]{JRC2019}
\bibinfo{author}{{Joint Research Center}}, \bibinfo{title}{{JRC Hydro-power
  database}}, \bibinfo{year}{2019}. \URLprefix
  \url{https://data.europa.eu/data/datasets/52b00441-d3e0-44e0-8281-fda86a63546d?locale=en},
  \bibinfo{note}{accessed: 15/08/2022}.
\bibitem[{{IRENA}(2022)}]{IRENA2022}
\bibinfo{author}{{IRENA}}, \bibinfo{title}{Data and statistics},
  \bibinfo{year}{2022}. \URLprefix \url{https://www.irena.org/statistics},
  \bibinfo{note}{accessed: 15/08/2022}.
\bibitem[{Powerplantmatching(2019)}]{Powerplantmatching2019}
\bibinfo{author}{Powerplantmatching}, \bibinfo{title}{Adding open street map
  (osm) as a source · issue \#12 · fresna/powerplantmatching},
  \bibinfo{year}{2019}. \URLprefix
  \url{https://github.com/FRESNA/powerplantmatching/issues/12},
  \bibinfo{note}{accessed: 15/08/2022}.
\bibitem[{H{\"{o}}rsch et~al.(2020)H{\"{o}}rsch, Neumann, Hofmann,
  Schlachtberger, and Brown}]{Horsch2020PyPSA-Eur:Code}
\bibinfo{author}{J.~H{\"{o}}rsch}, \bibinfo{author}{F.~Neumann},
  \bibinfo{author}{F.~Hofmann}, \bibinfo{author}{D.~Schlachtberger},
  \bibinfo{author}{T.~Brown},
\newblock \bibinfo{title}{{PyPSA-Eur: An Open Optimisation Model of the
  European Transmission System (Code)}}  (\bibinfo{year}{2020}). \URLprefix
  \url{https://doi.org/10.5281/zenodo.3885701#.XzKzdXCPvKE.mendeley}.
  \DOIprefix\doi{10.5281/ZENODO.3885701}.
\bibitem[{Fry(2022)}]{Frystacki2022comparison}
\bibinfo{title}{{A comparison of clustering methods for the spatial reduction
  of renewable electricity optimisation models of Europe}},
\newblock \bibinfo{journal}{Energy Informatics} \bibinfo{volume}{5}
  (\bibinfo{year}{2022}). \DOIprefix\doi{10.1186/s42162-022-00187-7}.
\bibitem[{Siala and Mahfouz(2019)}]{SIALA201975}
\bibinfo{author}{K.~Siala}, \bibinfo{author}{M.~Y. Mahfouz},
\newblock \bibinfo{title}{{Impact of the choice of regions on energy system
  models}},
\newblock \bibinfo{journal}{Energy Strategy Reviews} \bibinfo{volume}{25}
  (\bibinfo{year}{2019}) \bibinfo{pages}{75--85}. \URLprefix
  \url{https://www.sciencedirect.com/science/article/pii/S2211467X19300495}.
  \DOIprefix\doi{https://doi.org/10.1016/j.esr.2019.100362}.
\bibitem[{Kueppers et~al.(2020)Kueppers, Perau, Franken, Heger, Huber, Metzger,
  and Niessen}]{Kueppers2020Data-DrivenOptimization}
\bibinfo{author}{M.~Kueppers}, \bibinfo{author}{C.~Perau},
  \bibinfo{author}{M.~Franken}, \bibinfo{author}{H.~J. Heger},
  \bibinfo{author}{M.~Huber}, \bibinfo{author}{M.~Metzger},
  \bibinfo{author}{S.~Niessen},
\newblock \bibinfo{title}{{Data-Driven Regionalization of Decarbonized Energy
  Systems for Reflecting Their Changing Topologies in Planning and
  Optimization}},
\newblock \bibinfo{journal}{Energies} \bibinfo{volume}{13}
  (\bibinfo{year}{2020}) \bibinfo{pages}{4076}.
  \DOIprefix\doi{10.3390/en13164076}.
\bibitem[{Biener and {Garcia Rosas}(2020)}]{BIENER2020106349}
\bibinfo{author}{W.~Biener}, \bibinfo{author}{K.~R. {Garcia Rosas}},
\newblock \bibinfo{title}{{Grid reduction for energy system analysis}},
\newblock \bibinfo{journal}{Electric Power Systems Research}
  \bibinfo{volume}{185} (\bibinfo{year}{2020}) \bibinfo{pages}{106349}.
  \URLprefix
  \url{https://www.sciencedirect.com/science/article/pii/S0378779620301553}.
  \DOIprefix\doi{https://doi.org/10.1016/j.epsr.2020.106349}.
\bibitem[{Fioriti et~al.(2021)Fioriti, Lutzemberger, Poli, Duenas-Martinez, and
  Micangeli}]{Fioriti2021CouplingMicrogrid}
\bibinfo{author}{D.~Fioriti}, \bibinfo{author}{G.~Lutzemberger},
  \bibinfo{author}{D.~Poli}, \bibinfo{author}{P.~Duenas-Martinez},
  \bibinfo{author}{A.~Micangeli},
\newblock \bibinfo{title}{{Coupling economic multi-objective optimization and
  multiple design options: A business-oriented approach to size an off-grid
  hybrid microgrid}},
\newblock \bibinfo{journal}{International Journal of Electrical Power and
  Energy Systems} \bibinfo{volume}{127} (\bibinfo{year}{2021}).
  \DOIprefix\doi{10.1016/j.ijepes.2020.106686}.
\bibitem[{Oliphant(2007)}]{PythonOliphant}
\bibinfo{author}{T.~E. Oliphant},
\newblock \bibinfo{title}{Python for scientific computing},
\newblock \bibinfo{journal}{Computing in Science and Engineering}
  \bibinfo{volume}{9} (\bibinfo{year}{2007}) \bibinfo{pages}{10--20}.
  \DOIprefix\doi{10.1109/MCSE.2007.58}.
\bibitem[{Linopy(2022)}]{Linopy}
\bibinfo{author}{Linopy}, \bibinfo{title}{Benchmark of linopy: A package doing
  linear optimization with n-d labeled arrays in python}, \bibinfo{year}{2022}.
  \URLprefix \url{https://linopy.readthedocs.io/en/latest/benchmark.html},
  \bibinfo{note}{accessed: 01/07/2022}.
\bibitem[{Consortium(2022)}]{GpstBenchmark}
\bibinfo{author}{G.~P. S.~T. Consortium}, \bibinfo{title}{List of open-source
  tools for energy system modelling}, \bibinfo{year}{2022}. \URLprefix
  \url{https://g-pst.github.io/tools/}, \bibinfo{note}{accessed: 19/08/2022}.
\bibitem[{{ENTSO-E}(2020)}]{ENTSO-E2020ENTSO-EPlatform}
\bibinfo{author}{{ENTSO-E}}, \bibinfo{title}{{ENTSO-E Transparency Platform}},
  \bibinfo{year}{2020}. \URLprefix
  \url{https://transparency.entsoe.eu/dashboard/show}.
\bibitem[{{World Bank Group}(2022)}]{worldbank-network}
\bibinfo{author}{{World Bank Group}}, \bibinfo{title}{{Africa - Electricity
  Transmission and Distribution Grid Map }}, \bibinfo{year}{2022}. \URLprefix
  \url{https://energydata.info/dataset/africa-electricity-transmission-and-distribution-grid-map-2017},
  \bibinfo{note}{accessed: 06/05/2022}.
\bibitem[{{Nigerian Electricity Regulatory Commission}(2022)}]{NigerianTSO}
\bibinfo{author}{{Nigerian Electricity Regulatory Commission}},
  \bibinfo{title}{{Nigerian Transmission System Information}},
  \bibinfo{year}{2022}. \URLprefix
  \url{https://nerc.gov.ng/index.php/home/nesi/404-transmission},
  \bibinfo{note}{accessed: 06/05/2022}.
\bibitem[{Ritchie et~al.(2022)Ritchie, Rosado, Mathieu, and
  Roser}]{OurWorldInDataDemand}
\bibinfo{author}{H.~Ritchie}, \bibinfo{author}{P.~Rosado},
  \bibinfo{author}{E.~Mathieu}, \bibinfo{author}{M.~Roser},
  \bibinfo{title}{{Data on Energy by Our World in Data}}, \bibinfo{year}{2022}.
  \URLprefix \url{https://github.com/owid/energy-data/tree/master},
  \bibinfo{note}{accessed: 20/05/2022}.
\bibitem[{IRENA(2015)}]{IRENA}
\bibinfo{author}{IRENA}, \bibinfo{title}{{Africa 2030: Roadmap for a Renewable
  Energy Future.}}, \bibinfo{type}{Technical Report}, IRENA,
  \bibinfo{year}{2015}. \URLprefix \url{www.irena.org/remap}.
\bibitem[{Alova et~al.(2021)Alova, Trotter, and Money}]{Alova2021}
\bibinfo{author}{G.~Alova}, \bibinfo{author}{P.~A. Trotter},
  \bibinfo{author}{A.~Money},
\newblock \bibinfo{title}{A machine-learning approach to predicting africa’s
  electricity mix based on planned power plants and their chances of success},
\newblock \bibinfo{journal}{Nature Energy} \bibinfo{volume}{6}
  (\bibinfo{year}{2021}) \bibinfo{pages}{158--166}.
  \DOIprefix\doi{10.1038/s41560-020-00755-9}.
\bibitem[{Council(2022)}]{WindpotentialIFC}
\bibinfo{author}{G.~W.~E. Council}, \bibinfo{title}{{Exploring Africa's
  Untapped Wind Potential}}, \bibinfo{year}{2022}. \URLprefix
  \url{https://gwec.net/wp-content/uploads/2021/04/IFC-Africa-Wind-Technical-Potential-Oct-2020-1.pdf},
  \bibinfo{note}{accessed: 27/08/2022}.
\bibitem[{IRENA(2014)}]{IRENA2014}
\bibinfo{author}{IRENA}, \bibinfo{title}{Estimating the Renewable Energy
  Potential in Africa: A GIS-based approach}, \bibinfo{type}{Technical Report},
  \bibinfo{year}{2014}. \URLprefix
  \url{https://www.irena.org/publications/2014/Aug/Estimating-the-Renewable-Energy-Potential-in-Africa-A-GIS-based-approach},
  \bibinfo{note}{accessed: 20/05/2022}.
\bibitem[{Bolinger and Bolinger(2022)}]{Bolinger2022}
\bibinfo{author}{M.~Bolinger}, \bibinfo{author}{G.~Bolinger},
\newblock \bibinfo{title}{Land requirements for utility-scale pv: An empirical
  update on power and energy density},
\newblock \bibinfo{journal}{IEEE Journal of Photovoltaics} \bibinfo{volume}{12}
  (\bibinfo{year}{2022}) \bibinfo{pages}{589--594}.
  \DOIprefix\doi{10.1109/JPHOTOV.2021.3136805}.
\bibitem[{Wikipedia(2022)}]{WikipediaSolar}
\bibinfo{author}{Wikipedia}, \bibinfo{title}{{List of photovoltaic power
  stations}}, \bibinfo{year}{2022}. \URLprefix
  \url{https://en.wikipedia.org/wiki/List_of_photovoltaic_power_stations},
  \bibinfo{note}{accessed: 04/06/2022}.
\bibitem[{authors(2022)}]{GithubWindList}
\bibinfo{author}{P.-E. authors}, \bibinfo{title}{{Validation of the solar and
  wind potential}}, \bibinfo{year}{2022}. \URLprefix
  \url{https://github.com/pypsa-meets-africa/pypsa-africa/blob/3f01c936850ceaa48774ea21a37a9e0bb6462d92/notebooks/validation/renewable_potential_validation.ipynb},
  \bibinfo{note}{accessed: 27/08/2022}.
\bibitem[{Reich et~al.(2004)Reich, Numbem, Almaraz, and Eswaran}]{Reich2004}
\bibinfo{author}{P.~Reich}, \bibinfo{author}{S.~Numbem},
  \bibinfo{author}{R.~Almaraz}, \bibinfo{author}{H.~Eswaran},
\newblock \bibinfo{title}{Land resource stresses and desertification in
  africa},
\newblock \bibinfo{journal}{Agro-Science} \bibinfo{volume}{2}
  (\bibinfo{year}{2004}). \DOIprefix\doi{10.4314/as.v2i2.1484}.
\bibitem[{IRENA(2022)}]{IRENAPxWeb2022}
\bibinfo{author}{IRENA}, \bibinfo{title}{Installed electricity capacity (mw) by
  country/area, grid connection, technology and year. pxweb},
  \bibinfo{year}{2022}. \URLprefix
  \url{https://pxweb.irena.org/pxweb/en/IRENASTAT/IRENASTAT__Power\%20Capacity\%20and\%20Generation/ELECCAP_2022_cycle1.px/},
  \bibinfo{note}{accessed: 15/08/2022}.
\bibitem[{{USAID}(2022)}]{Usaid2022}
\bibinfo{author}{{USAID}}, \bibinfo{title}{{Nigeria -- POWER AFRICA FACT
  SHEET}}, \bibinfo{year}{2022}. \URLprefix
  \url{https://www.usaid.gov/powerafrica/nigeria}.
\bibitem[{Monitor(2022)}]{GlobalEnergyMonitor}
\bibinfo{author}{G.~E. Monitor}, \bibinfo{title}{{Global Energy Monitor Wiki}},
  \bibinfo{year}{2022}. \URLprefix \url{https://www.gem.wiki/Main_Page},
  \bibinfo{note}{accessed: 28/08/2022}.
\bibitem[{Analytics and Institute(2022)}]{ClimateActionTracker}
\bibinfo{author}{C.~Analytics}, \bibinfo{author}{N.~C. Institute},
  \bibinfo{title}{Climate action tracker. report on 11.02.2022 for nigeria},
  \bibinfo{year}{2022}. \URLprefix
  \url{https://climateactiontracker.org/countries/nigeria/},
  \bibinfo{note}{accessed: 15/07/2022}.
\bibitem[{Ritchie et~al.(2022)Ritchie, Rosado, Mathieu, and
  Roser}]{OurWorldInDataGeneration}
\bibinfo{author}{H.~Ritchie}, \bibinfo{author}{P.~Rosado},
  \bibinfo{author}{E.~Mathieu}, \bibinfo{author}{M.~Roser},
  \bibinfo{title}{{Data on Energy by Our World in Data}}, \bibinfo{year}{2022}.
  \URLprefix
  \url{https://ourworldindata.org/grapher/share-electricity-source-facet?country=~NGA},
  \bibinfo{note}{accessed: 19/08/2022}.
\bibitem[{Group(2017)}]{nesg2017}
\bibinfo{author}{N.~E.~S. Group}, \bibinfo{title}{Comparison of Costs of
  Electricity Generation in Nigeria}, \bibinfo{type}{Technical Report},
  \bibinfo{year}{2017}.
\bibitem[{Kim et~al.(2021)Kim, Abdel-Hameed, Joseph, Ramadhan, Nandutu, and
  Hyun}]{Kim2021}
\bibinfo{author}{J.~Kim}, \bibinfo{author}{A.~Abdel-Hameed},
  \bibinfo{author}{S.~R. Joseph}, \bibinfo{author}{H.~H. Ramadhan},
  \bibinfo{author}{M.~Nandutu}, \bibinfo{author}{J.~H. Hyun},
\newblock \bibinfo{title}{Modeling long-term electricity generation planning to
  reduce carbon dioxide emissions in nigeria},
\newblock \bibinfo{journal}{Energies} \bibinfo{volume}{14}
  (\bibinfo{year}{2021}). \DOIprefix\doi{10.3390/en14196258}.
\bibitem[{Parzen et~al.(2021)Parzen, Neumann, Van Der~Weijde, Friedrich, and
  Kiprakis}]{Parzen2021BeyondSystems}
\bibinfo{author}{M.~Parzen}, \bibinfo{author}{F.~Neumann},
  \bibinfo{author}{A.~H. Van Der~Weijde}, \bibinfo{author}{D.~Friedrich},
  \bibinfo{author}{A.~Kiprakis},
\newblock \bibinfo{title}{{Beyond cost reduction: Improving the value of energy
  storage in electricity systems}}  (\bibinfo{year}{2021}). \URLprefix
  \url{https://arxiv.org/abs/2101.10092}.
\bibitem[{{Hamisu Umar} et~al.(2021){Hamisu Umar}, Bora, Banerjee, Gupta, and
  Anjum}]{Umar2021}
\bibinfo{author}{N.~{Hamisu Umar}}, \bibinfo{author}{B.~Bora},
  \bibinfo{author}{C.~Banerjee}, \bibinfo{author}{P.~Gupta},
  \bibinfo{author}{N.~Anjum},
\newblock \bibinfo{title}{Performance and economic viability of the pv system
  in different climatic zones of nigeria},
\newblock \bibinfo{journal}{Sustainable Energy Technologies and Assessments}
  \bibinfo{volume}{43} (\bibinfo{year}{2021}) \bibinfo{pages}{100987}.
  \DOIprefix\doi{https://doi.org/10.1016/j.seta.2020.100987}.
\bibitem[{lisazeyen et~al.(2022)lisazeyen, euronion, Neumann, Brown, martavp,
  and lukasnacken}]{lisazeyen_2022_6885392}
\bibinfo{author}{lisazeyen}, \bibinfo{author}{euronion},
  \bibinfo{author}{F.~Neumann}, \bibinfo{author}{T.~Brown},
  \bibinfo{author}{martavp}, \bibinfo{author}{lukasnacken},
  \bibinfo{title}{Pypsa/technology-data: Technology data v0.4.0},
  \bibinfo{year}{2022}. \URLprefix
  \url{https://doi.org/10.5281/zenodo.6885392}.
  \DOIprefix\doi{10.5281/zenodo.6885392}.
\bibitem[{Wronkiewicz(2018)}]{devseed2017}
\bibinfo{author}{M.~Wronkiewicz}, \bibinfo{title}{Mapping the electricity grid
  from space}, \bibinfo{year}{2018}. \URLprefix
  \url{http://devseed.com/ml-grid-docs/}, \bibinfo{note}{accessed: 15/08/2022}.
\bibitem[{Arderne et~al.(2020)Arderne, Zorn, Nicolas, and
  Koks}]{Arderne2020PredictiveData}
\bibinfo{author}{C.~Arderne}, \bibinfo{author}{C.~Zorn},
  \bibinfo{author}{C.~Nicolas}, \bibinfo{author}{E.~E. Koks},
\newblock \bibinfo{title}{Predictive mapping of the global power system using
  open data},
\newblock \bibinfo{journal}{Scientific Data} \bibinfo{volume}{7}
  (\bibinfo{year}{2020}) \bibinfo{pages}{19}. \URLprefix
  \url{https://doi.org/10.1038/s41597-019-0347-4}.
  \DOIprefix\doi{10.1038/s41597-019-0347-4}.
\bibitem[{Huang et~al.(2022)Huang, Yang, Streltsov, Bradbury, Collins, and
  Malof}]{huang2021gridtracer}
\bibinfo{author}{B.~Huang}, \bibinfo{author}{J.~Yang},
  \bibinfo{author}{A.~Streltsov}, \bibinfo{author}{K.~Bradbury},
  \bibinfo{author}{L.~M. Collins}, \bibinfo{author}{J.~M. Malof},
\newblock \bibinfo{title}{Gridtracer: Automatic mapping of power grids using
  deep learning and overhead imagery},
\newblock \bibinfo{journal}{IEEE Journal of Selected Topics in Applied Earth
  Observations and Remote Sensing} \bibinfo{volume}{15} (\bibinfo{year}{2022})
  \bibinfo{pages}{4956--4970}. \DOIprefix\doi{10.1109/JSTARS.2021.3124519}.
\bibitem[{Talukdar et~al.(2018)Talukdar, Gupta, Rajpura, and
  Hegde}]{talukdar2018transfer}
\bibinfo{author}{J.~Talukdar}, \bibinfo{author}{S.~Gupta},
  \bibinfo{author}{P.~Rajpura}, \bibinfo{author}{R.~S. Hegde},
\newblock \bibinfo{title}{Transfer learning for object detection using
  state-of-the-art deep neural networks},
\newblock in: \bibinfo{booktitle}{2018 5th International Conference on Signal
  Processing and Integrated Networks (SPIN)}, \bibinfo{organization}{IEEE},
  \bibinfo{year}{2018}, pp. \bibinfo{pages}{78--83}.
\bibitem[{Deng et~al.(2021)Deng, Li, Chen, and Duan}]{deng2021unbiased}
\bibinfo{author}{J.~Deng}, \bibinfo{author}{W.~Li}, \bibinfo{author}{Y.~Chen},
  \bibinfo{author}{L.~Duan},
\newblock \bibinfo{title}{Unbiased mean teacher for cross-domain object
  detection},
\newblock in: \bibinfo{booktitle}{Proceedings of the IEEE/CVF Conference on
  Computer Vision and Pattern Recognition}, \bibinfo{year}{2021}, pp.
  \bibinfo{pages}{4091--4101}.
\bibitem[{{Maxar Technologies}(2010)}]{maxar}
\bibinfo{author}{{Maxar Technologies}}, \bibinfo{title}{Open data program},
  \bibinfo{year}{2010}. \bibinfo{note}{Accessed: 15/08/2022}.
\bibitem[{Ritchie(2021)}]{Ritchie2021}
\bibinfo{author}{H.~Ritchie},
\newblock \bibinfo{title}{Covid’s lessons for climate, sustainability and
  more from our world in data},
\newblock \bibinfo{journal}{Nature} \bibinfo{volume}{598}
  (\bibinfo{year}{2021}) \bibinfo{pages}{9--9}.
  \DOIprefix\doi{10.1038/d41586-021-02691-4}.
\bibitem[{Poli et~al.(2019)Poli, Pelacchi, Lutzemberger, Baffa~Scirocco, Bassi,
  and Bruno}]{Poli2019}
\bibinfo{author}{D.~Poli}, \bibinfo{author}{P.~Pelacchi},
  \bibinfo{author}{G.~Lutzemberger}, \bibinfo{author}{T.~Baffa~Scirocco},
  \bibinfo{author}{F.~Bassi}, \bibinfo{author}{G.~Bruno},
\newblock \bibinfo{title}{The possible impact of weather uncertainty on the
  dynamic thermal rating of transmission power lines: A monte carlo error-based
  approach},
\newblock \bibinfo{journal}{Electric Power Systems Research}
  \bibinfo{volume}{170} (\bibinfo{year}{2019}) \bibinfo{pages}{338--347}.
  \URLprefix \url{www.scopus.com}, \bibinfo{note}{cited By :17}.

\end{thebibliography}







\end{document}